\documentclass[reprint,amsmath,amssymb,aps,prb,superscriptaddress,longbibliography]{revtex4-1}
\usepackage{array}
\usepackage{bm}
\usepackage{color}
\usepackage{float}
\usepackage{graphicx}
\usepackage[colorlinks,citecolor=blue]{hyperref}
\usepackage{natbib}
\usepackage{newtxtext}
\usepackage{newtxmath}
\usepackage[caption=false]{subfig}
\usepackage{mathrsfs}

\usepackage{amsmath,amssymb}
\usepackage{verbatim}
\usepackage{amsfonts}
\usepackage{dcolumn}
\usepackage{bm}
\usepackage{color}

\newcommand{\beq}{\begin{eqnarray} }
\newcommand{\eeq}{\end{eqnarray} }
\newcommand{\Beq}{\begin{eqnarray*} }
\newcommand{\Eeq}{\end{eqnarray*} }
\newcommand{\Bmat}{\left(\begin{matrix}}
\newcommand{\Emat}{\end{matrix}\right)}
\newcommand{\up}{\uparrow}
\newcommand{\dn}{\downarrow}

\begin{document}

\title{Symmetry-protected gapless spin liquids on the strained honeycomb lattice}
\author{Jiucai Wang}
\affiliation{Department of Physics, Renmin University of China, Beijing 100872, China}

\author{Zheng-Xin Liu}
\email{liuzxphys@ruc.edu.cn}
\affiliation{Department of Physics, Renmin University of China, Beijing 100872, China}
\affiliation{Tsung-Dao Lee Institute \& School of Physics and Astronomy, Shanghai Jiao Tong University, Shanghai 200240, China}

\date{\today}

\begin{abstract}
By including a material-relevant off-diagonal interaction called the $\Gamma$ term into the Kitaev model and introducing spatial anisotropy in the interaction strength on the honeycomb lattice, we obtain a series of nodal Z$_2$ quantum spin liquids (QSLs) from parton approach. These QSLs share the same projective symmetry group and are characterized by certain numbers of symmetry-protected Majorana cones in their low-energy excitation spectrum. We illustrate that the physical properties of the QSLs are dependent on the information of the cones. Using the $\pmb k\cdot\pmb p$ method, we analyze the chirality of every cone with respect to mass generating perturbations. Especially, for an applied external magnetic field, we provide the maximum-mass field-orientation for every cone. Thus, for arbitrarily oriented weak magnetic fields, we can immediately read out the Chern number of the system and the properties of the resultant chiral spin liquids.  The new gapless QSLs predicted in our phase diagrams are promising to be realized experimentally by exerting uniaxial pressure to tune the anisotropy of the interaction strength.  We further show that all these QSLs can be distinguished by measurable quantities. Based on the study of these QSL phases, we conclude that a complete classification of nodal QSLs with certain symmetry should include not only the projective symmetry groups but also the information of the cones, {\it i.e.}, their total number, their chiralities, and the way in which they are symmetry-related.

\end{abstract}


\maketitle

\section{Introduction}

Quantum spin liquids (QSLs) are exotic phases of matter exhibiting no conventional long-range order down to the lowest temperatures. Resulting from strong quantum fluctuations, QSLs are characterized by long-range entanglement and the existence of intrinsic fractional  excitations called anyons. The low-energy physics of QSLs is beyond the Ginzburg-Landau-Wilson paradigm\cite{Balents,zhouyi} and is instead captured by topological quantum field theory or tensor category theory\cite{XGWen,Kitaev}. QSLs with full spin-rotation symmetry are also called resonating-valence-bond states where the spins pair up with each other and form singlets (like the electrons in superconductors except that the spins in QSLs cannot move)\cite{Anderson}. Anyons in certain gapped QSLs obey non-Abelian braiding statistics and have potential applications in topological quantum computations. However, since most spin systems in two or higher dimensions exhibit long-range magnetic order at zero temperature, it is challenging to obtain a true QSL ground state. Kitaev proposed a honeycomb-lattice spin model \cite{Kitaev} (with no continuous spin-rotation symmetry) in which the ground state is an exactly solvable QSL with the excitation spectrum either gapless or gapped. In a suitable magnetic field, the gapless Kitaev spin liquid (KSL) is turned into a gapped chiral spin liquid (CSL) that supports non-Abelian anyon excitations. It was further shown that this field-induced CSL belongs to a family of chiral phases which are classified in a 16-fold way depending on their Chern number mod 16\cite{Kitaev}.

\begin{figure}[t]
\includegraphics[width=8cm]{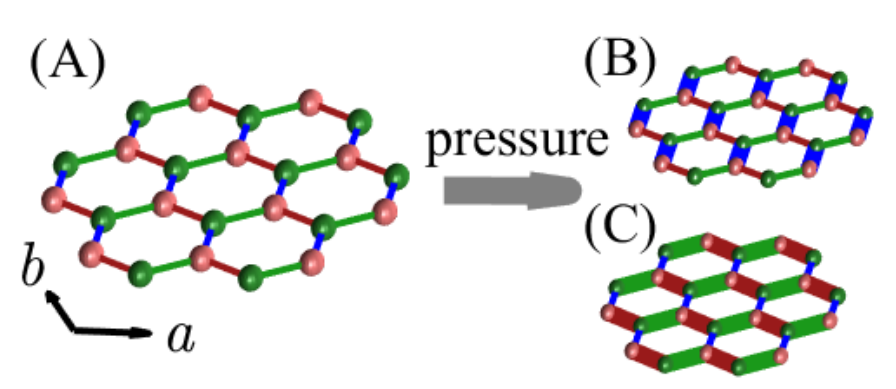} \
\caption{Intensity of interactions with and without strain. (A) Isotropic case, $a$ and $b$ are lattice constants; (B) The dimer-type anisotropy; (C) The zigzag-chain-type anisotropy.
}
\label{fig:Pressure}
\end{figure}

The KSL attracts lots of research interests. To realize the Kitaev model, a series of spin-orbit entangled candidate materials have been proposed and profoundly studied\cite{rjk,rcjk}, such as $\alpha$-RuCl$_3$\cite{rfgft,YJKim,rsea,rjea,rcaoetal}, $\alpha$-Li$_2$IrO$_3$\cite{incomm}, Na$_2$IrO$_3$\cite{singh,XLiu,ryea,rchoietal}, Cu$_2$IrO$_3$\cite{abra}, H$_3$LiIr$_2$O$_6$\cite{rhliiro} and Na$_2$Co$_2$TeO$_6$\cite{lefran,bera,LiYuan}. However, most of these materials manifest magnetic long-range order at low temperatures, indicating the existence of non-Kitaev interactions, including the Heisenberg term, the $\Gamma$-term \cite{HSKim, Li_theo, Jianxin_nuetron} and the $\Gamma'$-term \cite{lukas}. Nevertheless, these materials are proximate to a QSL phase, since the excitation spectra above the ordering temperature exhibit anomalous features\cite{Ji_neutron,rins_field,Bastien_raman_exp, Nasu_raman_theo,Hirobe_thermal}. Interestingly, the low-temperature magnetic order in some Kitaev materials can be suppressed under controllable situations. For instance, a suitable magnetic field can drive $\alpha$-RuCl$_3$ into a spin-liquid-like disordered phase \cite{rzea, thermal, Masuda, Baek, Jansa, Liu_KG}. High pressure is another method to control the magnetic transition \cite{pressure-tuned} as observed in $\alpha$-RuCl$_3$ or $\alpha$-Li$_2$IrO$_3$ where a dimer phase is obtained after the magnetic order being killed \cite{yu_pressure, sun_pressure, Bastien_pressure, Mei_pressure, HermannLiIrO}.  In contrast to forming strong dimers, pressure may also distort the lattice into coupled zigzag-chains where the intra-chain interactions are stronger. An example of this type is Li$_2$RhO$_3$ \cite{zigchain}.

\begin{figure}[t]
\includegraphics[width=8.8cm]{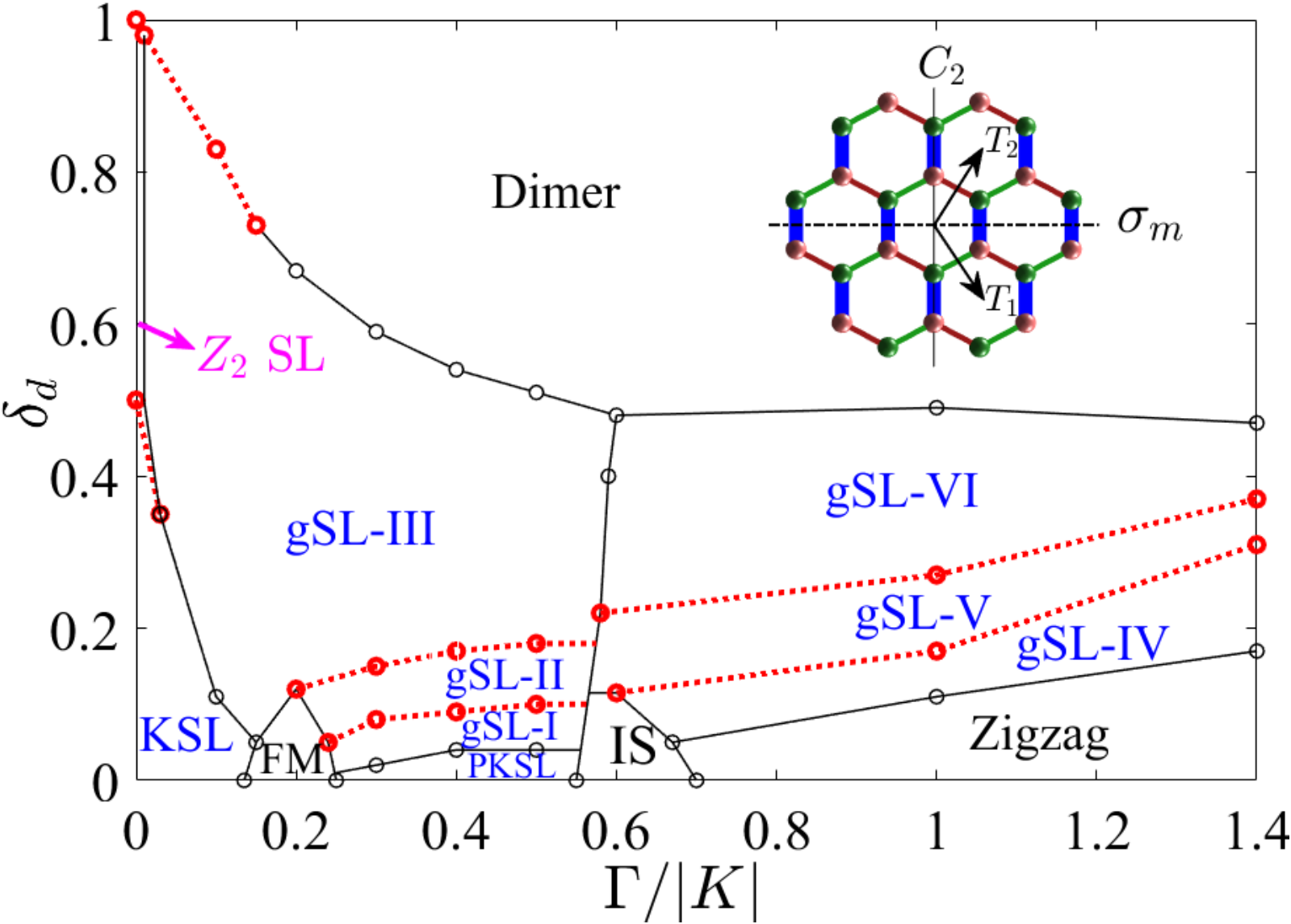} \
\caption{
Phase diagram of the anisotropic $K$-$\Gamma$ model (\ref{KG}) where the strong interacting bonds form dimers and the anisotropy is parametrized by $\delta_d$. The diagram contains three ordered phases (the FM, the IS, and the zigzag phase) and several QSLs (the KSL, the PKSL, the $Z_2$ QSL, and the gSL-I $\sim$ VI). The dashed-doted lines represent second-order phase transitions, and the black thin lines represent first-order phase transitions. The insert is a cartoon picture of the uniformly strained honeycomb lattice. $\sigma_m$ symbols the mirror reflection, $C_2$ represents a two-fold rotation along vertical bonds, and $T_1, T_2$ are the generators of the translation group.
}
\label{fig:AnisKGamma}
\end{figure}

\begin{figure}[t]
\includegraphics[width=8.8cm]{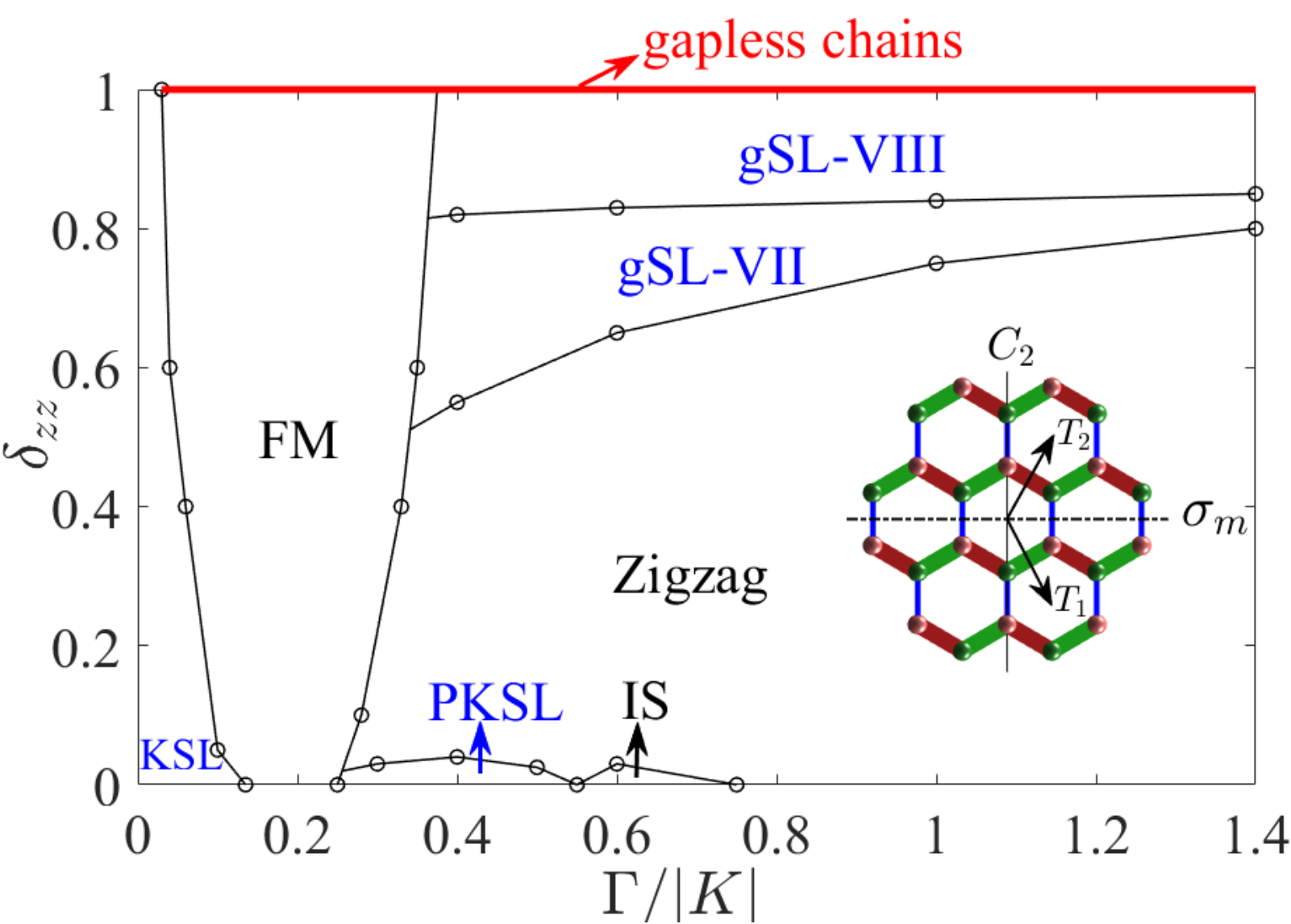} \
\caption{Phase diagram of the anisotropic $K$-$\Gamma$ model (\ref{KG}) where the strong interacting bonds form zigzag chains and the anisotropy is parametrized by $\delta_{zz}$. In contrast to the diagram in Fig.~\ref{fig:AnisKGamma}, the FM phase and the zigzag ordered phase are much more robust against the $\delta_{zz}$ anisotropy, while the IS phase becomes smaller. Two new gapless QSLs appear in the diagram, {\rm i.e.}, the gSL-VII, and the gSL-VIII. There are gapless chains marked by a thick red line in the one-dimensional limit ($\delta_{zz}=1$). All the phase transitions are of first order. The insert shows a cartoon picture of the uniformly strained honeycomb lattice where the strong bonds form zigzag chains.}
\label{fig:AnisKGamma_weak}
\end{figure}

In this work, we illustrate the possibility that QSL phases can be induced by (anisotropic) high pressure. To this end, we study a simple model on the uniformly strained honeycomb lattice, which contains both the Kitaev interactions and off-diagonal non-Kitaev interactions called the $\Gamma$ terms [see Eq.(\ref{KG})]. The strength of the interactions depends on the bond-directions, where the strong bonds either form dimers [Fig.~\ref{fig:Pressure}(B)] or form zigzag-chains  [Fig.~\ref{fig:Pressure}(C)]. From variational Monte Carlo (VMC) calculations by using Gutzwiller projected states as trial wave functions (see Appendix \ref{sm:VMC}), we obtain the phase diagrams as shown in  Figs.~\ref{fig:AnisKGamma} and \ref{fig:AnisKGamma_weak}. In the former case, the magnetic orders are suppressed quickly and a series of QSLs are observed at intermediate anisotropy. In the latter case, the ordered phases are much more robust while two QSL phases are generated by very large anisotropy. Most of the observed QSL phases are gapless and contain Majorana cones in their spinon dispersions. We demonstrate that the number of cones is closely related to the microscopic symmetry and can be reflected by their dynamical structure factors. Furthermore, by applying a magnetic field these gapless QSLs become gapped CSLs characterized by their quantized thermal Hall conductance. Using the $\pmb k\cdot\pmb p$ method, we obtain the maximum-mass field orientation for every cone, from which one can read the Chern number of the resultant CSLs for weak magnetic fields applied along arbitrary directions.

An important conclusion of our work is that a complete classification of nodal QSLs should include not only the symmetry fractionalization pattern (namely, the projective symmetry group), but also the information of the cones, including the total cone-number and the chiralities of every cone. On the other hand, since the strained lattice which causes the bond-dependence of the interaction strength can be achieved by exerting (uniaxial) pressure, our work suggests a promising way to realize the gapless QSL phases experimentally.

\section{The Model and The Method}

As mentioned, the $\Gamma$ interaction plays an important role in Kitaev materials. Therefore, we will start our discussion by considering an effective spin model that contains only the $K$ and $\Gamma$ interactions.
\begin{equation}\label{KG}
H = \sum_{\langle i,j \rangle \in\alpha\beta(\gamma)} K_{\gamma} S_i^\gamma S_j^\gamma + \Gamma_{\gamma} (S_i^\alpha S_j^\beta + S_i^\beta S_j^\alpha),
\end{equation}
where the $K_\gamma$ terms stand for the Kitaev interactions and the $\Gamma_\gamma$ terms represent the symmetric off diagonal exchange interactions. Most Kitaev materials are known to have $K_\gamma<0$ and $\Gamma_\gamma>0$, so we will only discuss this kind of interaction. In the isotropic case, namely $K_x=K_y=K_z$, $\Gamma_x =\Gamma_y = \Gamma_z$, the model has $D_{3d}\times Z_2^T$ symmetry and has been studied previously\cite{singleQ, Pollmann, Kee, PKSL}. However, little is known if the symmetry is lowered by the anisotropy in the intensities of spin-spin interactions. The anisotropy may be caused by, for instance, uniaxial pressure. In the following,  we will discuss two different situations according to the spatial dependence.

In the first case, we assume that the strong bonds form disconnected dimers, see Fig.~\ref{fig:Pressure}(B) for an illustration. A possible physical origin of the strong vertical bonds is the shortening of the length of the vertical bonds. For simplicity, we note $K_z=K,\ \Gamma_z=\Gamma$ and use only one variable $0\leq \delta_d\leq1$ to parametrize the degree of dimerization in the anisotropy interactions, namely, $K_x=K_y=(1-\delta_d) K$ and $\Gamma_x=\Gamma_y=(1-\delta_d)\Gamma$.

In the second case, we assume that the strong bonds form zigzag chains, as illustrated in Fig.~\ref{fig:Pressure}(C).  A possible physical origin of the weak vertical bond is the sharpening of angles between the spins and the intermediate atoms, which weakens the super-exchange interactions. For simplicity, we will note $K_x=K_y=K,\ \Gamma_x=\Gamma_y=\Gamma$ and use one parameter $0\leq \delta_{zz}\leq1$ to denote the degree of zigzag-chain type anisotropy such that $K_z=(1-\delta_{zz})K$, $\Gamma_z = (1-\delta_{zz})\Gamma$.

For general values of $\delta_d$ or $\delta_{zz}$, the model (\ref{KG}) has a $G=\tilde{\mathscr C}_{2h} \times Z_2^T$ magnetic point group symmetry, where $Z_2^T = \{ E,T \}$ is the time-reversal group and $\tilde{\mathscr C}_{2h}=\{E, P, C_2, \sigma_m\}$ is different from the usual point group $\mathscr C_{2h}$ since here the  two-fold axis $C_2$ is not perpendicular to the lattice plane, instead it lies in the plane along the vertical bonds. $P$ is the spatial inversion and $\sigma_m$ is the mirror reflection whose mirror plane is perpendicular to the $C_2$ axis. The $C_2$ and $\sigma_m$ operations are illustrated in the insets of Figs.~\ref{fig:AnisKGamma} and \ref{fig:AnisKGamma_weak}.

The model (\ref{KG}) can be mapped into an interacting fermionic model in the Majorana representation $S_i^m=ib_i^mc_i$ (under the constraint $b_i^xb_i^yb_i^zc_i=1$) introduced by Kitaev\cite{Kitaev}. One can combine the Majorana fermions into complex fermionic spinons $C_i=(c_{i \up}, c_{i \dn})^T$ such that the constraint is mapped into the particle-number constraint $C_i^\dag C_i=1$. The ground-state energy of the model can be calculated from VMC using the Gutzwiller projected mean-field ground states as trial wave functions. In constructing the mean-field Hamiltonian, we follow the guidance of projective symmetry group (PSG)\cite{igg,You_PSG} and construct different types of QSL ansatz as trial states. The magnetic order is treated as a background field, in which the ordering pattern is obtained from single-${\pmb Q}$ approximation\cite{singleQ,Liu_KG,PKSL}, and the amplitude is determined by minimizing the energy. The variational parameters include $\{ \rho_a^x$, $\rho_c^x$, $\rho_d^x$, $\phi_0^x$, $\phi_3^x$, $\phi_5^x$, $\phi_7^x$, $\phi_7^z$, $\theta\}$ whose meaning are interpreted in Appendix \ref{sm:differentPSG}.

%Our VMC calculations are performed on a lattice with 8 by 8 unit cells, namely, 128 sites.
Our VMC calculations are performed on tori of up to 10$\times$10 unit cells, i.e., of 200 lattice sites.
The spinon dispersion of the QSLs can be qualitatively obtained by diagonalizing the mean-field Hamiltonian in a larger system size with the optimized parameters from VMC. From this dispersion, we can locate the positions of the nodes in the gapless QSLs.

\section{The Phase diagrams and phase transitions}

The phase diagrams for the dimer-type anisotropy and the zigzag-chain-type anisotropy are shown in Figs.~\ref{fig:AnisKGamma} and \ref{fig:AnisKGamma_weak}, respectively.  All of the QSL phases have a $Z_2$ invariant gauge group and share the same PSG with the Kitaev's exact solution. We have tried a few ansatz with different PSGs and find that they are higher in energy comparing to the ones shown in our phase diagrams (for details see Appendix \ref{sm:differentPSG}).

\subsection{Dimer-type anisotropy}
The phase diagram of the first case (with $\delta_d$ anisotropy) is shown in Fig.~\ref{fig:AnisKGamma}. The line at $\Gamma=0$ is exactly solvable, where the gapless Kitaev QSL ceases at $\delta_d=0.5$ with a second-order phase transition to the gapped $Z_2$ QSL. As a benchmark, our phase diagram at $\Gamma=0$ is completely consistent with the above result. Since the gap of the $Z_2$ gauge-flux excitations ({\it i.e.} the vison gap) in the $Z_2$ QSL phase is much smaller than that in the KSL\cite{Motome}, it is no surprise that the gapped $Z_2$ QSL is very fragile and small $\Gamma$ interaction can cause a transition to another phase.

In the line $\delta_d=0$ the model has a higher symmetry and has been studied previously\cite{PKSL,Pollmann, Kee}, where three ordered phases [namely the ferromagnetic (FM) phase, the incommensurate spiral (IS) phase, and the zigzag phase] plus two QSL phases [namely the KSL and the proximate Kitaev spin liquid (PKSL)] were found\cite{PKSL}. Interestingly, the ordered phases are completely suppressed at anisotropy $\delta_d\sim0.15$, and several new gapless QSL phases, labeled as gSL-I$\sim$VI, are generated. When the anisotropy is very large (with $\delta_d$ greater than 0.5), the system enters a gapped disordered phase --- the dimer phase. Different from the gapped $Z_2$ QSL phase at very small $\Gamma$ whose ground states on a torus are four-fold degenerate, the dimer phase is trivial since it shows no such topological degeneracy (see Appendix \ref{app: GSD}). In other words, the $Z_2$ gauge field is confined in the trivial dimer phase while deconfined in the gapped $Z_2$ QSL phase.

The observation of the series of gapless spin liquid phases with different numbers of cones is the central results of this work. The phase showing up at $0.04<\delta_d<0.1$ above the PKSL is labeled as gSL-I, whose spinon dispersion contains 10 Majorana cones. With the increasing of $\delta_d$ (with $\Gamma<0.6$), the system undergoes successive continuous transitions to other gapless QSLs, namely the gSL-II and the gSL-III, which contain six Majorana cones and two Majorana cones respectively. At larger $\Gamma$ (with $\Gamma>0.6$), three more gapless QSLs,  namely, the gSL-IV, the gSL-V, and the gSL-VI are found in sequence with the increasing of $\delta_d$ between the zigzag phase (at $\delta_d<0.15$) and the dimer phase (at $\delta_d> 0.5$). These three QSLs contain 14, 10, 2 Majorana cones, respectively. Later we will discuss the significance of the number of Majorana cones.

The phase transitions from the magnetically ordered states to the QSLs are all of first order. The transition from the gSL-VI to the dimer phase is also first-order. Interestingly, the transition from gSL-III to the dimer phase is continuous at $\Gamma<0.16$ but becomes first-order at $0.16< \Gamma <0.6$. The phase transitions between different QSLs are either first-order or continuous. The first-order phase transitions are characterized by discontinuous jumps of some variational parameters, and sudden changes in the number of Majorana cones. For example, the transition from the PKSL to the gSL-I and the transition from the KSL to the gSL-III are both accompanied by a sudden growth of $\phi_7^z$, the transition from the gSL-III to the gSL-V or to the gSL-VI is accompanied by a sudden growth of $\rho_a^x$. The continuous transitions between QSLs, marked as red dashed-doted lines in Fig.~\ref{fig:AnisKGamma}, are characterized by smooth changes of the variational parameters and the merging and pairwise disappearance of the Majorana cones. A typical example is the transition from the KSL to the gapped $Z_2$ QSL, where the two cones merge and a gap opens. In the following continuous transitions, four of the cones merge in pairs and disappear simultaneously: the one from the gSL-I to the gSL-II, the one from the gSL-II to the gSL-III, and the one from the gSL-IV to the gSL-V. The transition from the gSL-V to the gSL-VI is special since eight cones merge and disappear simultaneously at the critical point.

\subsection{Zigzag-chain-type anisotropy}
Now we discuss the second case (with $\delta_{zz}$ anisotropy). The phase diagram is shown in  Fig.~\ref{fig:AnisKGamma_weak}, which is relatively simpler. At $\Gamma=0$, there is only one phase, the KSL. This result agrees with the exact solution. Different from the case with $\delta_d$ anisotropy, the ordered phases (except for the IS phase) are much more robust against the $\delta_{zz}$ anisotropy. The FM phase locates at the vicinity of $\Gamma=0.2$ and extends throughout the parameter region $0\leq \delta_{zz}\textless 1$, its width increases with $\delta_{zz}$. At larger $\Gamma$, the PKSL phase ($0.25<\Gamma<0.45$) and the IS phase ($0.45<\Gamma<0.75$) appear in sequence at small $\delta_{zz}$, but both are quickly suppressed at $\delta_{zz}\sim 0.05$. Above the PKSL and the IS phases, there is an overwhelming zigzag-order phase, which extends from $\Gamma\sim 0.25$ to the large $\Gamma$ limit below a critical $\delta_{zz}$. When the zigzag order is suppressed above $\delta_{zz}\sim 0.5$, two gapless QSLs show up in sequence, which are labeled as gSL-VII and gSL-VIII respectively and are separated by a first-order phase transition. The gSL-VII state contains 16 Majorana cones in the spinon excitation spectrum while the gSL-VIII state has 8. Owing to the strong $\delta_{zz}$ anisotropy, the dispersions of the spinon excitations in these two QSLs are fairly flat along the zigzag-chain direction.

In the limit $\delta_{zz}=1$, the system becomes decoupled zigzag chains. The system is solvable at $\Gamma=0$, where the ground state is a gapless $Z_2$ state with extensive degeneracy and the low-energy excitations are dominated by two Majorana "cones".  At $\Gamma>0$, it was shown that the system has a hidden $O_h$ point group symmetry and the ground state is a gapless state whose excitation spectrum approximately agrees with the $SU(2)_1$ Wess-Zumino-Witten model\cite{Affleck}. We obtain a similar result (for details see Appendix \ref{sm:chain}) with the difference that in our VMC calculation the (first-order) transition from the gapless $Z_2$ phase to the gapless $U(1)$ phase occurs at a small but finite $\Gamma$ (with $\Gamma/|K|\sim0.05$) while Ref.~\onlinecite{Affleck} concluded that $\Gamma=0$ is the first-order phase transition point. Since the energy difference between the $U(1)$ state and the $Z_2$ state is very small (or order $10^{-4}|K|$ persite), we infer that the phase boundary between the $U(1)$ state and the gSL-VIII phase is very close (if not equal) to $\delta_{zz}=1$, namely, the $U(1)$ state survives only if the coupling between the zigzag chains is very weak. \\
%couple with each other to form 2-dimensional lattice. \\

\section{Properties of the Majorana cones}

The gapless QSLs are characterized by the Majorana cones in the excitation spectrum.  In the following we will discuss the nature of the cones, including their robustness under perturbations, their interchanging under symmetry operations, and their chiralities with respect to mass-generating perturbations.

\subsection{Symmetry protection}

The gapless QSLs are obtained from Gutzwiller projection of superconducting states whose dispersions contain certain numbers of Majorana cones. The low-energy quantum fluctuations of these QSLs are described by $Z_2$ gauge fields coupling to the nodal fermionic spinons (see Appendix \ref{sm:VMC}). Generally, Gutzwiller projection does not change the qualitative behavior of the low-energy dispersion, {\it i.e.} the low-energy excitation spectrum of the QSLs qualitatively agree with that of the corresponding mean-field theory. For instance, Gutzwiller projected gapped superconductor corresponds to a gapped QSL (if the $Z_2$ gauge field does not suffer from confinement), while Gutzwiller projected nodal superconductor corresponds to a nodal $Z_2$ QSL. Especially, if the Majorana cones are robust at the mean-field level,  then after projection the number of cones in the corresponding QSL remains unchanged.

Indeed, the robustness of the Majorana cones in the mean-field theory are guaranteed by the  $PT$ symmetry, namely, the combination of spatial inversion $P$ and time reversal $T$. When acting on the mean-field Hamiltonian $H_{\rm mf}=\sum_{\pmb k} H_{\pmb k}$, each symmetry operation $g$ is promoted to the corresponding PSG element $\hat g$. In the PSG of the gapless QSLs, $\hat P^2=-1$, $\hat T^2=1$, $\{\hat P,\hat T\}=0$,  $(\widehat{PT})^2=1$ are satisfied (see Appendix \ref{sm:PSG}). The $PT$ symmetry of the QSLs requires the following relation at each $\pmb k$ point,
$$
M(\widehat{PT}) H_{\pmb k}^*M^\dag(\widehat{PT})=H_{\pmb k},
$$
where $M(\widehat{PT})$ is a real symmetric matrix and is the representation of $\widehat{PT}$. We can write $M(\widehat{PT})=W^2$, where $W$ is a symmetric unitary matrix $W^T=W, W^*=W^{-1}$. Therefore
$$
W H_{\pmb k}^*W^*=W^*H_{\pmb k}W,
$$
namely, with the $\pmb k$-independent constant unitary matrix $W$ the Hamiltonian $H_{\pmb k}$ is transformed into a real symmetric matrix $H_{\pmb k}'=W^*H_{\pmb k}W$ whose eigenstates are real.

For any closed loop $l$ in the Brillouin zone (BZ), the Berry phase $e^{i\theta_{l}}$ of the eigenstates of $H_{\pmb k}'$ is quantized to real numbers $1$ or $-1$ (to ensure that the Berry phase is well-defined, the loop $l$ is assumed to be fully gapped). Noticing that the transformation $W$ is $\pmb k$-independent, so the Berry phase of the eigenstates of $H_{\pmb k}$ in the loop $l$ is also quantized to 1 or $-1$. Actually, the Berry phase 1 or $-1$ indicates that there are even or odd number of singularities ({\it i.e.} Majorana cones) in the area enclosed by the loop $l$, respectively.

Noticing that any momentum $\pmb k$ in the BZ is invariant under $PT$, this loop $l$ reserves the $PT$ symmetry. The quantized Berry phase indicates that such gapped 1D loops have a $\mathbb Z_2$ topological classification \cite{Wen_class, Kitaev_class, ZhaoYuXin}. This means that the cones cannot be gapped without breaking the $PT$ symmetry unless they merge and disappear in pairs. The merging of the cones is essentially a continuous quantum phase transition. Therefore, the number of cones is not allowed to change unless the $PT$ symmetry is broken or a phase transition occurs \cite{SymmetricMass}.

The $PT$-symmetry protection of the Majorana cones in the mean-field description suggests that the conic dispersion of the spinon excitations in the corresponding QSL is robust against perturbations reserving the $PT$ symmetry.

\subsection{Distribution and relation under symmetry}

The symmetry group $G=\tilde{\mathscr C}_{2h}\times Z_2^T$ imposes a constraint on the number of Majorana cones.

Suppose that a cone is locating at momentum $\pmb k$, then a group element $\alpha\in G$ may transform $\pmb k$ to an inequivalent point $\alpha \pmb k$ which is also the location of a cone. The resultant momentum points form a set, called the star of wave vectors, noted as $\{^* \pmb k\}$. Noticing that a wavevector $\pmb k$ is invariant under its little co-group $G_k$ (a subgroup of $G$), the number of wave vectors in $\{^* \pmb k\}$ is equal to the number of cosets of the little co-group. If $\pmb k$ is a general momentum point, then the little co-group $G_k=\{E, PT\}$ contains 2 group elements, therefore the number of cosets, {\rm i.e.} the number of wave vectors in $\{^*\pmb k\}$, is 4. If $\pmb k$ is on the high symmetry line of $\sigma_m$ (excluding the BZ center which is not a cone), then the little co-group is enlarged to $G_k=\{E, PT, \sigma_m, C_2T\}$, and correspondingly the number of vectors in $\{^* \pmb k\}$ reduces to 2. To reserve the symmetry $G$, the set of wave vectors of the gapless Majorana points in a QSL state must be composed of several $\pmb k$ stars. Consequently, the total number of cones in a QSL reserving the $\tilde{\mathscr C}_{2h}\times Z_2^T$ symmetry is equal to $4n+2m$ where $(n,m)$ are integers (see Table.~\ref{tab:cones}).

\begin{table}[t]
\centering
\begin{tabular}{|c |c   c| c  c  c|}
\hline
 QSL  & $\ \ (n,  m)\ \ $   & No. of cones & $\nu$ of CSL  & GSD & anyon types   \\
\hline
$Z_2$ QSL            & (0,     0)    &  0  & 0& 4 & $e,m,\varepsilon,I$  \\
KSL                 & (0,     1)    &  2  & 1& 3 & $\sigma,\varepsilon,I$\\
PKSL                & (2,     3)    &  14 & 5& 3 & $\sigma,\varepsilon,I$\\
gSL-I               & (1,     3)    &  10 & 1& 3 & $\sigma,\varepsilon,I$\\
gSL-II              & (1,     1)    &  6  & 1& 3 & $\sigma,\varepsilon,I$\\
gSL-III             & (0,     1)    &  2  &-1& 3 & $\sigma,\varepsilon,I$\\
gSL-IV              & (3,     1)    &  14 & 3& 3 & $\sigma,\varepsilon,I$\\
gSL-V               & (2,     2)    &  10 & 1& 3 & $\sigma,\varepsilon,I$\\
gSL-VI              & (0,     1)    &   2 & 1& 3 & $\sigma,\varepsilon,I$\\
\hline
gSL-VII             & (2,     4)    &  16 &-2& 4 & $a,\bar a,\varepsilon,I$\\
gSL-VIII            & (1,     2)    &  8  & 0& 4 & $e,m,\varepsilon,I$\\
\hline
\end{tabular}
\caption{Information of all the QSLs appearing in Figs.~\ref{fig:AnisKGamma} and \ref{fig:AnisKGamma_weak}. The middle two columns give the information of the cones, where $n$ is the number of $\pmb k$ stars locating at general positions and $m$ is the number of $ \pmb k$ stars on the high symmetry line of $\sigma_m$, the total number of cones is equal to $4n+2m$. The last three columns list the information of the field-induced CSLs, where $\nu$ is the Chern number in a weak magnetic field along ${1\over\sqrt3}(\pmb x+\pmb y+\pmb z)$ direction, GSD abbreviates the ground-state degeneracy on a torus, and the anyon types are determined by $\nu$. $I$ denotes the vacuum and $\varepsilon$ is the fermion. The two different vortices $e$ and $m$ are anti-particles of themselves, and another two vortices $a$ and $\bar a$ are anti-particles of each other; $\sigma$ are the vortices in the non-Abelian phases when $\nu$ is odd. The self-statistical angles ({\it i.e.}, the topological spin) of the vortices are determined by the Chern number $e^{i{\nu\pi\over8}}$.
}\label{tab:cones}
\end{table}

\begin{figure}[t]
\includegraphics[scale=0.34]{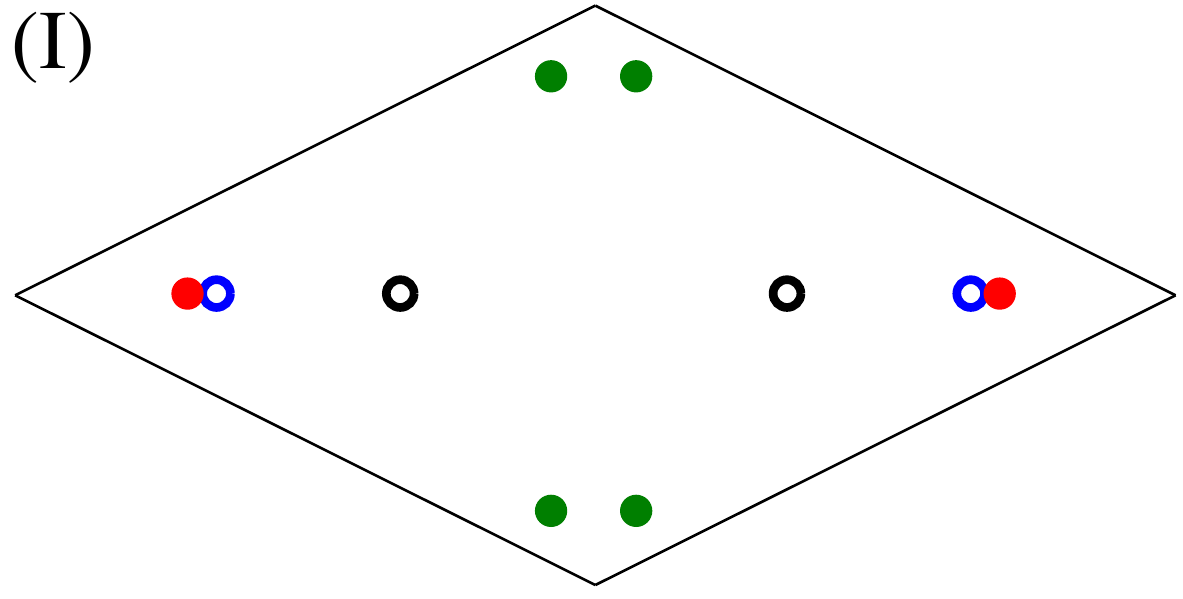}
\includegraphics[scale=0.34]{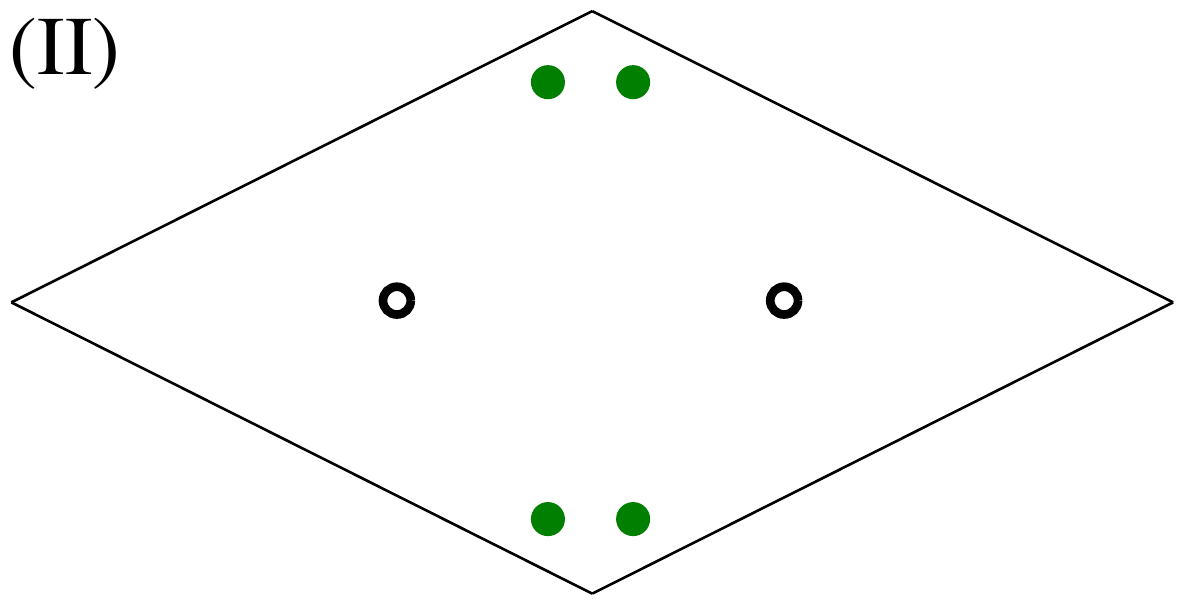}
\includegraphics[scale=0.34]{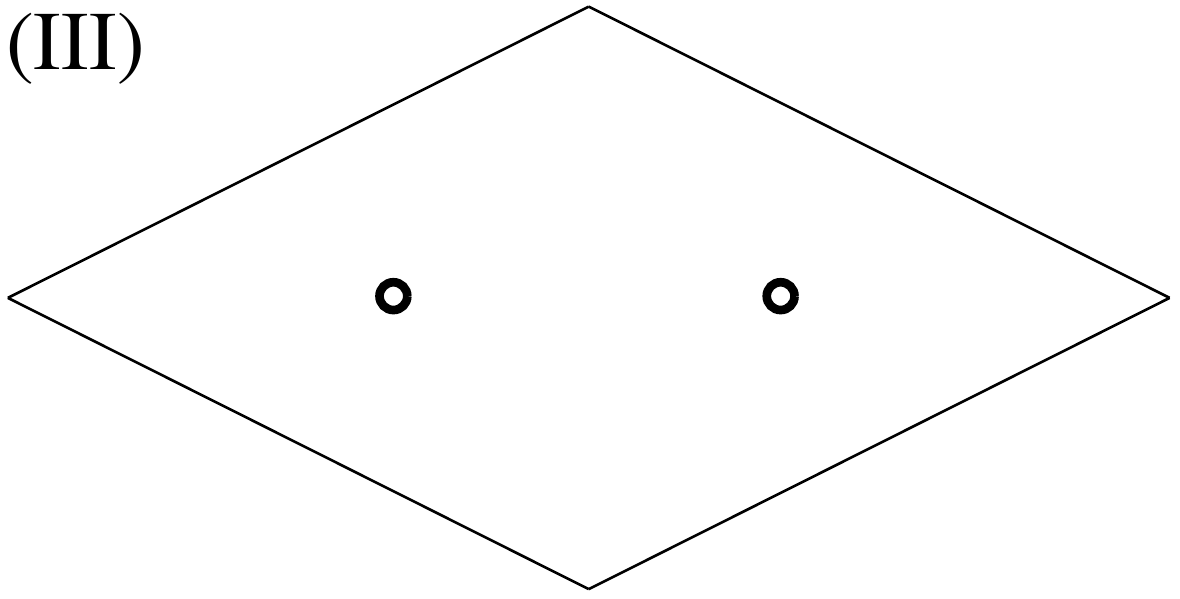}
\includegraphics[scale=0.34]{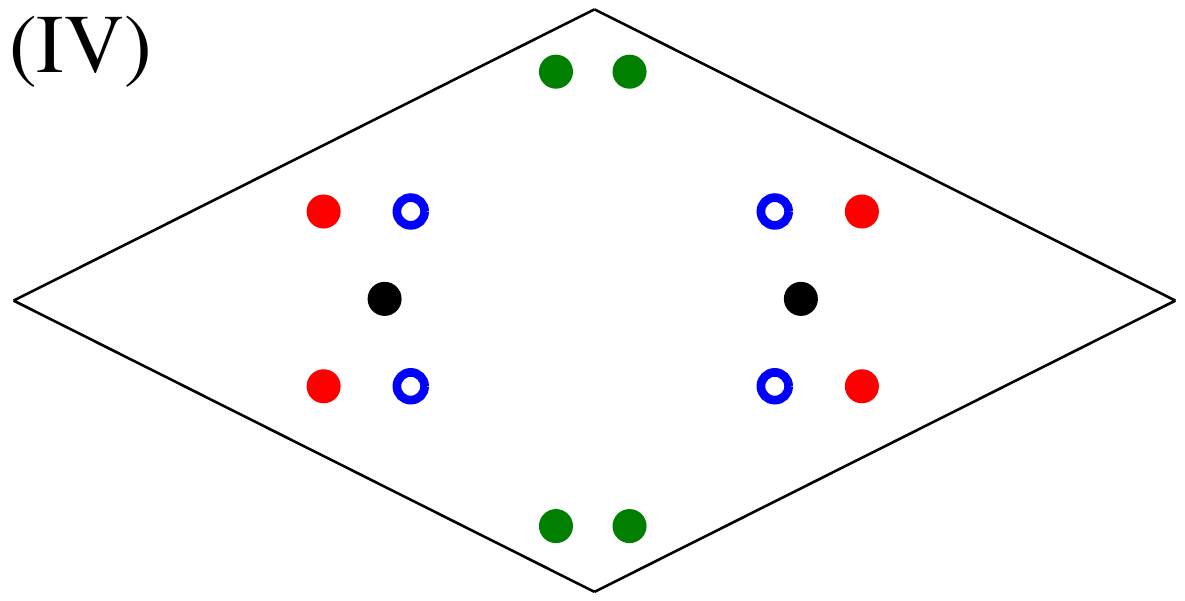}
\includegraphics[scale=0.34]{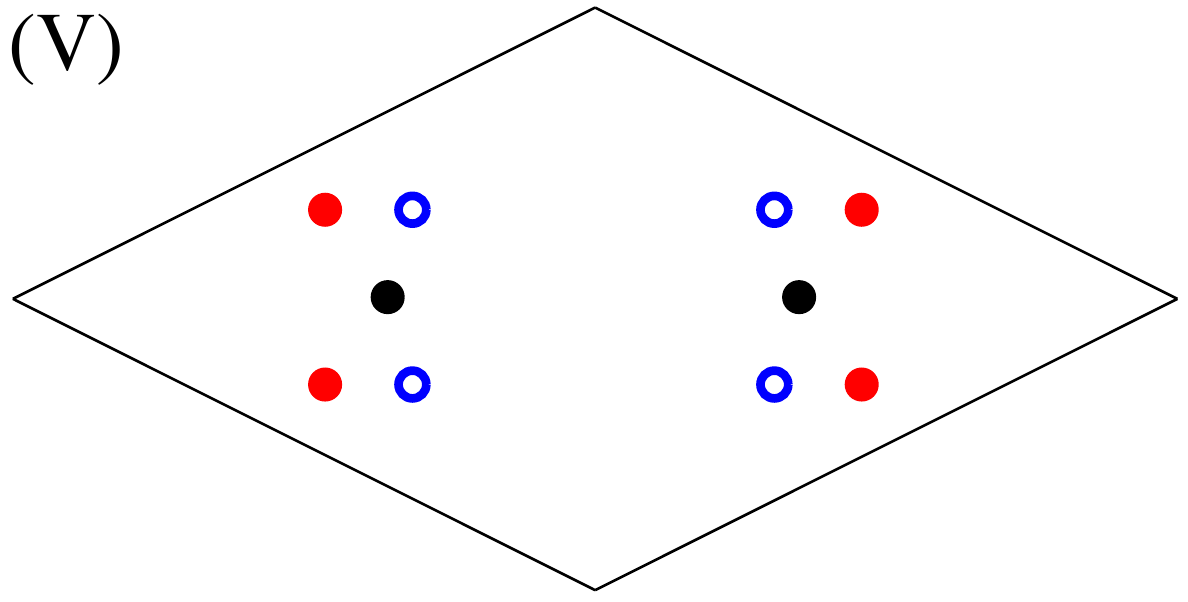}
\includegraphics[scale=0.34]{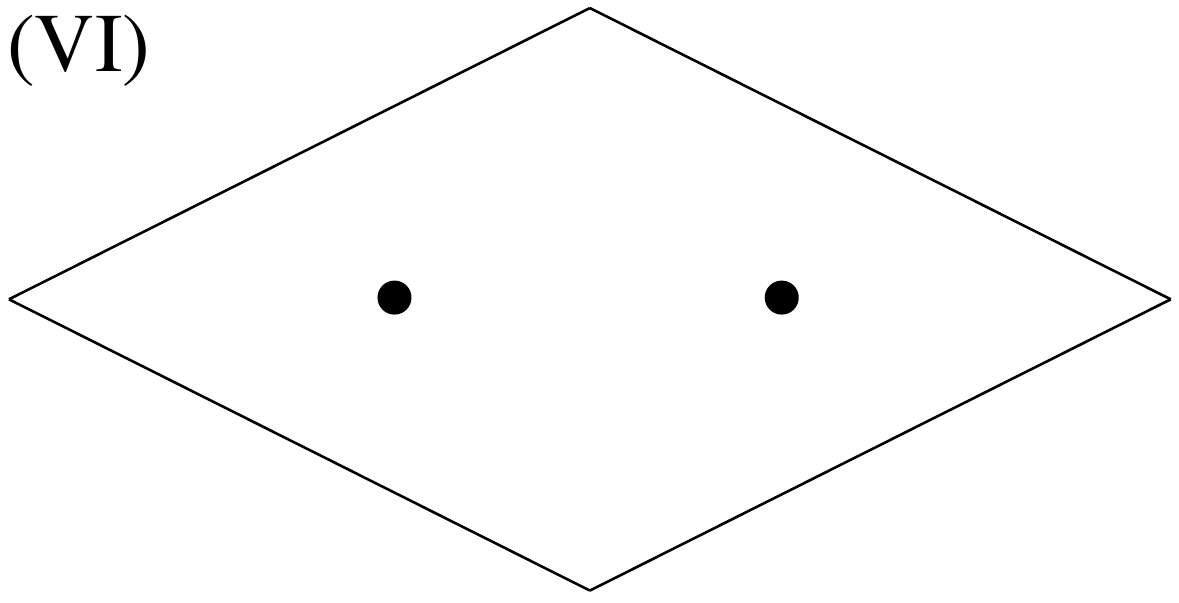}
\includegraphics[scale=0.34]{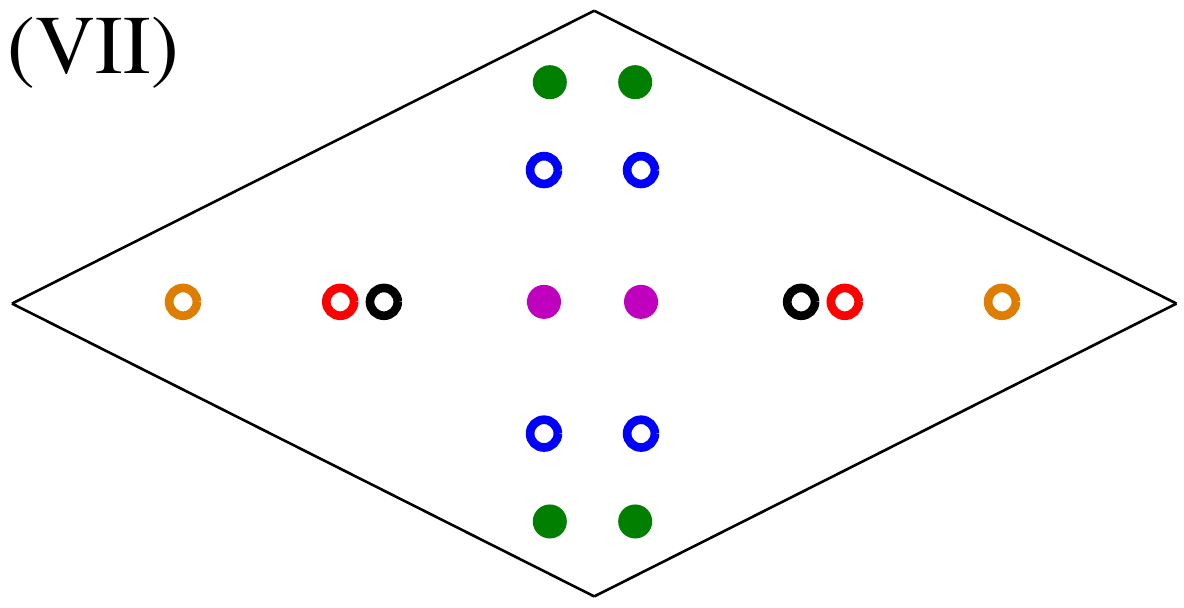}
\includegraphics[scale=0.34]{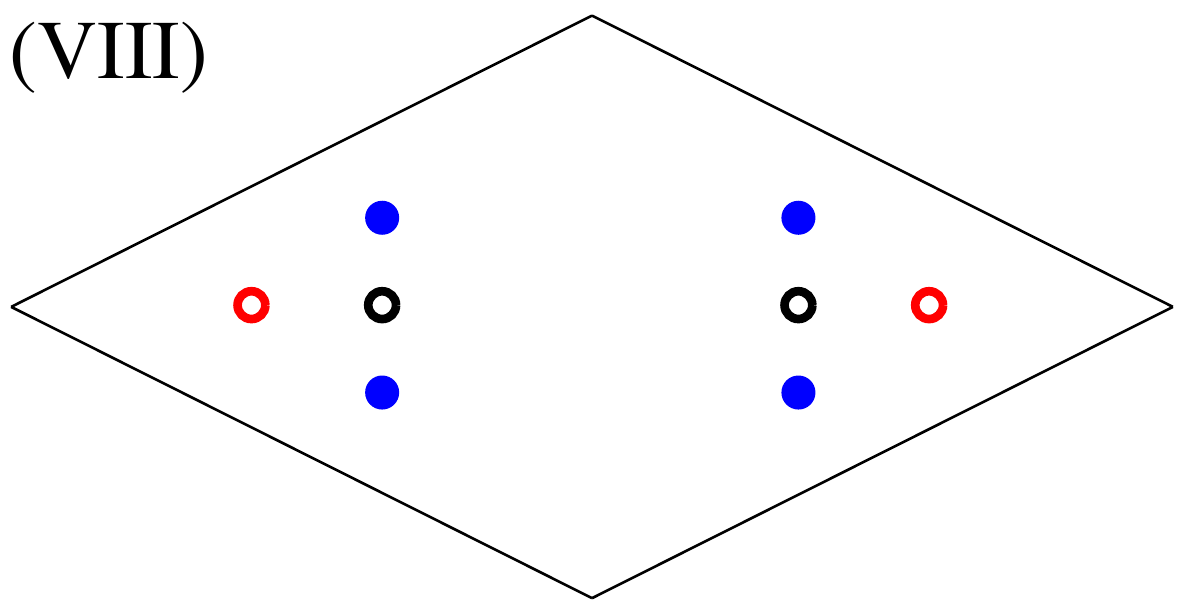}
\includegraphics[scale=0.34]{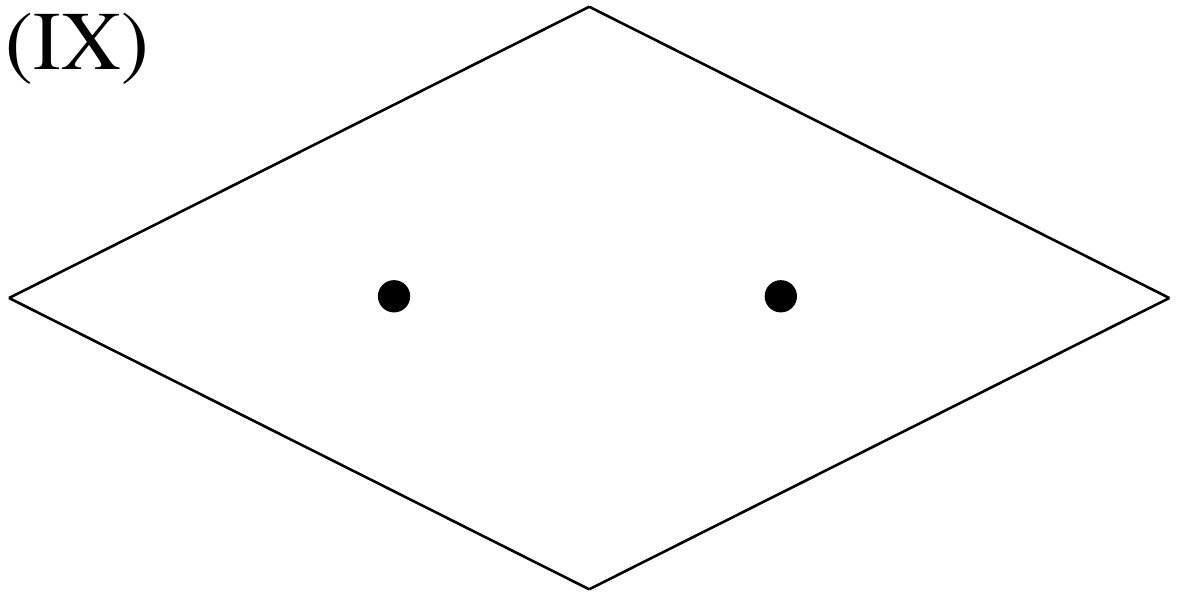}
\includegraphics[scale=0.34]{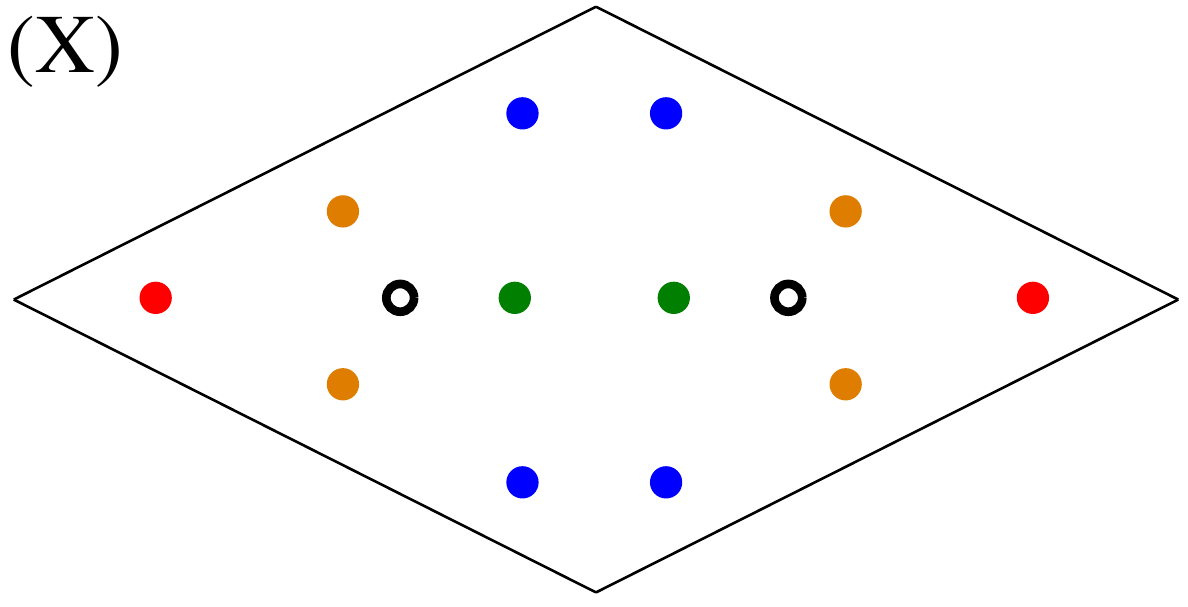}
\caption{Illustration of the location of the cones in the gapless QSLs.  Each dot stands for a Majorana cone, the dots with the same color are symmetry-related and belong to the same $\{^*\pmb k\}$. The cones locating at the solid dots have positive chirality while the ones at hollow dots have negative chirality.  (I) the 10-cone gSL-I state with $(n,m)=(1,3)$; (II) the 6-cone gSL-II state with $(n,m)=(1,1)$; (III) the 2-cone state gSL-III with $(n,m)=(0,1)$;(IV) the 14-cone gSL-IV state with $(n,m)=(3,1)$;   (V) the 10-cone gSL-V state with $(n,m)=(2,1)$; (VI) the 2-cone state gSL-VI with $(n,m)=(0,1)$; (VII) the 16-cone gSL-VII state with $(n,m)=(2,4)$; (VIII) the 8-cone gSL-VIII state with $(n,m)=(1,2)$; (IX) the 2-cone state KSL with $(n,m)=(0,1)$; (X) the 14-cone PKSL state with $(n,m)=(2,3)$. The low-energy DSFs of the above gSLs are shown in Fig.~\ref{fig:DSF} of the Appendix \ref{sm:DSF}.
}\label{fig:cones}
\end{figure}

The low-energy excitations in the gapless QSLs are dominated by the excitations around the Majorana cones. Therefore the number of cones can be reflected in the low-frequency spin dynamic structure factor (DSF).  In case that two QSLs have the same number of cones, the location of the cones, {\it i.e.} the values of $(n,m)$, will be important to distinguish their low energy spectra. For instance, the PKSL and the gSL-IV both contain 14 cones, but the former has $(n,m)=(2,3)$ while the latter has $(n,m)=(3,1)$, as shown in Fig.~\ref{fig:cones}(IV)\&(X); another example is the gSL-I versus the gSL-V, both of them contain 10 cones but the former has $(n,m)=(1,3)$ and the latter has $(n,m)=(2,2)$, as illustrated in Fig.~\ref{fig:cones}(I)\&(V).

\subsection{Mass and chirality}

The cones in the gapless QSLs can be gapped out by symmetry-breaking perturbations, such as the magnetic fields or the three-spin interactions. If time reversal symmetry is broken, then the resultant gapped state may be a CSL whose properties are determined by the Chern number $\nu$ in the mean-field description (see Table.~\ref{tab:cones}), given that the $Z_2$ gauge field keeps deconfined after Gutzwiller projection.

At mean-field level, each cone contributes $\pm{1\over2}$ to the Chern number when obtaining a mass, where the sign $\pm$ is defined as the sign of the mass of the cone, or the {\it chirality} of the cone with respect to the perturbation which generates the mass. Once the chiralities of all the cones are known, the total Chern number will be known. In the following, we study the mass of the cones using $\pmb k\cdot \pmb p$ expansion.

Every Majorana cone has doubly degenerate zero energy modes at the Majorana point. Suppose that a Majorana point is locating at $\pmb k$, and the two zero-energy modes span a two-dimensional Hilbert space $\mathscr L_{\pmb k}^0$. The physical properties of the cone can be obtained by projecting the Hamiltonian $H_{\pmb k+\delta \pmb k}' $ onto $\mathscr L_{\pmb k}^0$, where $H_{\pmb k+\delta \pmb k}' =W^*H_{\pmb k+\delta \pmb k} W$ is real. At the leading order, the projected Hamiltonian takes the form
\Beq
\mathscr H_{\pmb k+\delta \pmb k} = P_{\pmb k}H_{\pmb k+\delta \pmb k}' P_{\pmb k}^\dag
= V_1({\pmb k}) \delta k_1 + V_2({\pmb k}) \delta k_2
\Eeq
where $P_{\pmb k}$ is the projection operator onto $\mathscr L_{\pmb k}^0$, and $V_1({\pmb k}), V_2({\pmb k})$ are two real matrices which can be expanded as linear combinations of $\sigma_x$ and $\sigma_z$. When projecting the mass term $H_M'=W^* H_M W$ onto $\mathscr L_{\pmb k}^0$, it must contain a $M({\pmb k}) \sigma_y$ component. Therefore, the massive cone is approximately described by the effective $\pmb k\cdot\pmb p$ Hamiltonian
\Beq
\mathscr H_{\pmb k+\delta \pmb k}^M = V_1({\pmb k}) \delta k_1 + V_2({\pmb k}) \delta k_2 + M({\pmb k}) \sigma_y
\Eeq
whose Chern number is equal to $\nu = {1\over 2}{\rm Sgn} (\pmb V_1\times \pmb V_2\cdot\pmb M)$ \cite{fluxcrystal} where the matrices $V_1, V_2, M$ are expanded by Pauli matrices with $V_1=\pmb V_1 \cdot\pmb \sigma, V_2=\pmb V_2 \cdot\pmb \sigma, M = \pmb M \cdot\pmb \sigma$. The vectors $\pmb V_1$ and $\pmb V_2$ lie in the $xz$-plane and $\pmb M = [0,M,0]$.

The Zeeman splitting caused by the magnetic field $H_M=\sum_i \pmb S_i\cdot \pmb B$ is a simple way to gap out the cones. For every cone, we analyze the chirality ${\rm Sgn} (\pmb V_1\times \pmb V_2\cdot\pmb M)$ and the amplitude of the effective mass $|M|$ for unit magnetic field along $\pmb x$- or $\pmb y$- or $\pmb z$-direction. From this information, one can select a field orientation for a given cone such that the chirality is positive and the gap is the biggest. We call this special direction the "maximum-mass direction" (MMD) of that cone. If the magnetic field has positive component along the MMD, the chirality of the cone is positive; otherwise the chirality reverses its sign. If the field is perpendicular to the MMD, then the gap vanishes and the cone remains gapless. Therefore, once we know the MMDs for every cone, we can immediately know the total Chern number for an arbitrarily oriented field. This helps to study the response of gapless QSLs to magnetic fields along arbitrary directions\cite{Liu_KG}. For the states in the same QSL phase, the MMD of a cone varies continuously with interaction parameters. As an example, we list the MMDs for all the cones of a representative state {\it i.e.} the gSL-V phase in Table.\ref{tab:Mass_V} (more examples are discussed in Appendix \ref{sm:Chern}).

We pay special attention to the magnetic field oriented along ${1\over\sqrt3}(\pmb x+\pmb y+\pmb z) \equiv (111)$, which breaks the time reversal symmetry $T$ but reserves the inversion symmetry $P$ and combined symmetries $C_2T, \sigma_mT$ (see Appendix \ref{sm:Chern}). In this case the cones in the same $\{^*\pmb k\}$ are still symmetry-related and contribute the same amount to the total Chern number $\nu$. For field oriented along $\pmb B\parallel (111)$, the chirality of a cone obtained from above $\pmb k\cdot\pmb p$ method is consistent with the sign of the Berry curvature at the same cone. Therefore, the pattern of cones $(n,m)$ and the chirality of cones in every $\{^*\pmb k\}$ completely determine the value of $\nu$ in the resultant CSLs (see Appendix \ref{sm:Chern} for details).

The cones can also obtain mass from interactions, such as the three-spin ring-exchange interaction $\sum\limits_{\langle ijk\rangle} \pmb S_i\times\pmb S_j\cdot \pmb S_k = {1\over 12i}[\chi_{ij}\chi_{jk}\chi_{ki} + {\rm cyclic(ijk)}-H.c.]$ where $\langle i j k\rangle$ are the neighboring sites which form the smallest triangle and $\chi_{ij}=C_i^\dag C_j=\sum_\alpha c_{i\alpha}^\dag c_{j\alpha}$. The mean-field decouplings of the ring-exchange interaction contain a next-nearest neighbor hopping term $H_M=\sum\limits_{\langle\langle i,j\rangle\rangle} (t C_i^\dag C_j + H.c.)$ which gaps out the cone if $t$ is not real. In Table.\ref{tab:Mass_V}, we list the chirality of all the cones in the gSL-V phase according to next-nearest neighbor hopping in the case $t=i$. Similar discussion can also be applied to other time reversal breaking interactions such as $\sum\limits_{\langle ijk\rangle} S_i^x  S_j^y S_k^z$.

It should be noted that the above method is applicable only when the spinon gap in the resulting state is linear to the amplitude of the perturbation. If a perturbation contributes zero mass to a cone in linear order, it is still possible that the cone opens a gap via higher-order processes (for example, in the pure Kitaev model a weak Zeemann field gaps out the cones in $|\pmb B|^3$ law).

\begin{table}[t]
\centering
\begin{tabular}{|c|c|c|c|}
\hline
Cones &  MMD with $\pmb B\cdot \pmb S$ &  $\pmb B\parallel (111)$ & $\sum\limits_{\langle\langle i,j\rangle\rangle} (i C_i^\dag C_j + H.c.)$\\
\hline
D$_{\ }$       & (+0.45,+0.45,+0.78)     &  $+$  &  $+$\\  % (0.4494,0.4494,0.7721)
\hline
R$_1$       & (+0.77,-0.16,+0.61)     &  $+$  &  $-$   \\ %(0.7728,-0.1643,0.6130)
\hline
R$_2$      & (-0.16,+0.77,+0.61)     &  $+$  &  $+$   \\ %(-0.1643,0.7728,0.6130)
\hline
B$_1$       & (-0.88,+0.04,-0.47)     &  $-$  &  $+$  \\ %(-0.8828,0.0449,-0.4676)
\hline
B$_2$      & (+0.04,-0.88,-0.47)     &  $-$  &  $-$  \\ %(0.0449,-0.8828,-0.4676)
\hline
\end{tabular}
\caption{Mass information for the Majorana cones on the left half BZ in the gSL-V phase [see Fig.~\ref{fig:cones}(V)]. MMD stands for maximum mass directions (see the main text for definition). D stands for the dark cone on the high-symmetry line; R$_1$ (R$_2$) stands for the red upper (lower) cone;  B$_1$ (B$_2$)  stand for the blue upper(lower) cone. Cones related by inversion symmetry have the same MMD, therefore the cones on the right half BZ are not shown. The signs in columns 3 and 4 stand for the chiralities of the corresponding cones with respect to $\pmb B\parallel(111)$ and next-nearest-neighbor hopping, respectively.
}\label{tab:Mass_V}
\end{table}

\section{Physical detections}

The gapless nodal QSLs differ from the gapped ones by their low-temperature density of states, which are reflected in the temperature ($T$) dependence of the specific heat or the thermal conductance. The nodal ones have power-law $T$ dependence while the gapped ones show exponential $T$ dependence. The features of the cones discussed above can help us to further distinguish different nodal QSLs experimentally.\\

\subsection{Dynamic structure factor}
DSF can be measured from the neutron scattering experiment and provides useful information to distinguish different QSLs. In the following we calculate the DSFs of the gapless QSLs from their mean-field dispersions with the parameters determined from VMC calculations. The frequency (momentum) of the DSF is determined by the total energy (momentum) of a pair of spinon excitations\cite{PKSL}. At low frequency, the DSF is dominated by the spinons close to the Majorana cones.

As mentioned, if two QSLs have different $(n,m)$, they can be distinguished from their different DSFs. In this way, most of the gapless QSLs are distinguished from the others (see Appendix \ref{sm:DSF} for the DSFs of different QSLs).

But, three of the QSLs, the KSL, the gSL-III and the gSL-VI, all have $(n,m)=(0,1)$. It seems that they are not distinguishable. However, the KSL is special since the parameters $\eta_3^x,\eta_5^x,\eta_7^z,\eta_7^x$ which mix the $b^m$-fermions ($m=x,y,z$) and the $c$ fermions are zero\cite{Song, PKSL}. Consequently, the $b^m$ fermions are gapped while the $c$ fermions are gapless\cite{Kitaev,PKSL}. Therefore, the physical excitations excited by spin operators $S^m=ib^mc$ cost finite energy. Resultantly, the low-frequency DSF is vanishing below the gap of the $b^m$ bands. In other words, the KSL shows a gapped spin dynamics although the energy spectrum is gapless. In contrast, in the gSL-III, and the gSL-VI, the $b^\alpha$ and $c$ fermions are hybridized at the nodal points such that the spin dynamics are gapless.  But it remains a problem to distinguish the gSL-III and the gSL-VI. A solution is found in the following. \\

\subsection{Descendent chiral spin liquids}
The cones in the gapless QSLs can be gapped out by a magnetic field via Zeeman coupling.  When the Chern number $\nu$ is nontrivial, the resultant state is a CSL given that the field is not too strong. As Kitaev pointed out, the CSLs have a 16-fold classification depending on the Chern number $\nu$ mod 16 \cite{Kitaev}.

For a weak magnetic field\cite{note} along the $(111)$ direction ({\it i.e.} $B_x=B_y=B_z$), the Chern numbers we obtained are $\nu=0, \pm1, -2, 3, 5$. When $\nu$ is odd, the CSL is non-Abelian whose ground state degeneracy (GSD) on a torus is 3, while when $\nu$ is even, the CSL is Abelian and the GSD is 4. Except for the $Z_2$ QSL, all the CSLs descendent from the gapless QSLs in the phase diagram Fig.~\ref{fig:AnisKGamma} are non-Abelian (see Appendix \ref{sm:Chern}). This is an exciting result because it indicates that anisotropic interactions are plausible to generate non-Abelian CSLs which can be applied in topological quantum computations.

Our VMC calculation of the GSD on a torus is consistent with the theoretical predictions.  From the Chern number $\nu$ we can also read the information of the elementary anyon excitations. For instance, the topological spin of the vortex in a CSL with Chern number $\nu$ is $e^{i\nu {\pi\over 8}}$, see Table \ref{tab:cones}. The edge of a CSL is gapless and contains $\nu$ branches of chiral Majorana excitations, each branch carries a chiral central charge ${1\over2}$. The total chiral central charge is $c_-={\nu\over2}$, which gives rise to a measurable physical quantity --- the thermal Hall conductance which is quantized to $\kappa_{xy} /T=c_- \Lambda={\nu\over2}\Lambda$ with  $\Lambda = \pi k_B^2 / 6h$.

The mentioned gSL-III and the gSL-VI have opposite Chern numbers and can be distinguished from their different thermal Hall conductances. Thus, combining the DSF and the thermal Hall conductance of the weak field-induced CSL, we can completely distinguish all of the QSLs that appeared in the phase diagrams in Figs.~\ref{fig:AnisKGamma} and \ref{fig:AnisKGamma_weak}.

The Chern number $\nu$ can be tuned by changing the magnetic field\cite{note, PKSL}. For instance,  $\nu$ can be turned into $-\nu$ by reversing the direction of the magnetic field. Interestingly, our phase diagram Fig.~\ref{fig:AnisKGamma} provides an alternative way to change the Chern number $\nu$ without changing the magnetic field. For instance, the transition from the PKSL to the gSL-I can be achieved by increasing pressure (so as to increase $\delta_d$). In a weak magnetic field, the former has $\nu=5$ while the latter has $\nu=1$. Therefore, the $\nu=5$ phase can be driven to the $\nu=1$ phase by exerting proper pressure.  Specially, the continuous phase transition between the gSL-II to the gSL-III indicates that the Chern number can be changed from $\nu=1$ to $\nu=-1$ (and vice versa) with a continuous topological transition by tuning the pressure instead of reversing the direction of the magnetic field\cite{Qianghua,Nasu}. This continuous topological phase transition with Chern number changing by 2 is protected by the $\tilde{\mathscr C}_{2h}\times Z_2^T$ symmetry. Similarly, the phase transition from the CSL corresponding to the gSL-V to the one corresponding to the gSL-IV is also a topological transition where the Chern number changes by 2. \\
%and to the one corresponding to gSL-VI are, and 0, respectively

\section{Complete classification of nodal $Z_2$ QSLs}

Above we have shown that different gapless QSLs  in Figs.~\ref{fig:AnisKGamma} and \ref{fig:AnisKGamma_weak} belong to different phases since they have distinct physical properties. Generally, to distinguish different QSL phases, we need a classification theory of quantum phases.

The classification of QSLs is challenging due to the lack of local order parameters. Even though, things have been made clear for gapped QSLs, which are called the symmetry enriched topological (SET) phases. Based on the emergent gauge structure called the invariant gauge group (IGG) in the parton representation, PSG was proposed  to distinguish different mean-field spin liquid states\cite{igg}. The IGG describes the topological order carried by the QSL, while PSG distinguishes different QSL states that carry the same topological order and the same physical symmetry. It was realized that PSG actually describes the way how the spinon excitations ({\it i.e.} the gauge charge of the IGG) are symmetry fractionalized and are classified by the second group cohomology of the symmetry group\cite{Hemele,RanYing}. Furthermore, not only spionons but also the "magnetic" excitations ({\it i.e.} the gauge flux of the IGG) can carry fractional symmetry representations \cite{Hemele, ChengMengQiYang}. Therefore, depending on the fusion group of the Abelian anyons, a complete classification of gapped QSLs may need more than one set of PSGs to describe the fractional symmetry representations for both the "electronic" excitations and the "magnetic" excitations. More recently, it was realized that such classification may be over-complete, because it may include some anomalous classes which can only be realized on the surface of higher-dimensional phases \cite{ChengWang,ChenXie}. Therefore, the classification of true two-dimensional QSLs further requires an anomaly-free condition using group cohomology theory\cite{ChengWang}.

However, the complete classification of gapless QSLs is still an open question. Based on this work, we provide a clue to the answer. We do not go to the detail of the classification theory, which is beyond the scope of this work and is left for future study. We only list the key information that is necessary for a complete classification.

%(namely, the contribution of each cone to the Chern number when they are gapped by a magnetic field)
From our calculation (see Appendix \ref{sm:differentPSG}), we find that the PSG of all the QSLs in Figs.~\ref{fig:AnisKGamma} and \ref{fig:AnisKGamma_weak} is the same as that of the KSL. Namely, these QSLs have the same pattern of symmetry fractionalization in their spinon excitations. This indicates that PSG alone is not adequate to classify gapless QSLs. To distinguish these phases, we use the information of the cones, including the total number, the locations, the relation between them under symmetry operations, and their chiralities with respect to mass-generating perturbations. In summary, a complete classification of nodal $Z_2$ QSLs needs at least the following three ingredients: \\
\indent (1) the number of the cones and how they are symmetry-related (the positions of the cones form several $\pmb k$ stars);\\
\indent (2) the chirality of every cone; \\
\indent (3) the PSGs describing the symmetry fractionalization;\\
The indices of (1) and (2) for all the gapless $Z_2$ QSLs in our phase diagrams are shown in Fig.~\ref{fig:cones}. Noticing that the gauge flux excitations can also carry fractional symmetry representations which are included in the data in (3), we need to guarantee that the classified QSLs can be realized in two-dimensional lattice models with:\\
\indent (4) the anomaly-free condition. \\
The above approach can be generalized to classify nodal QSLs (including $U(1)$ Dirac QSLs) with other symmetry groups.
\\

\section{Discussion and Conclusion}

\subsection{Comparison to $\alpha$-RuCl$_3$ experiments}
Before concluding, we try to address the issue of the high-pressure experiments of $\alpha$-RuCl$_3$\cite{yu_pressure, sun_pressure, Bastien_pressure}, where the zigzag magnetic order is suppressed as the pressure goes above $0.8$ GPA. The resultant non-magnetic phase seem not to be a spin liquid because the low-energy spin fluctuations are very weak\cite{yu_pressure}.

Instead, a dimerized trivial ground state is enforced by the pressure accompanied by a structural phase transition \cite{Bastien_pressure}. To model this process, we adopt the dimer-type anisotropic $K$-$\Gamma$ interactions (with fixed $\Gamma/|K|=1.4$) plus additional antiferromagnetic Heisenberg interactions ($J$-terms) which exist on the short Ru-Ru bonds only. For simplicity, we assume that $J$ changes continuously and increases linearly with the anisotropy parameter $\delta_d$, namely $J/|K|=\xi\delta_d$, where $\xi$ is a phenomenological constant.

\begin{figure}[htbp]
\includegraphics[width=8.5cm]{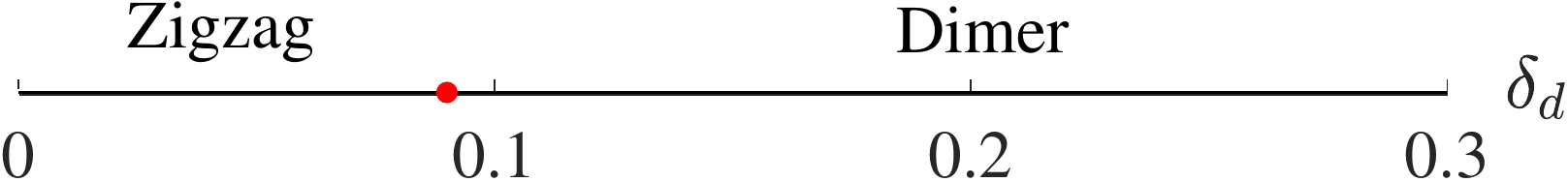} \
\caption{Phase diagram of the anisotropic $K$-$\Gamma$-$J$ model with $\Gamma/|K|=1.4$ and $J/|K|=20\delta_d$, where $J$ denotes the antiferromagnetic Heisenberg interactions existing on the $z$ bonds only. The phase transition is of first order.
}
\label{KGJz}
\end{figure}

If $\xi$ is very small, the phase diagram should remain qualitatively the same as $\xi=0$. In our VMC simulation, intermediate QSL phases which separate the zigzag ordered phase and the dimer phase are still found as long as $\xi<2$. However for $\xi>2$, the intermediate QSL phases disappear and a direct first-order transition from the zigzag phase to the dimer phase is observed. For instance, if $\xi=20$ the direct phase transition occurs at $\delta_d \simeq 0.09$, as illustrated in Fig.~\ref{KGJz}. This qualitatively agrees with the $\alpha$-RuCl$_3$ experimental results that a relatively low pressure can destroy the zigzag order. Our result indicates that in order to obtain QSLs by applying pressure, the fast increasing Heisenberg interactions on the strong bonds should be avoided. \\

\subsection{Conclusion}
In summary, we have studied the anisotropic $K$-$\Gamma$ model using variational Monte Carlo method. We consider two kinds of anisotropy, one is of the dimer type and the other is of the zigzag-chain type. The phase diagrams are given in Figs.~\ref{fig:AnisKGamma} and \ref{fig:AnisKGamma_weak}, respectively. Our calculations are benchmarked at $\Gamma=0$ where our result is completely consistent with the exact solution. When $\Gamma> 0$, we observe totally eleven QSLs, including the KSL, the proximate KSL, the new gapless spin liquid phases gSL-I$\sim$VIII plus one gapped $Z_2$ QSL phase.  Most of the phase transitions between the QSLs are of first order, but continuous phase transitions are also found which are characterized by the adiabatic merging and disappearing of the cones in pairs.

The gapless QSLs have symmetry protected gapless nodal points, where the number and the positions of the cones are important in distinguishing different QSL phases.  We find that the chiralities of the cones with respect to perturbations can be analyzed using $\pmb k\cdot\pmb p$ expansion. From the chirality of the cones, we can immediately obtain the total Chern number of the resultant CSLs induced by the weak magnetic fields applied along arbitrary directions. By applying magnetic fields along (111) direction and (-1,-1,-1)-direction, we could obtain 6 classes of non-Abelian $Z_2$ CSLs with Chern number $\nu=\pm1, \pm3, \pm5$, where the total number of such phases is 8 (up to edge chiral boson modes) according to Kitaev's seminal work\cite{Kitaev}.

From the above study, we further conclude that the PSG alone can not completely classify the gapless QSLs. For the $\tilde{\mathscr C}_{2h} \times Z_2^T$ symmetry we studied, a complete classification of all possible nodal $Z_2$ QSLs should also include the pattern of the cones $(n,m)$ and the chirality of every cone.

Since the strain in the lattice may be tuned by external pressure, our results are instructive for experimental realization of new gapless QSLs in related materials. Although pressure may also introduce additional antiferromagnetic Heisenberg exchanges on the short bonds, as occurred in $\alpha$-RuCl$_3$, the intermediate gapless QSLs phase can still exist as long as the Heisenberg terms are not very strong. Recently, new Kitaev materials are keeping being discovered\cite{LiYuan,NiNi,Sala,JQYan}, we predict that pressure induced gapless QSL phases may be observed in some of the candidates. Tuning pressure also provides a practical way for tuning possible phase transitions between different QSLs.

\section*{Acknowledgements}
The authors thank Y. Li, J. Ma and W.-Q. Yu for informative discussions. This work was supported by the Ministry of Science and Technology of China (Grant No. 2016YFA0300504), the NSF of China (Grants No. 11574392 and No. 11974421), and the Fundamental Research Funds for the Central Universities and the Research Funds of Renmin University of China (No. 19XNLG11).\\

%%%%%%%%%%%%%%%%%%%%%%%%%%%%
\appendix

\section{Variational Monte Carlo approach}\label{sm:VMC}

\subsection{Fermionic representation and variational Monte Carlo}
Variational Monte Carlo (VMC) is a powerful method to study quantum magnetism, especially quantum spin liquids (QSLs). It uses Gutzwiller projected mean-field states as trial wave functions. In this approach, the fermionic slave-particle representation $S_i^m =\frac{1}{2} C_i^\dagger \sigma^m C_i$ is introduced, where $C_i^\dagger = (c_{i\uparrow}^\dagger, c_{i\downarrow}^\dagger)$, $m \equiv x,y,z$, and $\sigma^m$ are Pauli matrices. The particle-number constraint $\hat{N_i} = c_{i\uparrow}^\dagger c_{i\uparrow} +c_{i\downarrow}^\dagger c_{i\downarrow} = 1$ should be imposed at every site to ensure that the size of the fermionic Hilbert space is the same as that of the original physical spin. The complex fermion operators can be seen as linear combinations of Kitaev's Majorana fermion operators, namely, $c_\up={1\over2}(b^z+ic), c_\dn={1\over2}(b^x+ib^y)$. The spin interactions in the Hamiltonian are rewritten in terms of interacting fermionic operators and  are decoupled into a noninteracting mean-field Hamiltonian.

The general mean-field Hamiltonian for a spin-orbit coupled spin liquid can be expressed
\begin{eqnarray}\label{Hmf}
H_{\rm mf}^{\rm SL} &=&  \sum_{\langle i,j \rangle\in\alpha\beta(\gamma)}  {\rm Tr} [U_{ji}^{(0)^{\gamma}} \psi_i^\dagger
\psi_j ] + {\rm Tr} [U_{ji}^{(1)^{\gamma}} \psi_i^\dagger (iR_{\alpha\beta}) \psi_j ]  \nonumber \\
& & + {\rm Tr} [U_{ji}^{(2)^{\gamma}} \psi_i^\dagger \sigma^\gamma \psi_j ] + {\rm Tr} [U_{ji}^{(3)^{\gamma}}
\psi_i^\dagger \sigma^\gamma R_{\alpha\beta} \psi_j ]  + {\rm H.c.}  \nonumber \\
& & + \sum_i {\rm Tr}  (\pmb \lambda_i\cdot  \psi_i \pmb \tau\psi_i^\dagger ),
\end{eqnarray}
where only the nearest-neighbor couplings are considered, $\psi_i = (C_i\ \bar C_i)$, $\bar C_i = (c_{i\downarrow}^\dag, -c_{i\uparrow}^\dag)^T$, and $R_{\alpha\beta} = - \frac{i}{\sqrt{2}} (\sigma^\alpha +\sigma^\beta)$ is a rotation matrix. The matrices $U_{ji}^{(0)^{\gamma}}, U_{ji}^{(1)^{\gamma}}, U_{ji}^{(2)^{\gamma}}, U_{ji}^{(3)^{\gamma}}$ can be expanded using the bases $\tau^0,\tau^x,\tau^y,\tau^z$ where $\tau^{x,y,z}$ are the Pauli matrices and $\tau^0$ is the identity matrix. In principle, all the expanding coefficients should be treated as variational parameters. However, as will be seen later, the number of variational parameters can be reduced since they should satisfy the symmetry requirements.
$\pmb \lambda^{x,y,z}$ are Lagrangian multipliers to ensure the $SU(2)$ gauge invariance\cite{Anderson} which can generally be set to zero in the VMC calculations unless there exists an external magnetic field.

To describe the magnetic order of the spin rotation symmetry breaking phases, we introduce a background field $\pmb M_i$ whose direction is adopted from the single-${\pmb Q}$ approximation\cite{singleQ} and whose amplitude (together with a canting angle) is determined by VMC. Therefore, the full mean-field Hamiltonian for the anisotropic $K$-$\Gamma$ model on the honeycomb lattice is
\begin{equation}\label{TotalHami}
H_{\rm mf}^{\rm total} = H_{\rm mf}^{\rm SL} - {\textstyle \frac{1}{2}} \sum_i
(\pmb {M}_i \cdot C_i^\dagger \pmb \sigma C_i + {\rm H.c.}).
\end{equation}
The essence of the VMC approach is that the local constraint is enforced by Gutzwiller projection. The Gutzwiller projected mean-field ground states provide a series of trial wave functions $|\Psi(x) \rangle = P_G |\Psi_{\rm mf}(x) \rangle$, where $x$ denotes the variational parameters. The energy of the trial state $E (x) = \langle \Psi(x) |H| \Psi(x) \rangle / \langle \Psi(x)| \Psi(x) \rangle$ is computed using Monte Carlo sampling, and the variational parameters $x$ are determined by minimizing the energy $E(x)$. Our calculations are performed on tori of up to 10$\times$10 unit cells, i.e.~of 200 lattice sites.

\subsection{Projective symmetry groups}
The number of variational parameters can be reduced if the symmetry of the mean-field Hamiltonian, namely, the projective symmetry group (PSG), is considered. The fermionic representation has a local $SU(2)$ gauge symmetry\cite{Anderson}. In the mean-field Hamiltonian (\ref{Hmf}), the $SU(2)$ "gauge symmetry" is broken and only its subgroup Z$_2$ is still a symmetry. This $Z_2$ symmetry is called the invariant gauge group (IGG). The PSG is the central extension of the physical symmetry group by the IGG\cite{igg}.

Aside from $\tilde{\mathscr C}_{2h} \times Z_2^T$, the anisotropic $K$-$\Gamma$ model is also invariant under the translational group generated by $\{T_1, T_2 \}$. After some calculations, we obtain 192 different PSGs given that the IGG is $Z_2$ (see Appendix \ref{sm:PSG}). The PSGs partially classify the possible spin liquid phases with the given symmetry group and IGG. A spin liquid mean-field Hamiltonian of the present model should respect one of the PSGs. For instance, Kitaev spin liquid (KSL) phase belongs to the class (I-B) \cite{You_PSG}.

The PSG reduces the number of allowed parameters and the exact forms of $U_{ji}^{(m)^\gamma}$ are given in Appendix \ref{sm:PSG}. We have adopted several different PSGs to construct different classes of trial spin liquid Hamiltonians. In class (I-B) case, 9 variational parameters $\{\rho_a^x,\rho_c^x,\rho_d^x,\phi_0^x,\phi_3^x,\phi_5^x,\phi_7^x,\phi_7^z,\theta\}$ are adopted, with
\Beq
&&U_{ji}^{(0)^z}=i\theta(\phi_0^x+\rho_a^x+\rho_c^x)\tau^0, \\
&&U_{ji}^{(0)^{x,y}}=i(\phi_0^x+\rho_a^x +\rho_c^x)\tau^0, \\
&&U_{ji}^{(1)^z}=i\theta\phi_3^x(\tau^x+\tau^y+\tau^z) +i\theta (\rho_a^x - \rho_c^x  + \rho_d^x) (\tau^x + \tau^y),\\
&&U_{ji}^{(1)^{x,y}}=i\phi_3^x(\tau^x+\tau^y+\tau^z) +i(\rho_a^x - \rho_c^x  + \rho_d^x) (\tau^{x,y} + \tau^z),\\
&&U_{ji}^{(2)^z}=i\theta\phi_5^x(\tau^x+\tau^y +\tau^z)+ i \theta(\rho_a^x + \rho_c^x ) \tau^z,\\
&&U_{ji}^{(2)^{x,y}}=i\phi_5^x(\tau^x+\tau^y +\tau^z)+ i(\rho_a^x + \rho_c^x ) \tau^{x,y},\\
&&U_{ji}^{(3)^z}=i(\phi_7^z+\theta \rho_c^x- \theta \rho_a^x - \theta \rho_d^x) (\tau^x - \tau^y),\\
&&U_{ji}^{(3)^{x,y}}=\pm i\phi_7^x(\tau^x+\tau^y+\tau^z) \pm i(\rho_c^x- \rho_a^x - \rho_d^x) (\tau^{y,x} - \tau^z).
\Eeq
In class (I-A) case, 9 variational parameters $\{\phi_1^x, \phi_1^{''x}, \phi_3^{x}, \phi_3^{''x}, \phi_5^{x}, \phi_5^{''x}, \phi_7^{x}, \phi_7^{''x}, \theta\}$ are adopt, with
\Beq
&&U_{ji}^{(0)^z}= \theta\phi_1^x(\tau^x-\tau^y), \ \ U_{ji}^{(0)^{x,y}}= \phi_1^x(\tau^x-\tau^y) +\phi_1^{''x}\tau^z,  \\
&&U_{ji}^{(1)^z}=i\theta\phi_3^{''x} \tau^z ,\ \ U_{ji}^{(1)^{x,y}}=\pm i\phi_3^x(\tau^x-\tau^y) + i\phi_3^{''x} \tau^z,\\
&&U_{ji}^{(2)^z}=i\theta\phi_5^{''x}\tau^z,\ \ U_{ji}^{(2)^{x,y}}=\pm i\phi_5^x(\tau^x-\tau^y) + i\phi_5^{''x} \tau^z,\\
&&U_{ji}^{(3)^z}=i\theta\phi_7^x (\tau^x - \tau^y),\ \ U_{ji}^{(3)^{x,y}}=i\phi_7^x(\tau^x-\tau^y) \pm i\phi_7^{''x}\tau^z.
\Eeq
The optimal parameters are determined variationally by minimizing the energy. We also use different magnetic ordered states as trial wave functions in VMC calculation.  The state with lowest energy is treated as the ground state. It turns out that all the QSLs in the phase diagrams (Figs.~\ref{fig:AnisKGamma} and \ref{fig:AnisKGamma_weak}) share the same PSG as the KSL.

\section{Classification of PSG with lattice anisotropy} \label{sm:PSG}
Here we present the classification of $Z_2$ PSG on the distorted honeycomb lattice with consideration of spin-orbit coupling. The full symmetry group (SG) is the direct product of wallpaper group and time reversal, with the presentation SG=$\{T,T_1,T_2,P,\sigma_m |T^2=1,P^2=1,\sigma_{m}^2=1 \}$ subject to 13 definition relations. The generators of the symmetry group are illustrated in Fig.~\ref{fig:AnisKGamma} or Fig.~\ref{fig:AnisKGamma_weak}. The four generators of the wallpaper group act on the honeycomb lattice in the following way:
\Beq
&&T_1(x_1,x_2,A)=(x_1+1,x_2,A), \\
&&T_1(x_1,x_2,B)=(x_1+1,x_2,B), \\
&&T_2(x_1,x_2,A)=(x_1,x_2+1,A), \\
&&T_2(x_1,x_2,B)=(x_1,x_2+1,B), \\
&&P(x_1,x_2,A)=(-x_1,-x_2,B), \\
&&P(x_1,x_2,B)=(-x_1,-x_2,A), \\
&&\sigma_m(x_1,x_2,A)=(x_2,x_1,B), \\
&&\sigma_m(x_1,x_2,B)=(x_2,x_1,A),
\Eeq
where each unit cell is labeled by integer coordinates $x_1$ and $x_2$ along the translation axes of $T_1$ and $T_2$. In the following, the index of sublattices $A,B$ will be omitted if an equation is independent of sublattices. Considering time reversal, the full symmetry group of the system contains 5 generators, $T,T_1,T_2,P,\sigma_m$, satisfying the following 13 definition relations:
\beq
&&T_1 T_2 T_1^{-1} T_2^{-1}=1 \label{b9}\\
&&T T_1 T T_1^{-1}=1 \label{b10}\\
&&T T_2 T T_2^{-1}=1 \label{b11}\\
&&P T_1 P T_1=1 \label{b12}\\
&&P T_2 P T_2=1 \label{b13}\\
&&\sigma_m T_1 \sigma_m^{-1} T_2^{-1}=1 \label{b14}\\
&&\sigma_m T_2 \sigma_m^{-1} T_1^{-1}=1 \label{b15}
\eeq
\beq
&&T^2=1 \label{b16}\\
&&P^2=1 \label{b17}\\
&&\sigma_m^2 =1 \label{b18}\\
&&T P T P=1 \label{b19}\\
&&T \sigma_m T \sigma_m=1 \label{b20}\\
&&P \sigma_m P \sigma_m =1\label{b21}
\eeq
All the pure gauge operations that leave the mean-field ansatz invariant form a subgroup of the PSG, known as the invariant gauge group (IGG). Since spinon pairing is non-vanishing in all the spin liquids obtained from our VMC, we only consider the case IGG = $Z_2$. Thus for each definition relation $g_n \dots g_2g_1 = 1$, there is a corresponding PSG representation
$$
G_{g_n}(g_{n-1}... g_1(i)) \dots G_{g_2}(g_1(i))G_{g_1}(i)=\eta_m,
$$
where $\eta_m = \pm\tau^0,\ m=1,2,...13$ are group elements in the IGG and these parameters determines the classification of the PSG. Gauge equivalent solutions of $G_g$ are considered to belong to the same class of PSG. To reduce the gauge redundancy, we will fix part of the gauge degrees of freedom in later discussion.

Firstly, after gauge transformations, one can set $G_{T_2}(x_1,x_2)=\tau^0$, and $G_{T_1}(x_1,0)=\tau^0$. Then Eq.~(\ref{b9}) can be represented as $G_{T_1}(x_1,x_2+1)=\eta_1 G_{T_1}(x_1,x_2)$, which yields the following solution:
\begin{gather}\label{b22}
G_{T_1}(x_1,x_2)=\eta_1^{x_2},\ \ G_{T_2}(x_1,x_2)=\tau^0.
\end{gather}
Substitute Eq.~(\ref{b22}) into the PSG representation of Eq.~(\ref{b10}) and Eq.~(\ref{b11}): $G_TKG_{T_1}K=\eta_2G_{T_1}$ and $G_TKG_{T_2}K=\eta_3G_{T_2}$, we obtain
\Beq
&& G_T(x_1+1,x_2)KG_T(x_1,x_2)=\eta_2,\\
&& G_T(x_1,x_2+1)KG_T(x_1,x_2)=\eta_3.
\Eeq
Combining Eq.~(\ref{b16}), namely $G_TKG_TK=\eta_8$, we obtain the solution of $G_T$
\begin{gather}\label{b23}
G_T(x_1,x_2)=\eta_2^{x_1}\eta_3^{x_2}\eta_8^{x_1+x_2}G_T(0,0).
\end{gather}
From Eqs.~(\ref{b12})$\sim$(\ref{b15}), we obtain the following equations:
{\small
\Beq
&&G_P(x_1+1,x_2,\alpha)G_{T_1}(x_1,x_2,\alpha)=\eta_4G_{T_1}^{-1}(-x_1,-x_2,\bar\alpha)G_P(x_1,x_2,\alpha),\\
&&G_P(x_1,x_2+1,\alpha)G_{T_2}(x_1,x_2,\alpha)=\eta_5G_{T_2}^{-1}(-x_1,-x_2,\bar\alpha)G_P(x_1,x_2,\alpha),\\
&&G_{\sigma_m}(x_1+1,x_2,\alpha)G_{T_1}(x_1,x_2,\alpha)=\eta_6G_{T_2}^{-1}(x_2,x_1,\bar\alpha)G_{\sigma_m}(x_1,x_2,\alpha),\\
&&G_{\sigma_m}(x_1,x_2+1,\alpha)G_{T_2}(x_1,x_2,\alpha)=\eta_7G_{T_1}^{-1}(x_2,x_1,\bar\alpha)G_{\sigma_m}(x_1,x_2,\alpha),
\Eeq
}
where $\alpha=A,B$ and $\bar\alpha$ stands for the opposite sub-lattice of $\alpha$.

Not all the parameters $\eta_m$ are independent. Some of them can be transformed into each other by certain gauge transformation $G_g (x_1,x_2,\alpha)\to \mu(x_1,x_2,\alpha)G_g(x_1,x_2,\alpha)$,where $\mu(x_1,x_2,\alpha)=\pm\tau^0$. It turns out that if some $G_g$ appears twice in an equation, the parameter $\eta_m$ in that equation is gauge invariant, otherwise that $\eta_m$ is not gauge independent and can be fixed to $\tau^0$ by some gauge transformation. For instance, $G_{T_1}$ or $G_{T_2}$ only shows up once in the equations of $\eta_6$ and $\eta_7$, so we can fix $\eta_6=\tau^0$ and $\eta_7=\tau^0$ by tuning the gauge of $G_{T_1}$ and $G_{T_2}$, respectively.

Therefore, the equations obtained from Eqs.~(\ref{b12})$\sim$(\ref{b15}), can be further simplified into the following sublattice-independent form
\Beq
&&G_P(x_1+1,x_2)=\eta_4G_P(x_1,x_2),\\
&&G_P(x_1,x_2+1)=\eta_5G_P(x_1,x_2),\\
&&G_{\sigma_m}(x_1+1,x_2)=\eta_1^{x_2}G_{\sigma_m}(x_1,x_2),\\
&&G_{\sigma_m}(x_1,x_2+1)=\eta_1^{x_1}G_{\sigma_m}(x_1,x_2),
\Eeq
which yields the solution
\beq
&&G_P(x_1,x_2)=\eta_4^{x_1}\eta_5^{x_2}G_P(0,0),\label{b24}\\
&&G_{\sigma_m}(x_1,x_2)=\eta_1^{x_1x_2}G_{\sigma_m}(0,0).\label{b25}
\eeq
The Eq.~(\ref{b25}) is consistent with Eq.~(\ref{b18}), $G_{\sigma_m}(A)G_{\sigma_m}(B)=G_{\sigma_m}(B)G_{\sigma_m}(A)=\eta_1^{2x_1x_2}\eta_{10}$. Furthermore, from Eqs.~(\ref{b19}) and (\ref{b20}) we find $\eta_2=\eta_3$, and from Eq.~(\ref{b21}) we find $\eta_4=\eta_5$.

Now all the $G_g(x_1,x_2,\alpha)$ has been reduced to $G_g(0,0,\alpha)$ with in a single unit cell. In later discussion $G_g(\alpha)$ will be used to denote $G_g(0,0,\alpha)$. The remaining task is to determine $G_T(\alpha)$, $G_P(\alpha)$ and $G_{\sigma_m}(\alpha)$.  Eqs.(\ref{b16})$\sim$(\ref{b21}) yield the following constraints
\beq
&&G_T(A)KG_T(A)K=G_T(B)KG_T(B)K=\eta_8 \label{b26}, \\
&&G_P(A)G_P(B)=G_P(B)G_P(A)=\eta_9 \label{b32},\\
&&G_T(B)KG_P(A)G_T(A)K=\eta_{11}G_P(A) \label{b27}, \\
&&G_T(A)KG_P(B)G_T(B)K=\eta_{11}G_P(B) \label{b28}, \\
&&G_T(B)KG_{\sigma_m}(A)G_T(A)K=\eta_{12}G_{\sigma_m}(A) \label{b29}, \\
&&G_T(A)KG_{\sigma_m}(B)G_T(B)K=\eta_{12}G_{\sigma_m}(B) \label{b30}, \\
&&(G_P(A)G_{\sigma_m}(B))^2=(G_P(B)G_{\sigma_m}(A))^2=\eta_{13} \label{b31}.
\eeq
We start from the solution of $G_T$. Suppose $G_T = a_0\tau^0+ia_l\sigma_l$, $l=1,2,3$ to be the most general $SU(2)$ matrix. From Eq.~(\ref{b26}), we find
$$
G_TG_T^*=(a_0^2+a_1^2-a_2^2+a_3^2)\tau^0+2ia_2(a_3\tau^x+a_0\tau^y-a_1\tau^z){}{}
$$
If $\eta_8=-\tau^0$, the solution is $a_2=\pm 1, a_0=a_1=a_3=0$, i.e.  $G_T= \pm i\tau^y$. While if $\eta_8=\tau^0$, then $a_2=0, \ \ a_0^2+a_1^2+a_3^2=1,$ we can choose a solution $G_T(A)=G_T(B)=\tau^0$. In the following we will discuss these two cases separately.

Class (I): $\eta_8=-\tau^0$. We choose
\Beq
G_T(A)=i\tau^y,\ \ G_T(B)=i\eta_{14}\tau^y,
\Eeq
where $\eta_{14}=\pm \tau^y$. Substituting into Eqs.~(\ref{b27})$\sim$(\ref{b30}), we find $\eta_{11}=\eta_{12}=\eta_{14}$. Without losing generality, we can fix
\Beq
G_{\sigma_m}(A)=\tau^0,\ \ G_{\sigma_m}(B)=\eta_{10}.
\Eeq
Plugging into Eq.~(\ref{b31}), we obtain $(G_P(A))^2=(G_P(B))^2=\eta_{13}$. According to the sign of $\eta_{13}$, the class (I) is divided into two subclasses. Notice that $\eta_2=\eta_8=-\tau^0$ because time reversal operation ($(G_TK)^2=-1$) is independent on coordinates and sub-lattices \cite{igg}.

Class (I-A): $\eta_{13}=\tau^0$. The solution of Eq.~(\ref{b32}) is
\Beq
G_P(A)=\tau^0,\ \ G_P(B)=\eta_9.
\Eeq
The solutions in the class (I-A) are summarized as
\beq
&&G_{T_1}(x_1,x_2)=\eta_1^{x_2} ,\\
&&G_{T_2}(x_1,x_2)=\tau^0  ,\\
&&G_T(x_1,x_2,A)=i\tau^y  ,\\
&&G_T(x_1,x_2,B)=i\eta_{11}\tau^y  ,\\
&&G_P(x_1,x_2,A)=\eta_4^{x_1+x_2}\tau^0 \label{b35},\\
&&G_P(x_1,x_2,B)=\eta_4^{x_1+x_2}\eta_9 \label{b36},\\
&&G_{\sigma_m}(x_1,x_2,A)=\eta_1^{x_1x_2}\tau^0 \label{b37},\\
&&G_{\sigma_m}(x_1,x_2,B)=\eta_1^{x_1x_2}\eta_{10} \label{b38},
\eeq
which are controlled by $\eta_1, \eta_4, \eta_9, \eta_{10}, \eta_{11}$, providing $2^5=32$ PSG's.\\

Class (I-B): $\eta_{13}=-\tau^0$. Because the global gauge freedom has not been fixed, we can set
\Beq
G_P(A)=i\tau^x,\ \ G_P(B)=-i\eta_9\tau^x.
\Eeq
The solutions in the class (I-B) are summarized as
\beq
&&G_{T_1}(x_1,x_2)=\eta_1^{x_2}, \\
&&G_{T_2}(x_1,x_2)=\tau^0,  \\
&&G_T(x_1,x_2,A)=i\tau^y,  \\
&&G_T(x_1,x_2,B)=i\eta_{11}\tau^y,  \\
&&G_P(x_1,x_2,A)=i\eta_4^{x_1+x_2}\tau^x \label{b43}, \\
&&G_P(x_1,x_2,B)=-i\eta_4^{x_1+x_2}\eta_9\tau^x \label{b44}, \\
&&G_{\sigma_m}(x_1,x_2,A)=\eta_1^{x_1x_2}\tau^0 \label{b45}, \\
&&G_{\sigma_m}(x_1,x_2,B)=\eta_1^{x_1x_2}\eta_{10} \label{b46},
\eeq
which are controlled by $\eta_1, \eta_4, \eta_9, \eta_{10}, \eta_{11}$, providing $2^5=32$ PSG's.\\

Class (II): $\eta_8=\tau^0$. In this case
\Beq
G_T(A)=G_T(B)=\tau^0.
\Eeq
Therefore, Eqs.~(\ref{b27})$\sim$(\ref{b30}) become $KG_P(A)K=\eta_{11}G_P(A)$, $KG_P(B)K=\eta_{11}G_P(B)$, $KG_{\sigma_m}(A)K=\eta_{12}G_{\sigma_m}(A)$ and $KG_{\sigma_m}(B)K=\eta_{12}G_{\sigma_m}(B)$. The general solution of $KG_gK = G_g$ is $G_g = e^{i\tau^y\theta}$, while the general solution of $KG_gK = -G_g$ is $G_g = i\tau^ze^{i\tau^y\theta}$. According to the sign of $\eta_{11}$ and $\eta_{12}$, the class (II) is divided into four subclasses.

Class (II-A1): $\eta_{11}=\eta_{12}=\tau^0$. Then we can obtain the general solution of Eqs.~(\ref{b27})$\sim$(\ref{b30}). The solutions in the class (II-A1) are summarized as
\Beq
&&G_{T_1}(x_1,x_2)=\eta_1^{x_2} ,\\
&&G_{T_2}(x_1,x_2)=\tau^0 ,\\
&&G_T(x_1,x_2,A)=\eta_2^{x_1+x_2}\tau^0  ,\\
&&G_T(x_1,x_2,B)=\eta_2^{x_1+x_2}\tau^0  ,\\
&&G_P(x_1,x_2,A)=\eta_4^{x_1+x_2} e^{i\tau^y\theta_1} ,\\
&&G_P(x_1,x_2,B)=\eta_4^{x_1+x_2}\eta_9 e^{-i\tau^y\theta_1} ,\\
&&G_{\sigma_m}(x_1,x_2,A)=\eta_1^{x_1x_2} e^{i\tau^y\theta_3} ,\\
&&G_{\sigma_m}(x_1,x_2,B)=\eta_1^{x_1x_2} \eta_{10} e^{-i\tau^y\theta_3},
\Eeq
which are determined by $\eta_1, \eta_2, \eta_4, \eta_9, \eta_{10}$, providing $2^5=32$ PSG's. Here $\theta_1$ can be any angle  and $\theta_3$ is dependent on $\theta_1$ according to the sign of $\eta_{13}$. More precisely, $\theta_3=\theta_1$ if $\eta_{13}=\tau^0$ and $\theta_3=\theta_1- {\pi\over2}$ if $\eta_{13}=-\tau^0$. \\

Class (II-B1): $-\eta_{11}=\eta_{12}=\tau^0$. Then we can obtain the general solution of Eqs.~(\ref{b27})$\sim$(\ref{b30}). The solutions in the class (II-B1) are summarized as
\Beq
&&G_{T_1}(x_1,x_2)=\eta_1^{x_2}, \\
&&G_{T_2}(x_1,x_2)=\tau^0 , \\
&&G_T(x_1,x_2,A)=\eta_2^{x_1+x_2}\tau^0 , \\
&&G_T(x_1,x_2,B)=\eta_2^{x_1+x_2}\tau^0  ,\\
&&G_P(x_1,x_2,A)=i\eta_4^{x_1+x_2} \tau^z e^{i\tau^y\theta_1} ,\\
&&G_P(x_1,x_2,B)=-i\eta_4^{x_1+x_2}\eta_9 \tau^z e^{-i\tau^y\theta_1} ,\\
&&G_{\sigma_m}(x_1,x_2,A)=\eta_1^{x_1x_2} e^{i\tau^y\theta_3} ,\\
&&G_{\sigma_m}(x_1,x_2,B)=\eta_1^{x_1x_2} \eta_{10} e^{-i\tau^y\theta_3},
\Eeq
which are determined by $\eta_1, \eta_2, \eta_4, \eta_9, \eta_{10}$, providing $2^5=32$ PSG's. Here $\theta_1$ and $\theta_3$ can be any angle.\\

Class (II-A2): $-\eta_{11}=-\eta_{12}=\tau^0$. Then we can obtain the general solution of Eqs.~(\ref{b27})$\sim$(\ref{b30}). The solutions in the class (II-A2) are summarized as
\Beq
&&G_{T_1}(x_1,x_2)=\eta_1^{x_2}, \\
&&G_{T_2}(x_1,x_2)=\tau^0 , \\
&&G_T(x_1,x_2,A)=\eta_2^{x_1+x_2}\tau^0 , \\
&&G_T(x_1,x_2,B)=\eta_2^{x_1+x_2}\tau^0 , \\
&&G_P(x_1,x_2,A)=\eta_4^{x_1+x_2} e^{i\tau^y\theta_1} ,\\
&&G_P(x_1,x_2,B)=-i \eta_4^{x_1+x_2}\eta_9 \tau^z e^{-i\tau^y\theta_1} ,\\
&&G_{\sigma_m}(x_1,x_2,A)=\eta_1^{x_1x_2} e^{i\tau^y\theta_3} ,\\
&&G_{\sigma_m}(x_1,x_2,B)=-i \eta_1^{x_1x_2} \eta_{10} \tau^z e^{-i\tau^y\theta_3},
\Eeq
which are determined by $\eta_1, \eta_2, \eta_4, \eta_9, \eta_{10}$, providing $2^5=32$ PSG's. Here $\theta_1$ can be any angle while $\theta_3$ is dependent on $\theta_1$ according to the sign of $\eta_{13}$). More precisely, $\theta_3=\theta_1$ if $\eta_{13}=-\tau^0$ and $\theta_3=\theta_1-\pi/2$ if $\eta_{13}=\tau^0$.\\

Class (II-B2): $\eta_{11}=-\eta_{12}=\tau^0$. Then we can obtain the general solution of Eqs.~(\ref{b27})$\sim$(\ref{b30}). The solutions in the class (II-B2) are summarized as
\Beq
&&G_{T_1}(x_1,x_2)=\eta_1^{x_2}, \\
&&G_{T_2}(x_1,x_2)=\tau^0 , \\
&&G_T(x_1,x_2,A)=\eta_2^{x_1+x_2}\tau^0 , \\
&&G_T(x_1,x_2,B)=\eta_2^{x_1+x_2}\tau^0 , \\
&&G_P(x_1,x_2,A)=\eta_4^{x_1+x_2} e^{i\tau^y\theta_1} ,\\
&&G_P(x_1,x_2,B)=\eta_4^{x_1+x_2}\eta_9 e^{-i\tau^y\theta_1} ,\\
&&G_{\sigma_m}(x_1,x_2,A)=i \eta_1^{x_1x_2} \tau^z e^{i\tau^y\theta_3} ,\\
&&G_{\sigma_m}(x_1,x_2,B)=-i \eta_1^{x_1x_2} \eta_{10} \tau^z e^{-i\tau^y\theta_3},
\Eeq
which are determined by $\eta_1, \eta_2, \eta_4, \eta_9, \eta_{10}$, providing $2^5=32$ PSG's. Here $\theta_1$ and $\theta_3$ can be any angle.

Finally, the number of algebraic PSG's in our classification is 192.

The PSG of Kitaev's exact spin liquid solution (we will call it Kitaev PSG in later discussion) is belonging to one of the above 192 classifications. Noticing that the $c$-fermions in Kitaev's solution\cite{Kitaev} is never mixed with other flavors, i.e. $\{b_x,b_y,b_z\}$, the corresponding PSG should keep $c$ fermions invariant. Under this condition, it is easy to figure out the gauge operations $G_g$\cite{You_PSG},
\beq
&&G_{T_1}=G_{T_2}=1, \label{PSG_T12} \\
&&G_P(A)=-G_P(B)=-\tau^0 ,\label{PSG_P}\label{KitaevPSG_P}\\
&&G_{\sigma_m}(A)=-G_{\sigma_m}(B)=e^{-i\frac{\pi}{2\sqrt{2}}(\tau^x-\tau^y)} ,\label{PSG_M}\\
&&G_T(A)=-G_T(B)=i\tau^y.\label{PSG_T}
\eeq
It turns out that the PSG of Kitaev spin liquid can be identified with one of the PSG in class (I-B), with invariants $\eta_1=\eta_4=\tau^0$ and $\eta_9=\eta_{10}=\eta_{11}=-\tau^0$ upon $SU(2)$ gauge transformations on $G_P, G_{\sigma_m}$ in Eqs.~(\ref{b43})$\sim$(\ref{b46}).
The gauge transformations are $W_A=ie^{i\frac{3\pi}{8}\tau^z} \tau^x$ on A-sublattice and $W_B=e^{i\frac{3\pi}{8}\tau^z}$ on B-sublattice.

\section{Spin-liquid states based on PSG}\label{sm:differentPSG}
Although we obtain 192 algebraic PSGs, it is impractical to study all the spin-liquid ansatz respecting all the different PSGs. In our VMC calculations, we only consider several PSGs which are close to the Kitaev's PSG class.

If only nearest-neighbor coupling terms are considered, the most general mean-field Ansatz takes the form Eq.~(\ref{Hmf}).
% \begin{eqnarray}\label{eq:MF}
% H_{\rm mf} &= & \sum_{\langle i,j \rangle\in\alpha\beta(\gamma)} {\rm Tr} [U_{ji}^{(0)^{\gamma}} \psi_i^\dagger
% \psi_j ] + {\rm Tr} [U_{ji}^{(1)^{\gamma}} \psi_i^\dagger (iR_{\alpha\beta}) \psi_j ] \nonumber \\
% & & + {\rm Tr} [U_{ji}^{(2)^{\gamma}} \psi_i^\dagger \sigma^\gamma \psi_j ] + {\rm Tr} [U_{ji}^{(3)^{\gamma}}
% \psi_i^\dagger \sigma^\gamma R_{\alpha\beta} \psi_j ].
% \end{eqnarray}
If the above ansatz describes a QSL state it should preserve certain PSG. On the other hand, the PSG restricts the number of parameters in the mean-field ansatz $U_{ji}^{(m)^\gamma}$. In the following we will firstly give the explicit form of spin liquid ansatz for several special PSGs. And we will perform Gutzwiller projection to these ansatz and pick up the one with the lowest energy as the ground state.

\subsection{Spin-liquid mean field ansatz}\label{mfansatz}
We first consider a special solution in class (I-B), namely the Kitaev class with invariants $\eta_1=\eta_4=\tau^0$ and $\eta_9=\eta_{10}=\eta_{11}=-\tau^0$.

To preserve the mirror symmetry (\ref{PSG_M}) and time reversal symmetry (\ref{PSG_T}), it requires that $U_{ji}^{(m)^\gamma}, m=0,1,2,3$ take the following form: on the z-bonds
\beq
&&U_{ji}^{(0)^z}=i\phi_0^z+\phi_1^z(\tau^x-\tau^y), \label{c2}\\
&&U_{ji}^{(1)^z}=i\phi_3^z(\tau^x+\tau^y) +i\phi_3^{'z}\tau^z ,\\
&&U_{ji}^{(2)^z}=i\phi_5^z(\tau^x+\tau^y) +i\phi_5^{'z}\tau^z ,\\
&&U_{ji}^{(3)^z}=\phi_6^z+i\phi_7^z(\tau^x-\tau^y);
\eeq
and on the x-bonds and y-bonds
\beq
&&U_{ji}^{(0)^{x}}=i\phi_0^x+\phi_1^x\tau^x-\phi_1^{'x}\tau^y +\phi_1^{''x}\tau^z, \\
&&U_{ji}^{(1)^{x}}=i\phi_3^x\tau^x+i\phi_3^{'x}\tau^y +i\phi_3^{''x}\tau^z ,\\
&&U_{ji}^{(2)^{x}}=i\phi_5^x\tau^x+i\phi_5^{'x}\tau^y +i\phi_5^{''x}\tau^z ,\\
&&U_{ji}^{(3)^{x}}=\phi_6^x+i\phi_7^x\tau^x+i\phi_7^{'x}\tau^y +i\phi_7^{''x}\tau^z,\\
&&U_{ji}^{(0)^{y}}=i\phi_0^x+\phi_1^{'x}\tau^x-\phi_1^x\tau^y - \phi_1^{''x}\tau^z, \\
&&U_{ji}^{(1)^{y}}=i\phi_3^{'x}\tau^x+i\phi_3^{x}\tau^y +i\phi_3^{''x}\tau^z , \\
&&U_{ji}^{(2)^{y}}=i\phi_5^{'x}\tau^x+i\phi_5^{x}\tau^y +i\phi_5^{''x}\tau^z, \\
&&U_{ji}^{(3)^{y}}=\phi_6^x-i\phi_7^{'x}\tau^x-i\phi_7^{x}\tau^y -i\phi_7^{''x}\tau^z. \label{c13}
\eeq

The inversion symmetry (\ref{PSG_P}) further requires that the parameters $\phi_1^x, \phi_1^{x'}, \phi_1^{x''}$ and $\phi_6^x, \phi_6^z$ must be vanishing. If the full symmetry group $G=\tilde{\mathscr C}_{2h} \times Z_2^T$ is considered, then the allowed parameters include $ \phi_0^z, \phi_3^z, \phi_3^{'z}, \phi_5^z, \phi_5^{'z}, \phi_7^z, \phi_0^x, \phi_3^x, \phi_3^{'x}, \phi_3^{''x}, \phi_5^x, \phi_5^{'x}, \phi_5^{''x}, \phi_7^{x}, \phi_7^{'x}, \phi_7^{''x} $. Besides these, the Kitaev decoupling further contribute a few parameters, namely
\Beq
H_{\rm mf} = \sum_{\langle i,j\rangle\in \gamma } \left[i\rho_a^\gamma c_ic_j + i\rho_c^\gamma b_i^\gamma b_j^\gamma + i\rho_d^\gamma (b_i^\alpha b_j^\beta + b_i^\beta b_j^\alpha)\right].
\Eeq
These parameters can be transformed into matrix as
\beq
&&{\tilde U}_{ji}^{(0)^{\gamma}} = i (\rho_a^\gamma + \rho_c^\gamma),\label{tildeU0}\\
&&{\tilde U}_{ji}^{(1)^\gamma} = i (\rho_a^\gamma - \rho_c^\gamma  + \rho_d^\gamma) (\tau^\alpha + \tau^\beta),\\
&&{\tilde U}_{ji}^{(2)^\gamma} = i (\rho_a^\gamma + \rho_c^\gamma ) \tau^\gamma,\\
&&{\tilde U}_{ji}^{(3)^\gamma} = i (\rho_c^\gamma - \rho_a^\gamma - \rho_d^\gamma) (\tau^\alpha - \tau^\beta).\label{tildeU3}
\eeq
In principle, all these parameters are used as independent variational parameters in the VMC calculation. For simplicity, we let $\phi_3^z=\phi_3^{'z}=\phi_3^{''z}$, $\phi_5^z=\phi_5^{'z}=\phi_5^{''z}$, $\phi_7^x=\phi_7^{'x}=\phi_7^{''x}$, $\rho_a^x=\rho_a^y$, $\rho_c^x=\rho_c^y$, $\rho_d^x=\rho_d^y$ and $\theta=\frac{\rho_a^z}{\rho_a^x}=\frac{\rho_c^z}{\rho_c^x}=\frac{\rho_d^z}{\rho_d^x}=\frac{\phi_0^z}{\phi_0^x}=\frac{\phi_3^z}{\phi_3^x}=\frac{\phi_5^z}{\phi_5^x}$. It is reasonable since the values of these parameters are small and have small contribution to the energy. Therefore,  in our VMC calculation we adopt the following parameters: $\rho_a^x$, $\rho_c^x$, $\rho_d^x$, $\phi_0^x$, $\phi_3^x$, $\phi_5^x$, $\phi_7^x$, $\phi_7^z$ and $\theta$.

As another example in class (I-B), we give the ansatz with uniform $\pi$-flux\cite{SSZhang}. The invariants are slightly different from the above: $\eta_4=\tau^0$ and $\eta_1=\eta_9=\eta_{10}=\eta_{11}=-\tau^0$.  The general form preserving the mirror symmetry and time reversal symmetry reads $\overline{U}_{ji}^{(m)^\gamma} = (-\tau^0)^{ji} (U_{ji}^{(m)^\gamma} + {\tilde U}_{ji}^{(m)^{\gamma}})$, where $U_{ji}^{(m)^\gamma}$ are given in Eq.~(\ref{c2})$\sim$(\ref{c13}) and ${\tilde U}_{ji}^{(m)^\gamma}$ are given by Eq.~(\ref{tildeU0})$\sim$(\ref{tildeU3}). We use $(-\tau^0)^{ji}$ to note the sign pattern of the uniform $\pi$-flux in each hexagon with doubled unit cell. Therefore, the $\pi$-flux state also contains 9 variational parameters: $\rho_a^x$, $\rho_c^x$, $\rho_d^x$, $\phi_0^x$, $\phi_3^x$, $\phi_5^x$, $\phi_7^x$, $\phi_7^z$ and $\theta$. The $\pi$-flux state is generally gapped.

Secondly, we provide another example in class (I-A), with invariants $\eta_1=\eta_4=\eta_9=\tau^0$ and $\eta_{10}=\eta_{11}=-\tau^0$.

The general form of $U_{ji}^{(m)^\gamma}$ that reserves the mirror symmetry (\ref{b37})$\sim$(\ref{b38}) is similar to the form in the Kitaev PSG. The inversion symmetry (\ref{b35})$\sim$(\ref{b36}) further requires that the parameters $\phi_0^z, \phi_3^z, \phi_5^z, \phi_6^z, \phi_0^x$ and $\phi_6^x$ must be vanishing. To reduce the number of variational parameters, we let $\phi_1^{x}=\phi_1^{'x}$, $ \phi_3^{x}=-\phi_3^{'x}$, $\phi_5^{x}=-\phi_5^{'x}$, $\phi_7^{x}=-\phi_7^{'x}$ and $\theta=\frac{\phi_1^z}{\phi_1^x}=\frac{\phi_3^{'z}}{\phi_3^{''x}}=\frac{\phi_5^{'z}}{\phi_5^{''x}}=\frac{\phi_7^z}{\phi_7^x}$. Therefore, in our VMC calculation we adopt the following parameters for the given PSG in class (I-A): $\phi_1^x$, $\phi_1^{''x}$, $\phi_3^{x}$, $\phi_3^{''x}$, $\phi_5^{x}$, $\phi_5^{''x}$, $\phi_7^{x}$, $\phi_7^{''x}$ and $\theta$.

\begin{figure}[t]
\includegraphics[width=8.5cm]{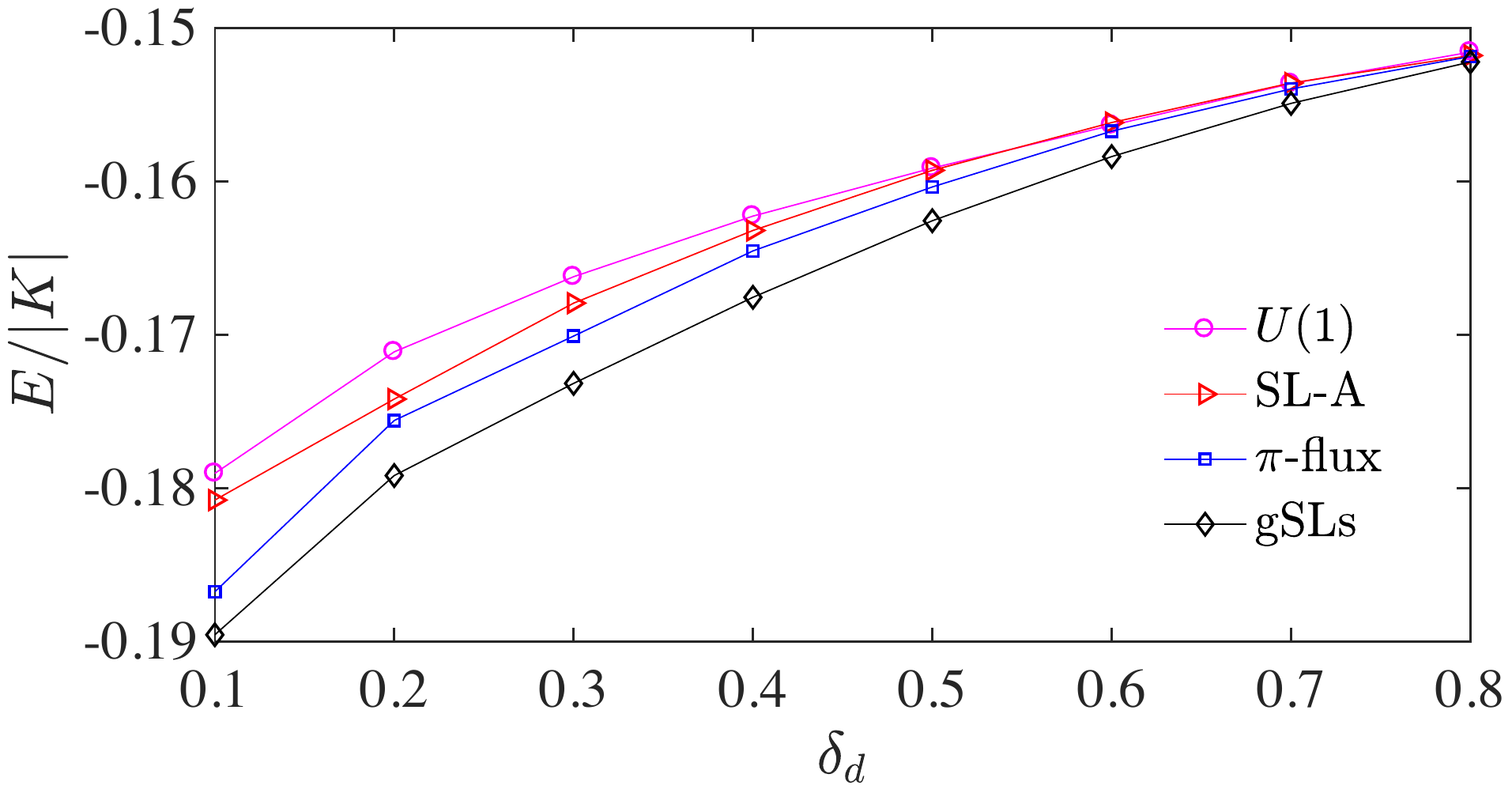} \
\caption{Energy (per site) vs $\delta_d$ with interactions $\Gamma/|K|=0.1$. The gSLs are always lowest in energy. }
\label{Gamma01}
\end{figure}

\begin{figure}[t]
\includegraphics[width=8.5cm]{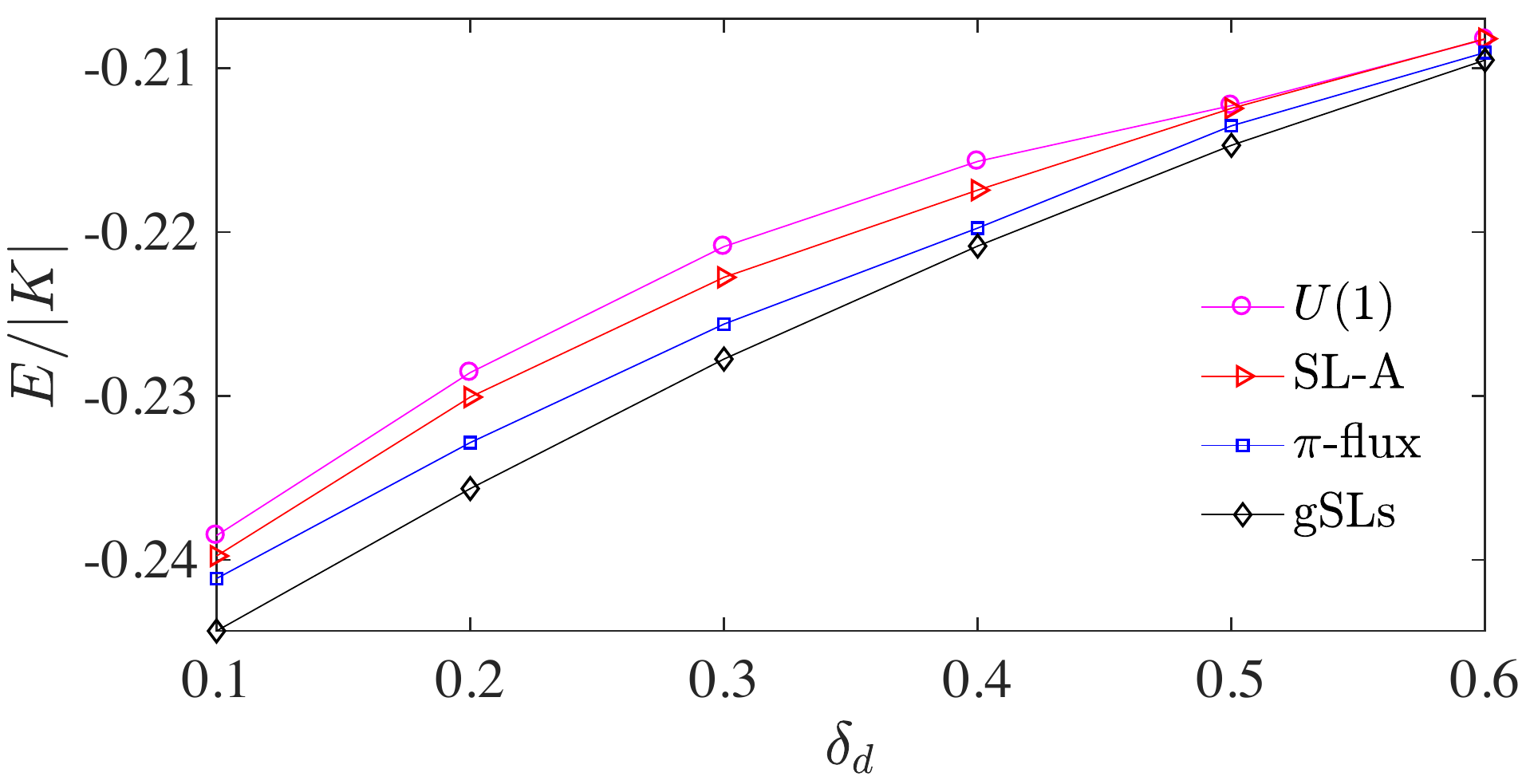} \
\caption{Energy (per site) vs $\delta_d$ with interactions $\Gamma/|K|=0.3$. The gSLs are always lowest in energy. }
\label{Gamma03}
\end{figure}

\subsection{Gutzwiller projection and the VMC-selected ground states}\label{gutzVMC}
In this section, we only consider the parameter interval where magnetically ordered states are not favored in energy. We perform Gutzwiller projection to the ansatz given in Appendix \ref{mfansatz} to calculate the energy and further determine the ground state using VMC.  In the following discussion the Gutzwiller projected gapless $Z_2$ ansatz's will be noted as `gSL's, especially the spin liquid  belonging to PSG class (I-A) is noted as SL-A. In addition to these $Z_2$ spin liquids, we also consider competing $U(1)$ spin liquid ansatz's.

Our VMC calculations are performed on a lattice of up to $10\times10$ unit cells ($i.e.$200 sites). It turns out that all the spin liquid states appearing in phase diagrams Figs.~\ref{fig:AnisKGamma} and \ref{fig:AnisKGamma_weak} belong to the same PSG class --- the Kitaev's PSG class. It should be noted that, as illustrated in Ref.\onlinecite{PKSL}, QSL states preserving the same PSG can fall into different phases. Here we only compare the projected states from different PSGs and will leave the discussion of distinguishing different quantum phases (with the same PSG) to Appendices \ref{sm:DSF} and \ref{sm:Chern}.

We first consider the case with dimer-anisotropy. The data with $\Gamma/|K|=0.1$ and $\Gamma/|K|=0.3$ are shown in  Fig.~\ref{Gamma01}  and Fig.~\ref{Gamma03}, respectively. It can be seen that the gSLs are always the lowest in energy comparing to all the other ansatz's.  We have confirmed that these gSLs are all $Z_2$ deconfined. The $Z_2$ deconfinement can be reflected in the ground state degeneracy (on a torus) of the resultant gapped state in a magnetic field, as listed in Table.~\ref{tab:GSD} in Appendix \ref{sm:Chern}.

\begin{figure}[t]
\includegraphics[width=8.5cm]{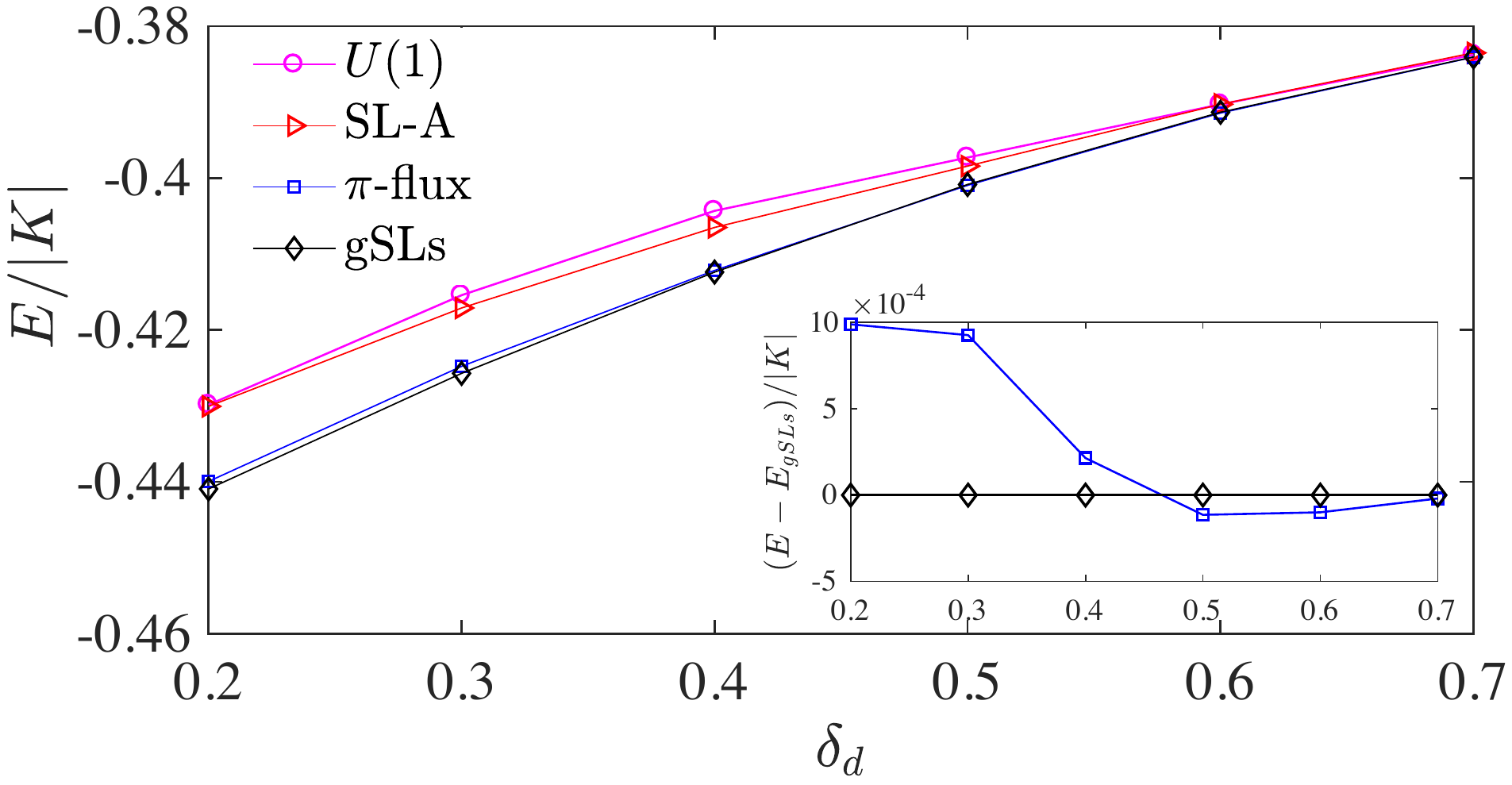} \
\caption{Energy (per site) vs $\delta_d$ for interactions $\Gamma/|K|=1$. A first-order phase transition between gSLs and $\pi$-flux state is found. The insert illustrates the energy difference between the two lowest ones.}
\label{Gamma10}
\end{figure}

\begin{figure}[t]
\includegraphics[width=8.5cm]{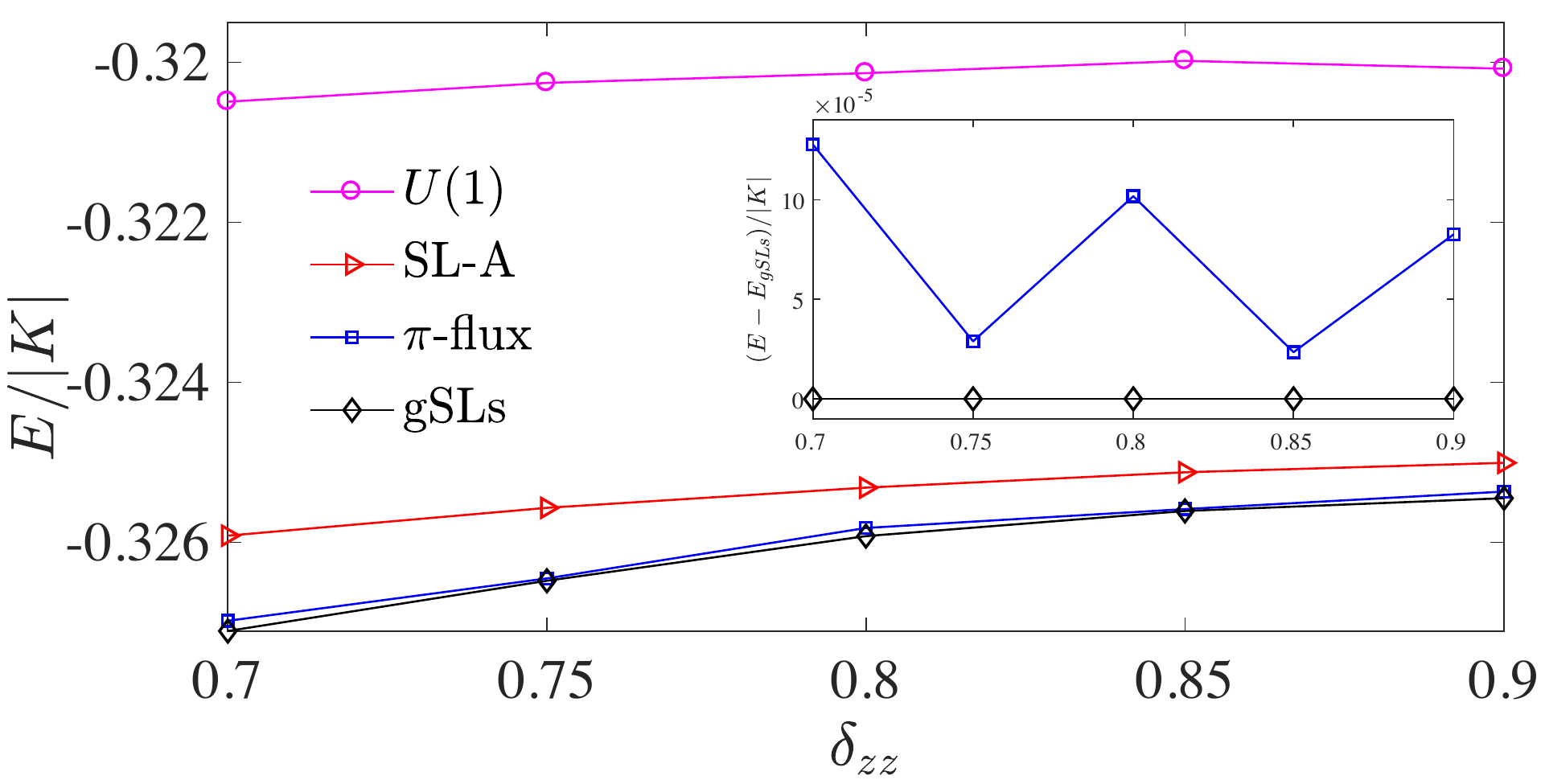} \
\caption{Energy (per site) vs $\delta_{zz}$ for interactions $\Gamma/|K|=0.6$ with zigzag anisotropy. The gSLs are always lowest in energy. The insert illustrates the energy difference between the two lowest ones.}
\label{Gamma06}
\end{figure}

The VMC calculations seem to indicate that at big $\Gamma$ there is a phase transition from the gSLs to a $\pi$-flux state (belonging to another PSG in class (I-B)) before the system enters the trivial dimer phase. The data with $\Gamma/|K|=1$ are shown in Fig.~\ref{Gamma10}. The $\pi$-flux state is very competing in energy at a large range of $\delta_d$ and beat the gSL-VI state as $\delta_d>0.47$. However, this $\pi$-flux state is $Z_2$ confined since it shows no degeneracy on a torus (see Appendix \ref{sm:Chern}). This means that the Gutzwiller projected $\pi$-flux state is trivial and belongs to the dimer phase.

We next consider the case with zigzag-type anisotropy. In Fig.~\ref{Gamma06}, the data for $\Gamma/|K|=0.6$ are shown. The gSLs are still favored among all the trial states. The $\pi$-flux state is very competing but still a little bit higher in energy comparing with the gSL-VII or gSL-VIII. To verify the $Z_2$ deconfinement of the projected states, we put the gSL-VII and gSL-VIII into a small magnetic field along ${1\over\sqrt3}(\pmb x+\pmb y+\pmb z)\equiv (111)$ direction. For both cases, the resultant gapped states are "almost" fourfold degenerate on a torus (for details see Appendix \ref{sm:Chern}).

\section{Finite-size effect}
%In view of all of these complexities, we performed a large number of studies to ensure
In this appendix, we illustrate that the phase diagrams presented in Figs.~\ref{fig:AnisKGamma} and \ref{fig:AnisKGamma_weak} are qualitatively the same in the thermal dynamic limit.

\begin{figure}[htbp]
\includegraphics[width=8.cm]{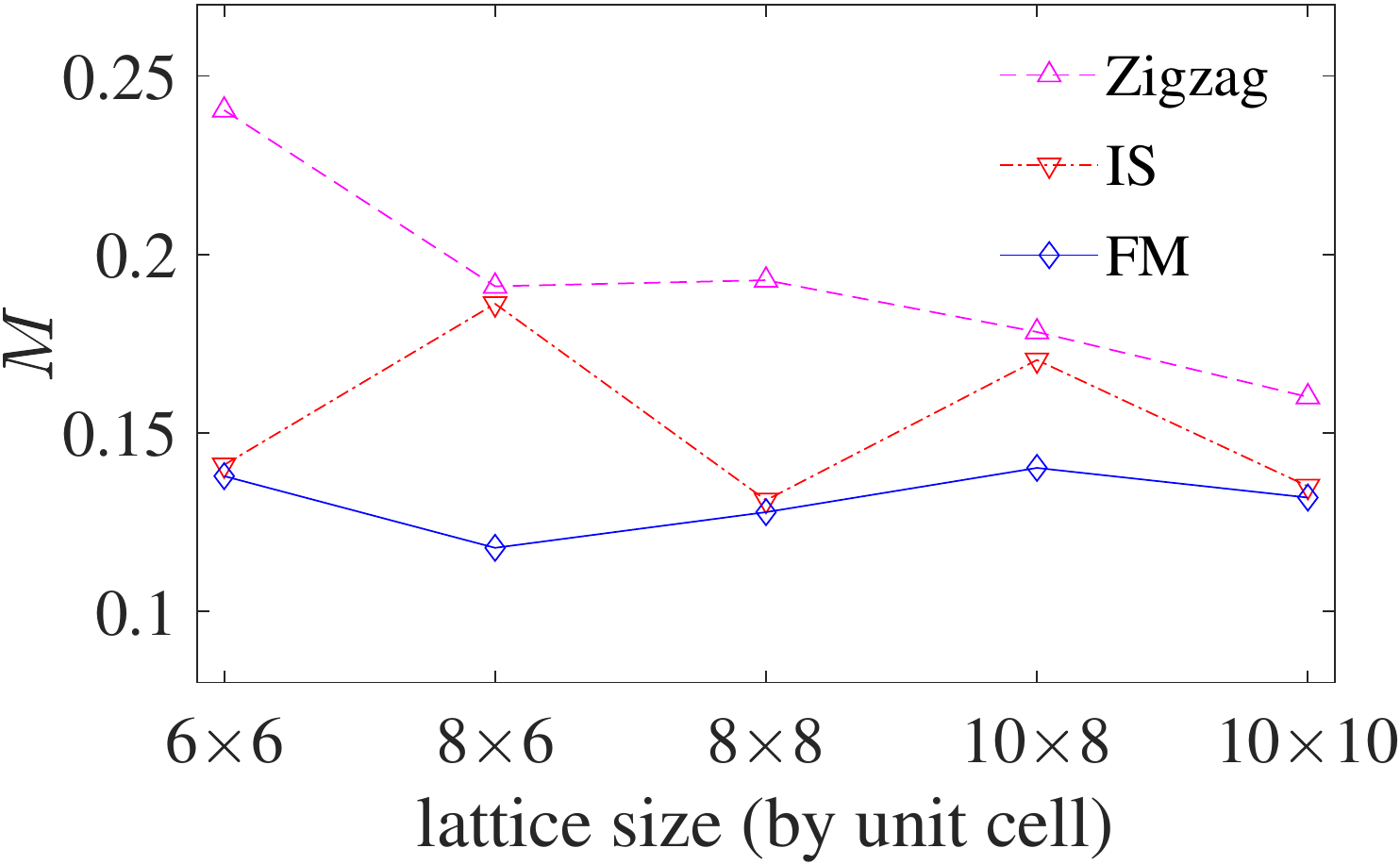}
\caption{Size scaling of the magnitude of $M$ for ordered states in Fig.~\ref{fig:AnisKGamma},
such as FM ($\Gamma/|K|=0.2$, $\delta_d=0.05$), IS ($\Gamma/|K|=0.6$, $\delta_d=0.05$), and zigzag ($\Gamma/|K|=1$, $\delta_d=0.05$).
}
\label{orders}
\end{figure}

To see the finite-size effect, we perform a size-scaling of the magnitude of $M$ in Eq.~(\ref{TotalHami}). As shown in Fig.~\ref{orders}, the magnitude $M$ of all the ordered phases (the FM, the IS, and the zigzag) is finite in the large size limit. We also performed finite-size scaling for the phase boundaries between different phases, and find that the phase boundaries change very slightly as the number of unit cells exceeds $8\times 8$ (not shown).

%To prove again that our phase diagrams are completely reliable.

%and check whether it goes to zero or remains finite in the larger lattice size,
%we determined every point in the phase diagram by examining its evolution with system size.

%completely reliable. To eliminate finite-size effects to the best of our ability, we determined every point in the phase diagram by examining its evolution with system size.
%We illustrate the results of these investigations in Fig.~\ref{boundary}.
%We find that the intermediary gSL-V and gSL-VI states disappear when system size is less than $8\times8$ unit cells, as shown in Fig.~\ref{boundary}(a).
%However, the intermediary gapless QSLs (gSL-IV, gSL-V, and gSL-VI) are robust when the system size is larger, as shown in Fig.~\ref{boundary}(b).

%\begin{figure}[t]
%\includegraphics[width=7.3cm]{Boundary_scaling_small_a}
%\includegraphics[width=7.5cm]{Boundary_scaling_large_b}
%\caption{Dependence on system size of the phase boundarys at fixed $\Gamma/|K|=1.4$ in Fig.~\ref{fig:AnisKGamma}.}
%\label{boundary}
%\end{figure}

\begin{figure*}[t]
\includegraphics[scale=0.359]{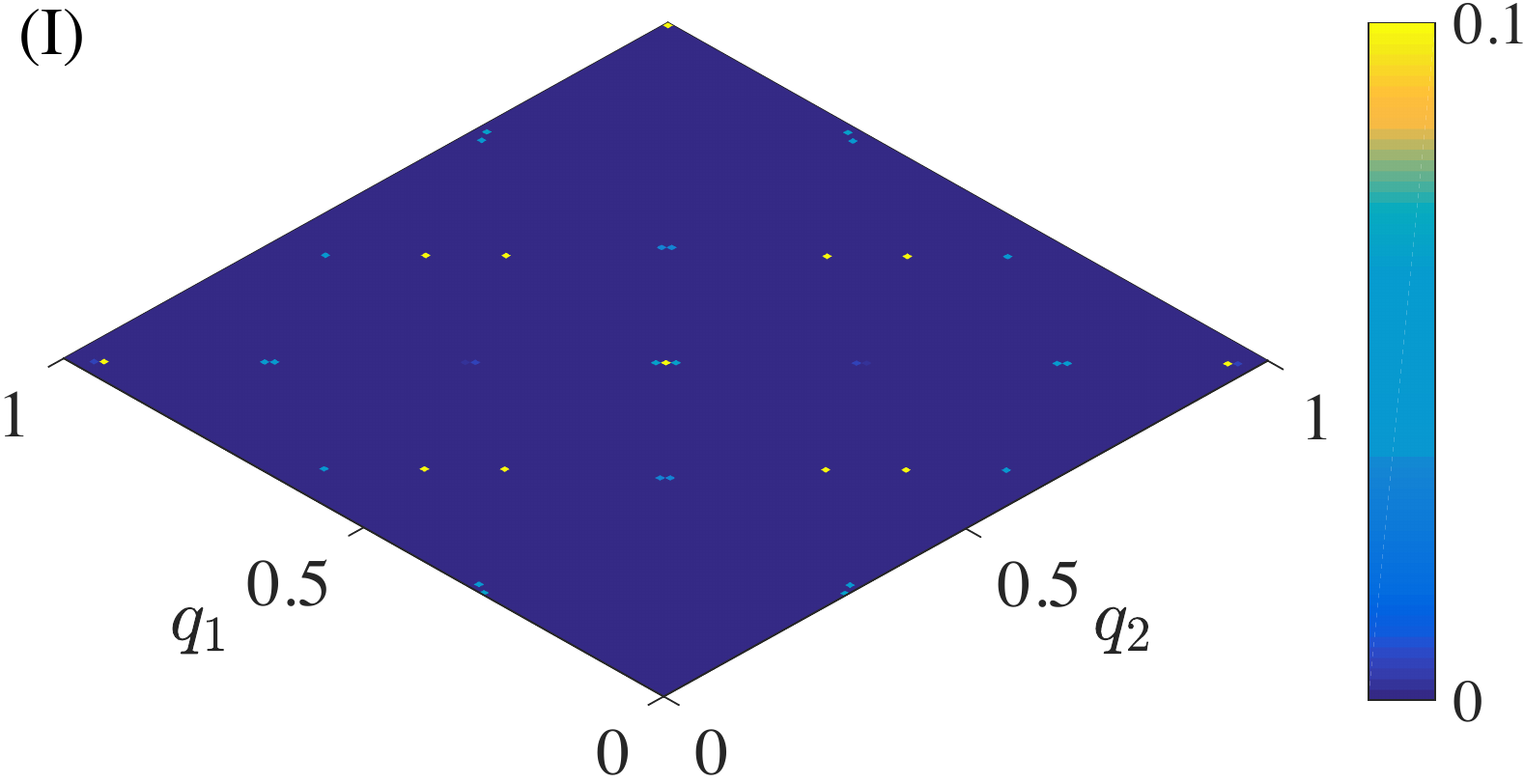}
\includegraphics[scale=0.35]{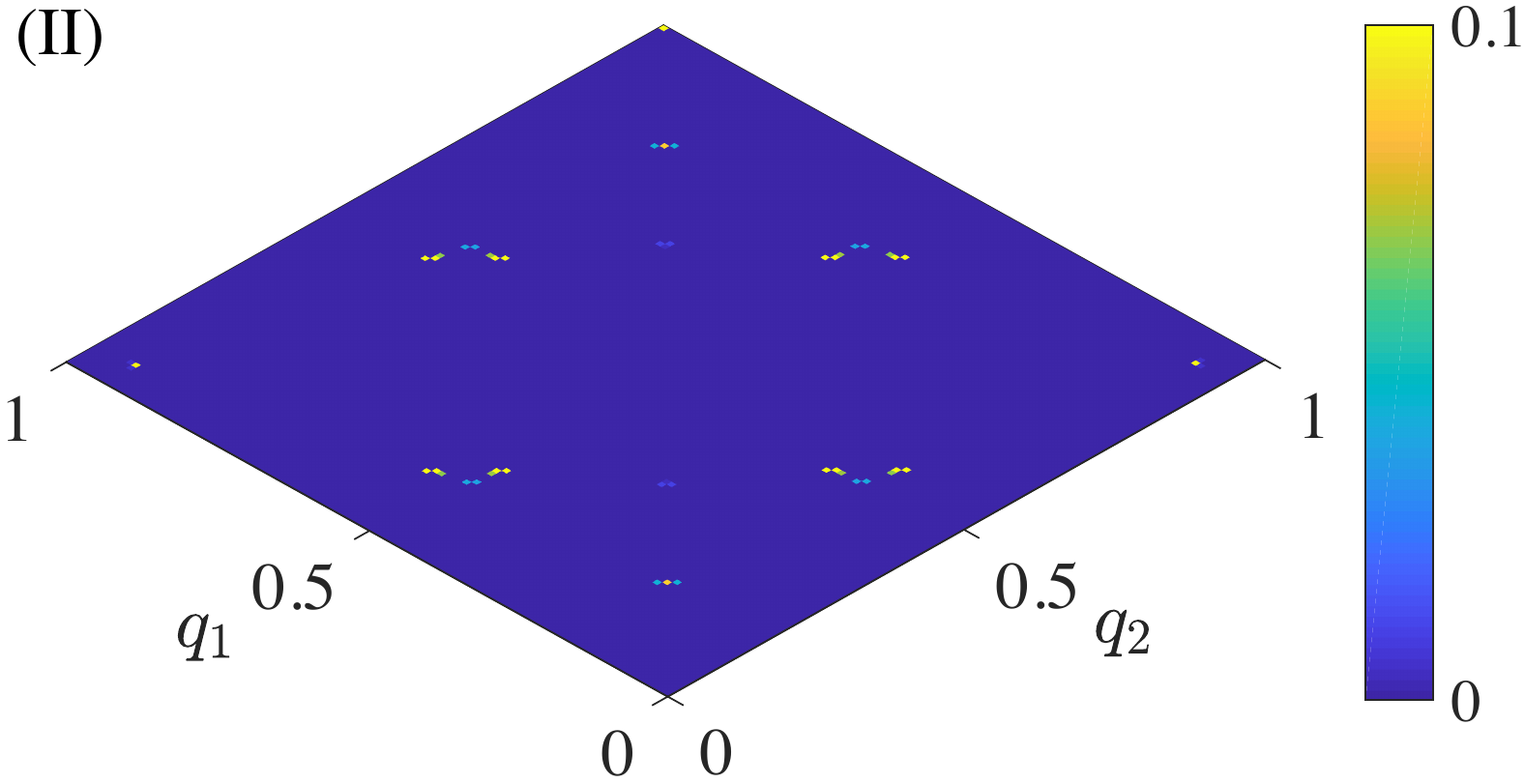}
\includegraphics[scale=0.35]{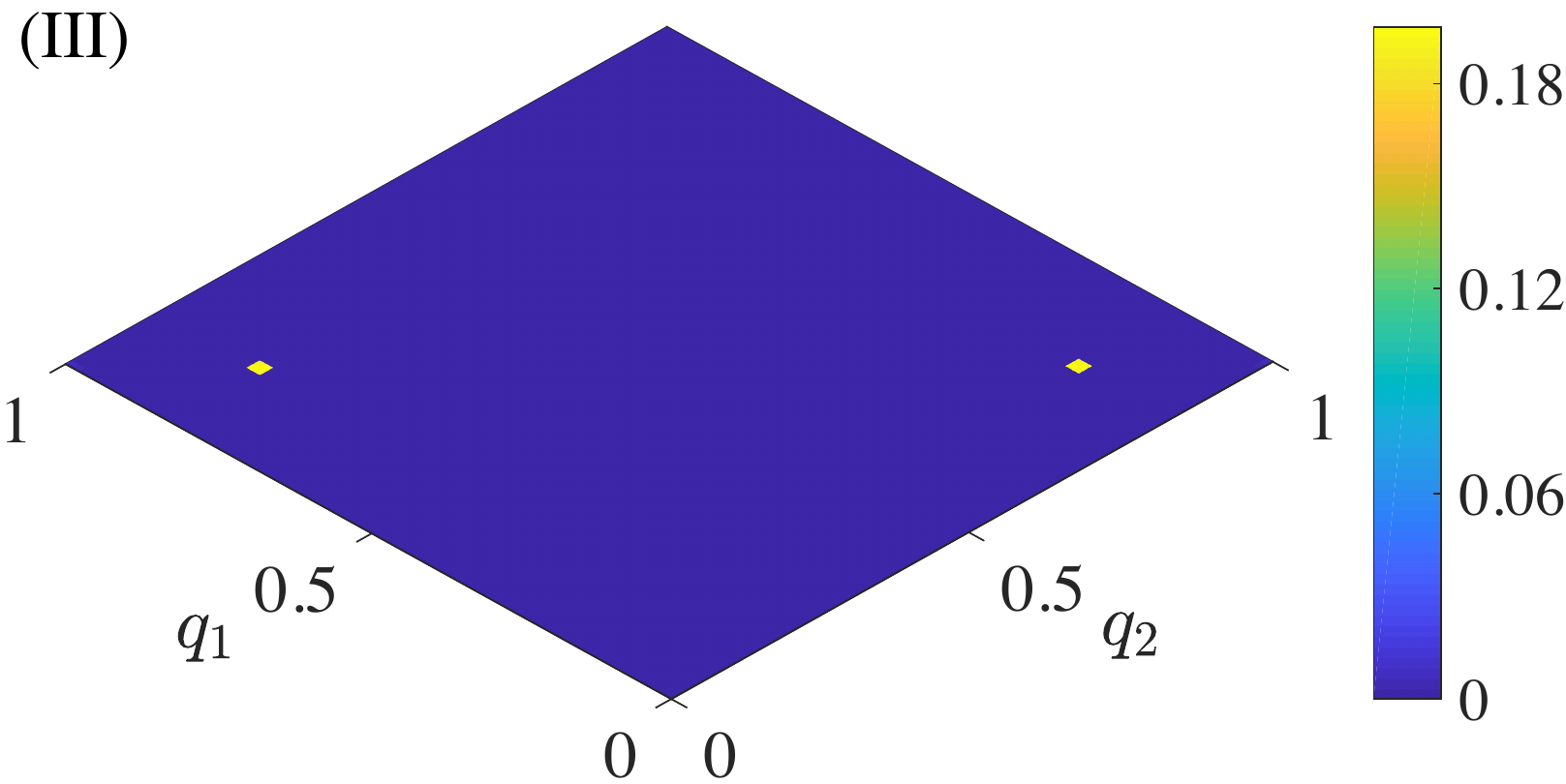}
\includegraphics[scale=0.35]{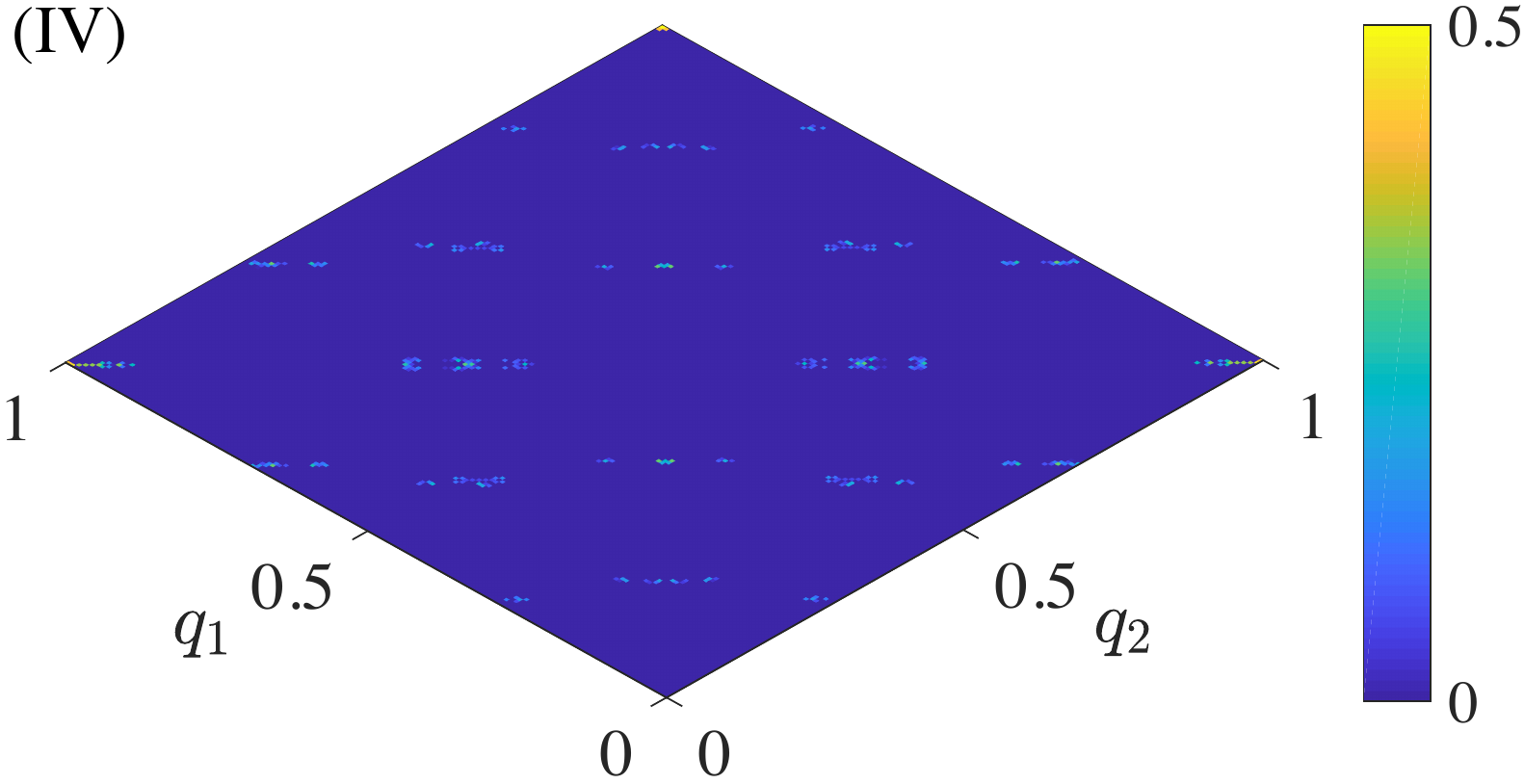}
\includegraphics[scale=0.35]{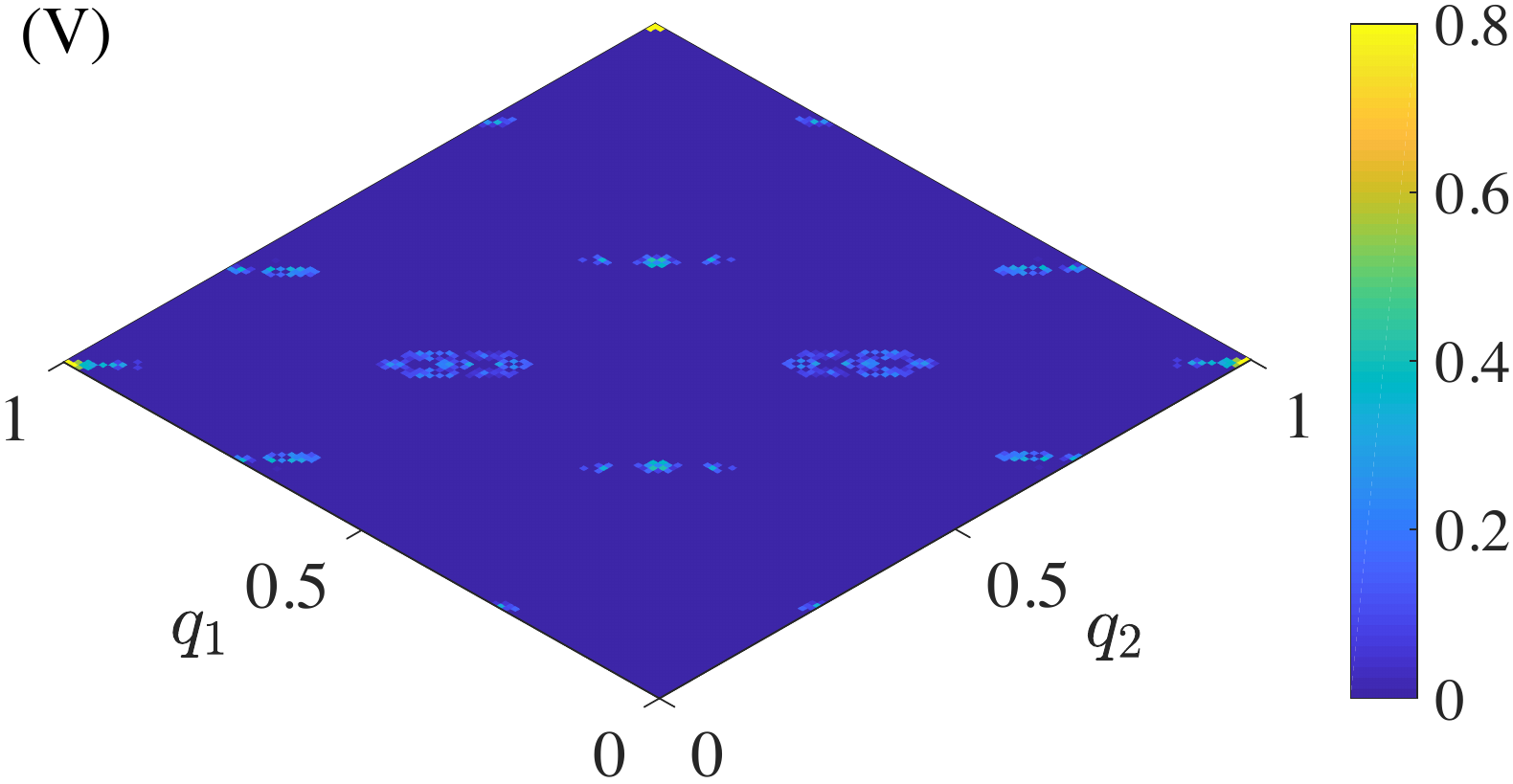}
\includegraphics[scale=0.35]{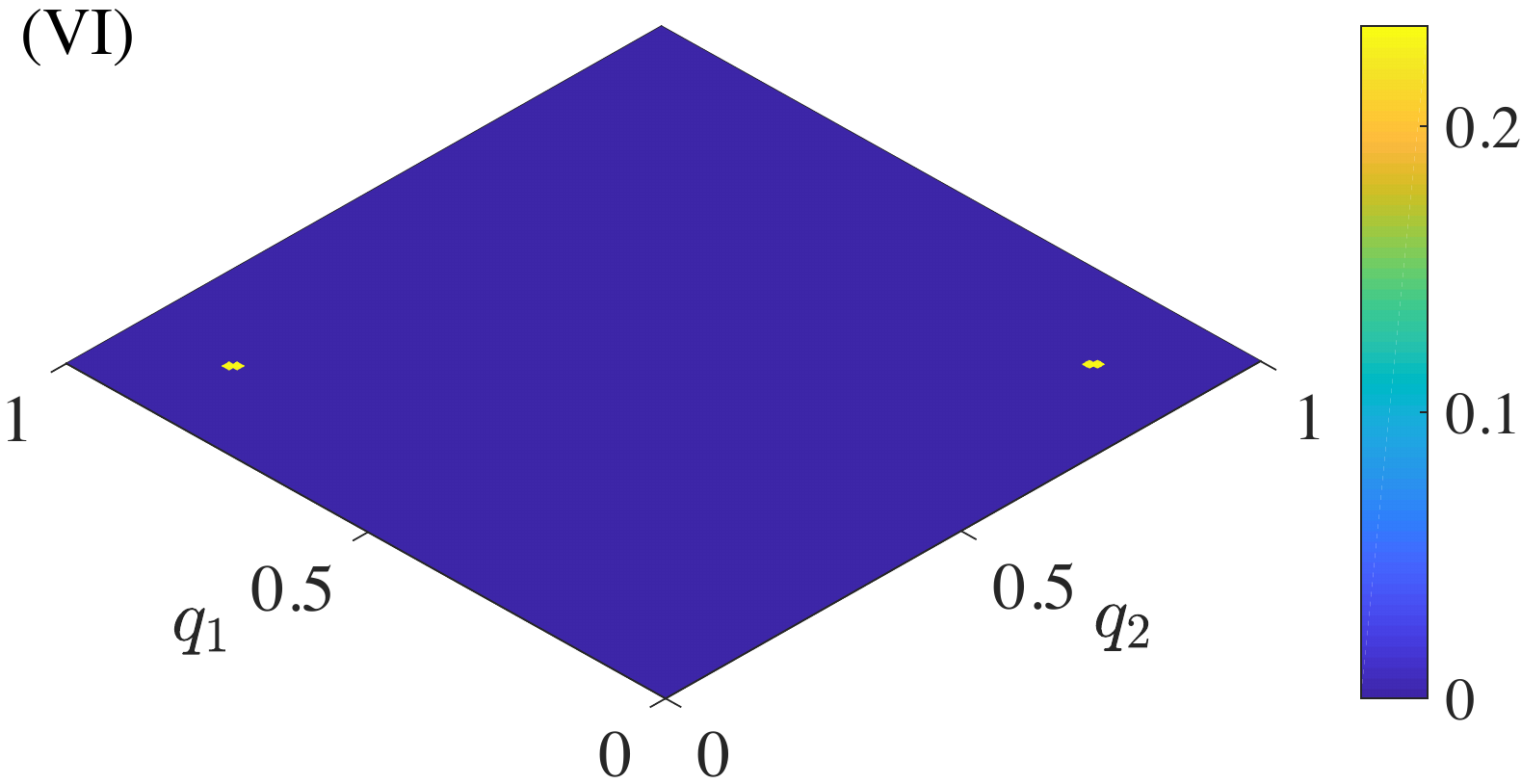}
\includegraphics[scale=0.35]{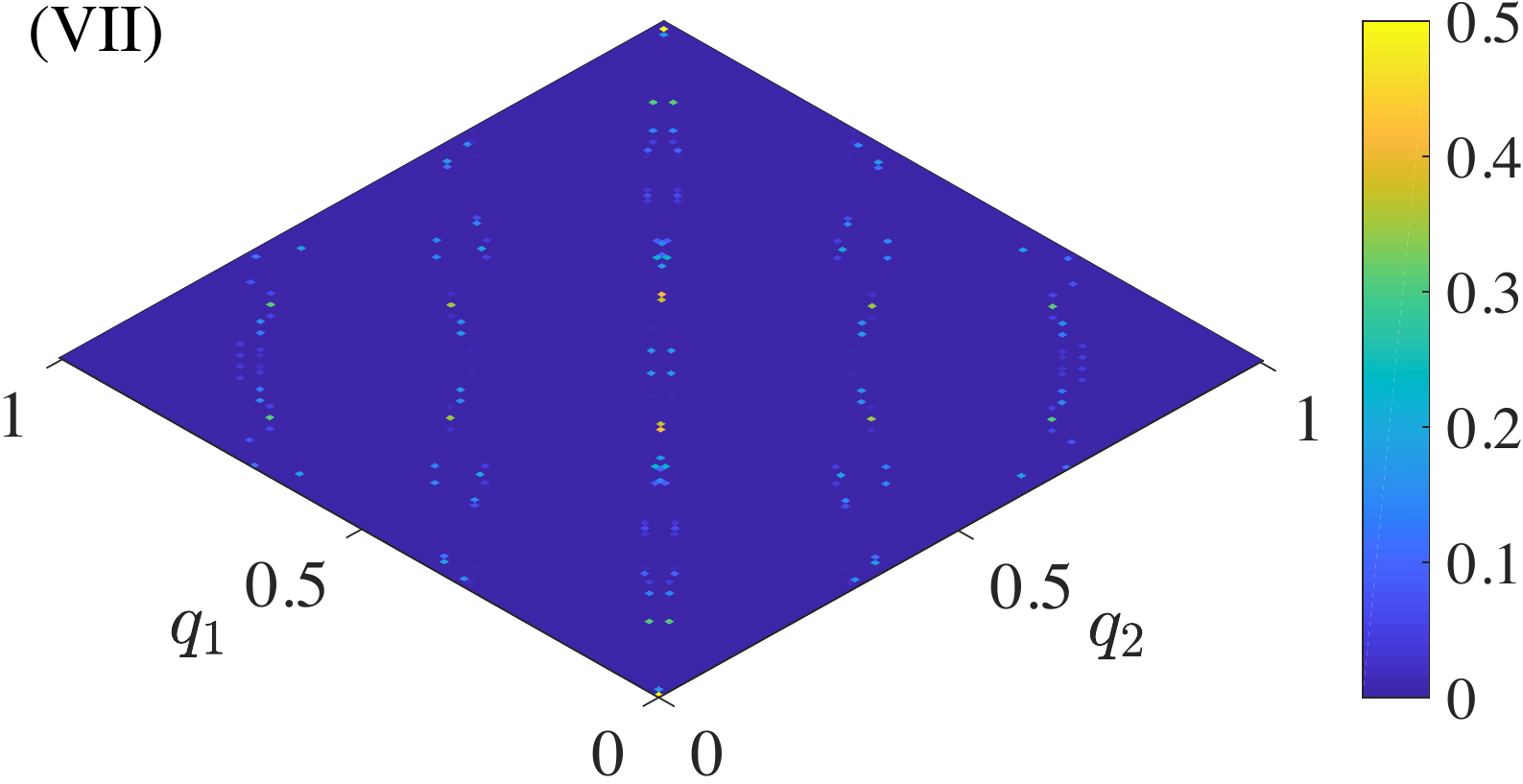}
\includegraphics[scale=0.35]{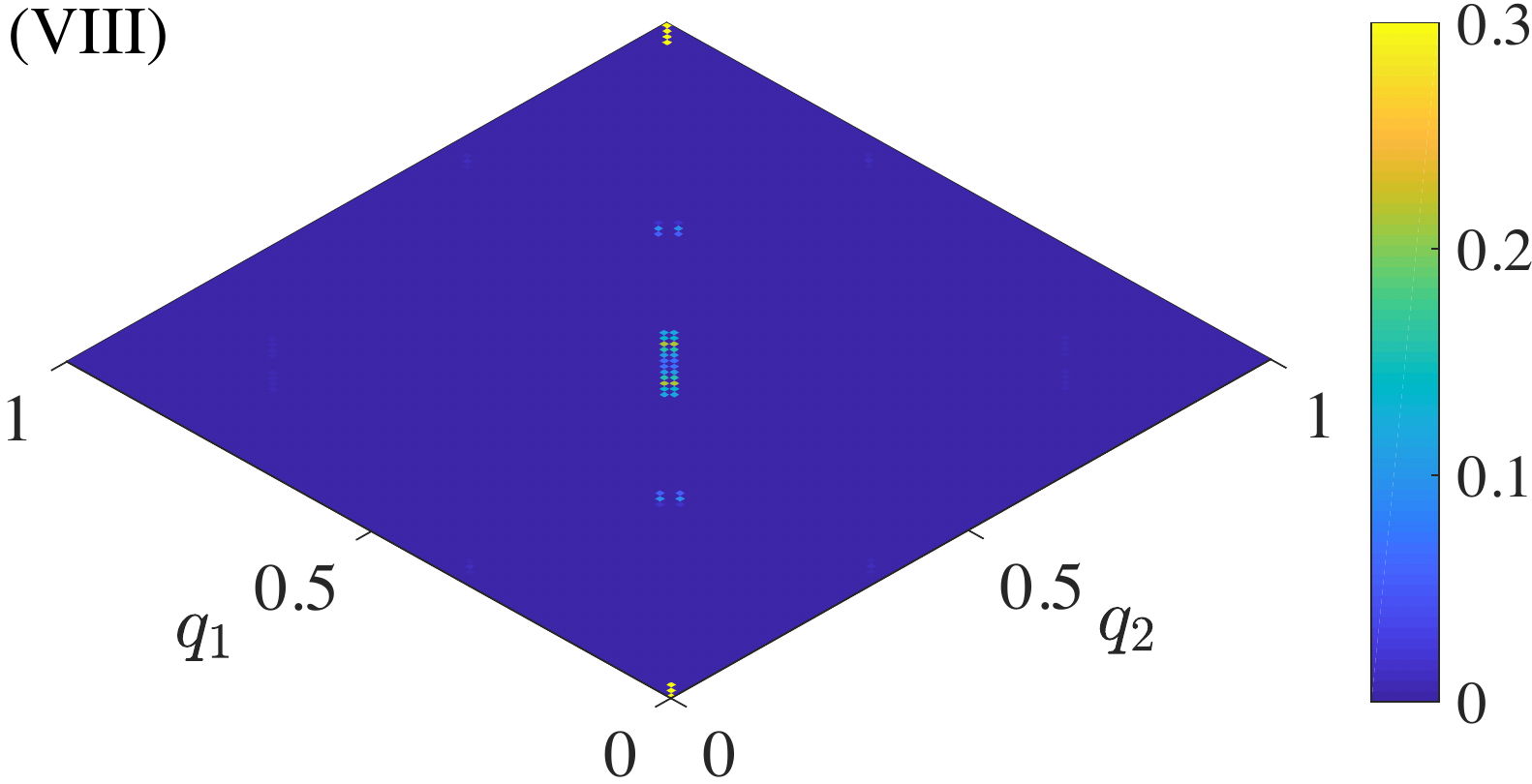}
\includegraphics[scale=0.35]{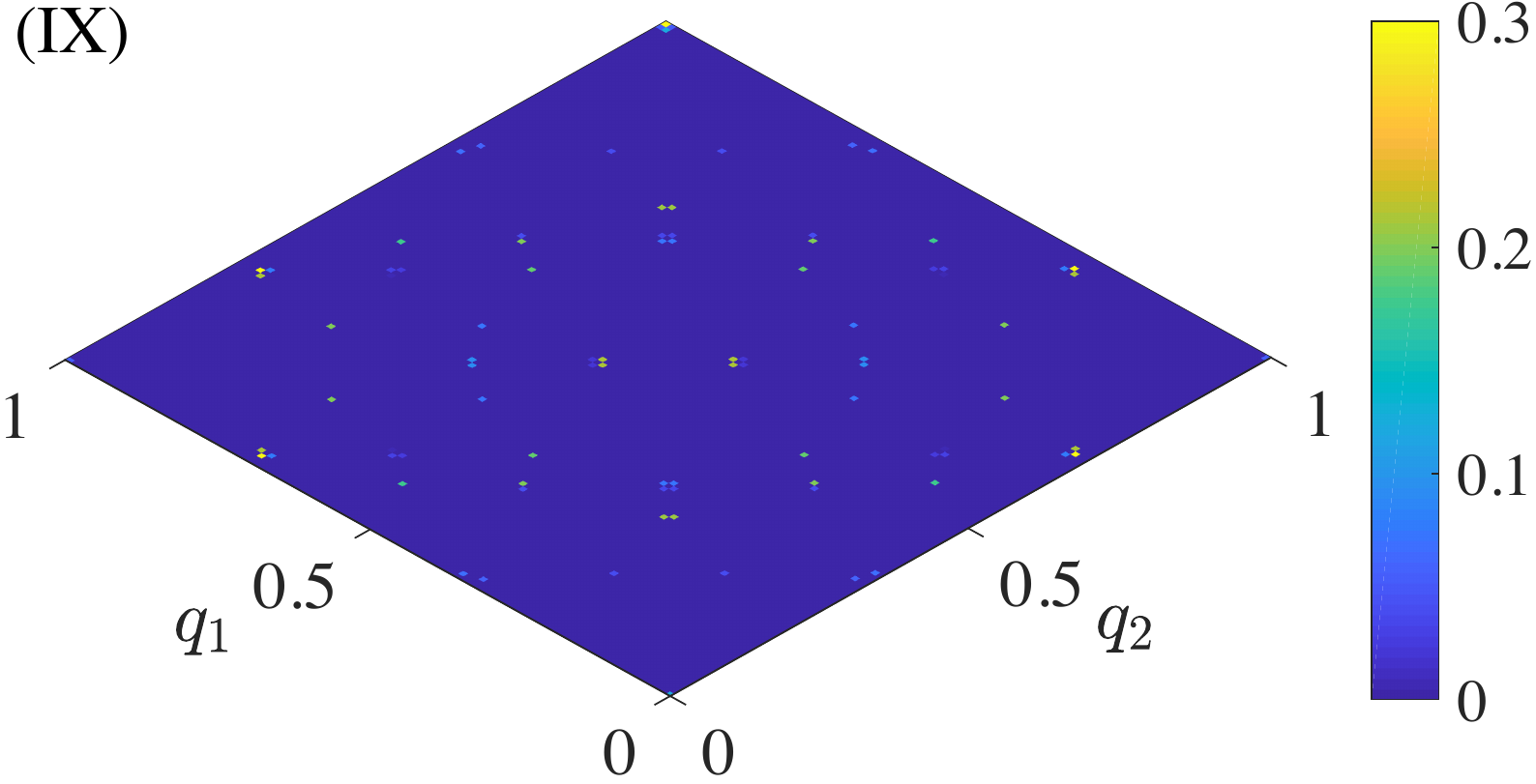}\
\caption{
Dynamic structure at zero frequency for (I) gSL-I ($\Gamma/|K|=0.4$, $\delta_d=0.05$), integrated over $0\leq \omega/|K| \leq 0.01$; (II) gSL-II ($\Gamma/|K|=0.4$, $\delta_d=0.1$), integrated over $0\leq \omega/|K| \leq 0.012$; (III) gSL-III ($\Gamma/|K|=0.4$, $\delta=0.4$), integrated over $0\leq \omega/|K| \leq 0.05$; (IV) gSL-IV ($\Gamma/|K|=1$, $\delta_d=0.15$), integrated over $0\leq \omega/|K| \leq 0.008$; (V) gSL-V ($\Gamma/|K|=1$, $\delta_d=0.2$), integrated over $0\leq \omega/|K| \leq 0.005$; (VI) gSL-VI ($\Gamma/|K|=1$, $\delta_d=0.4$), integrated over $0\leq \omega/|K| \leq 0.01$; (VII) gSL-VII ($\Gamma/|K|=0.6$, $\delta_{zz}=0.7$)  integrated over $0\leq \omega/|K| \leq 0.02$; (VIII) gSL-VIII ($\Gamma/|K|=0.6$, $\delta_{zz}=0.9$), integrated over $0\leq \omega/|K| \leq 0.01$; (IX) PKSL ($\Gamma/|K|=0.3$, $\delta_d=0.01$), integrated over $0 \leq \omega/|K| \leq 0.01$. The low-energy DSF of the KSL is not shown here, because the data are zero below the gap of the $b^{x,y,z}$ fermions.
}\label{fig:DSF}
\end{figure*}

\section{Dynamic structure factor of QSL phases}\label{sm:DSF}
The spin dynamic structure factor (DSF) reflects the low-energy excitations in a spin system and can be measured by neutron scattering experiments. In this appendix, the DSFs of QSL phases are calculated in the mean-field level.

The data for the gapless QSLs (PKSL, gSL-I$\sim$VIII) are in Fig.~\ref{fig:DSF} in sequence. We find that the DSFs are nonzero at very low frequency, indicating that these QSLs all have the gapless spin response, in contrast to KSL. Generally, the states with different numbers of cones have qualitatively different DSFs and are easily distinguished.

Furthermore, the DSF is sensitive to the locations of gapless points. Especially, if two gapless spin liquids contain the same number of Majorana cones but the cones are located at different positions in the BZ, then their DSFs will be different. For example, gSL-IV and PKSL both have 14 cones, their DSFs shown in Figs.~\ref{fig:DSF}(IV) and (IX) are distinguishable.

The gSL-III and gSL-VI both have two cones and the cones are located at similar positions in the BZ. So it is no wondering that their DSFs are qualitatively the same, as shown in Figs.~\ref{fig:DSF}(III) and (VI). These two phases can be distinguished by other methods, see Appendix \ref{sm:Chern}.

\section{Symmetry, Chern number, and GSD}\label{sm:Chern}

We have shown that the gapless points in the gapless QSL phases (namely, the PKSL and the gSL-I$\sim$VIII, see Fig.~\ref{fig:cones}) are a set of several $\{^* \pmb k\}$. The points in the same $\{^* \pmb k\}$ are related by $P$ or $C_2$ symmetry operation and are marked by the same color. This relation restricts the number of cones to be $4n+2m$. In the following, we will illustrate that in an applied magnetic field the remaining symmetry also constrains the Chern number in the resultant gapped chiral spin liquid.

\subsection{Mass information of the cones in magnetic fields}
The chirality of the Majorana cones with respect to magnetic fields can be obtained using $\pmb k\cdot \pmb p$ expansion method. From the chiralities of a cone with respect to the magnetic field along $\pmb x, \pmb y, \pmb z$ directions, we can figure out the maximum mass direction (MMD) of that cone. Once we have known the MMD for a cone, the chirality of this cone with respect to a weak magnetic field just depends on the sign of the component along the MMD. For states in the same QSL phase, the MMD of a cone varies continuously with interaction parameters. In Table.~\ref{tab:Mass_II_VIII}, we provide the detailed information for the cones in representative states in the nodal QSL phases gSL-I, gSL-II, gSL-VI and gSL-VIII which will be further discussed in Appendix \ref{SM:gM}.

From the MMD information of all the cones, we can immediately know the total Chern number for arbitrarily oriented magnetic fields. In the following sections, we will discuss in detail for the cases $\pmb B\parallel (111)$ and $\pmb B$ is oriented along non-symmetric directions.

\begin{table}[t]
\begin{tabular}{c|c|c|c|c}
\hline
\hline
       & Cone &  MMD with $\pmb B\cdot \pmb S$ &  $\pmb B\parallel(111)$ & $\sum\limits_{\langle\langle i,j\rangle\rangle} (i C_i^\dag C_j + H.c.)$\\
\hline & D$_{\ }$       & (-0.56,-0.56,+0.62)     &  $-$   &  $+$ \\ %(-0.5550   -0.5550    0.6196)
       & R$_{\ }$       & (-0.10,-0.10,+0.99)     &  $+$   &  $+$ \\ %(-0.0919   -0.0919    0.9915)
 gSL-I & B$_{\ }$       & (-0.05,-0.05,-0.99)     &  $-$   &  $-$ \\ %(-0.0460   -0.0460   -0.9979)
       &G$_1$       & (+0.85,-0.52,-0.12)     &  $+$   &  $+$ \\  %(0.8490    -0.5156   -0.1153)
       &G$_2$       & (-0.52,+0.85,-0.12)     &  $+$   &  $-$ \\  %(-0.5156   0.8490    -0.1153)
\hline
       & D$_{\ }$       & (-0.68,-0.68,+0.25)     &  $-$   &  $+$ \\   %(-0.6843,-0.6843,0.2520)
gSL-II &G$_1$          & (+0.82,-0.57,-0.08)     &  $+$   &  $+$ \\   %(0.8186,-0.5681,-0.0841)
          &G$_2$       & (-0.57,+0.82,-0.08)     &  $+$   &  $+$ \\  %(-0.5681,0.8186,-0.0841)
\hline
gSL-VI &  D$_{\ }$     & (+0.26,+0.26,+0.93)     &  $+$   &  $+$ \\ %(0.2598    0.2598    0.9300)
\hline
               &  D$_{\ }$  & (-0.56,-0.57,+0.60)     &  $-$   &  $+$ \\   %(-0.5643   -0.5700    0.5972)
gSL-VIII  & R$_{\ }$       & (-0.67,-0.69,+0.26)     &  $-$   &  $-$ \\   %(-0.6719   -0.6953    0.2551)
               & B$_1$      & (+0.51,+0.77,-0.38)      &  $+$   &  $+$ \\  %(0.5139    0.7670   -0.3843)
               & B$_2$      & (+0.77,+0.51,-0.38)     &  $+$   &  $-$ \\  %(0.7670   0.5139   -0.3843)
\hline
\hline
\end{tabular}
\caption{Mass information for the Majorana cones on the left half BZ in the gSL-I phase [see Fig.~\ref{fig:cones}(I)], the gSL-II phase [see Fig.~\ref{fig:cones}(II)], the gSL-IV phase [see Fig.~\ref{fig:cones}(IV)], and the gSL-VIII phase [see Fig.~\ref{fig:cones}(VIII)]. MMD stands for maximum mass directions (see the main text for definition). D,G,R,B stand for the color of the cones and denote dark, green, red, blue respectively, and $_1$ ($_2$) label the upper (lower) cone with the same color.
}\label{tab:Mass_II_VIII}
\end{table}

\subsection{Chern number in a field with $\pmb B\parallel (111)$}
When a magnetic field is added along the (111) direction, the time-reversal symmetry is violated, and the symmetry group reduces from $G=\tilde{\mathscr C}_{2v}\times Z_2^T$ to $G_c=\{ E, C_2T, P, \sigma_mT\}$. Noticing that one $\{^*\pmb k\}$ contains at most 4 wave vectors, the wave vectors in the same $\{^*\pmb k\}$ can be transformed into each other by the group elements in $G_c$.

Since the Chern number comes from the cones (each cone contribute a Chern number $1\over2$ or $-{1\over2}$ when it opens a gap), in the following we show that the cones in the same $\{^*\pmb k\}$ contribute the same Chern number (all equal to $1\over2$ or all equal to $-{1\over2}$). For example, if a state contains 6 cones with $n=1$ and $m=1$, then the total Chern number in a weak field $\pmb B\parallel (111)$ can only be one of the following: $3=2+1,\ \ 1=2-1,\ \ -1=-2+1,\ \ -3=-2-1$. The results listed in Table.~\ref{tab:cones} are consistent with the symmetry constraint.

We firstly consider the inversion symmetry $P$. The symmetry operator on the fermions in the following way,
\begin{gather}
P B_{\pmb {k}} P^{-1}=M_P B_{-\pmb k},
\end{gather}
where $B_{\pmb k}=(c_{\pmb k\up A}, c_{\pmb k\dn A},c_{\pmb k\up B},c_{\pmb k\dn B},  c_{-\pmb k\up A}^\dag, c_{-\pmb k\dn A}^\dag, c_{-\pmb k\up B}^\dag, c_{-\pmb k\dn B}^\dag)^T$, and $M_P=I\otimes (-i\sigma_y)\otimes I$ ($I$ is the 2 by 2 identity matrix) is determined by the Kitaev PSG (\ref{KitaevPSG_P}).

Since the total mean-field Hamiltonian $H=\sum_{\pmb k} B_{\pmb k}^\dag H_{\pmb k} B_{\pmb k}$ is invariant under inversion operator (\ref{KitaevPSG_P}), therefore
\begin{gather}
H_{-\pmb k}=M_p H_{\pmb k} M_p ^{\dagger}.
\end{gather}
If $|\varphi_{\pmb k}\rangle$ is the eigenvector of $H_{\pmb k}$ with eigenvalue $\varepsilon_k$, $H_{\pmb k}|\varphi_{\pmb k}\rangle =\varepsilon_k|\varphi_{\pmb k}\rangle$, then $H_{-\pmb k}M_P|\varphi_{\pmb k}\rangle=\varepsilon_k M_P|\varphi_{\pmb k}\rangle$, namely, $M_P|\varphi_{\pmb k}\rangle= e^{i\theta_{-\pmb k}} |\varphi_{-\pmb k}\rangle$ is the eigenvector of $H_{-\pmb k}$ with eigenvalue $\varepsilon_k$, where $e^{i\theta_{-\pmb k}} $ is a $\pmb k$-dependent gauge transformation. The Berry connections at $-\pmb k$ and $\pmb k$ in the same energy band are related as the following,
\Beq
A_{\pmb k}= \langle \varphi_{\pmb k}|\partial_{\pmb k} |\varphi_{\pmb k}\rangle& =& \langle \varphi_{\pmb k}|\left(|\varphi_{\pmb k+\delta\pmb k}\rangle-|\varphi_{\pmb k}\rangle\right)/\delta {\pmb k}  \\
& =& \langle \varphi_{\pmb k}|M_P^\dag \left(M_P|\varphi_{\pmb k+\delta\pmb k}\rangle-M_P|\varphi_{\pmb k}\rangle\right)/\delta {\pmb k}  \\
& =& \langle \varphi_{-\pmb k}|\left(e^{i\theta_{-\pmb k-\delta \pmb k}-i\theta_{-\pmb k}} |\varphi_{-\pmb k-\delta\pmb k}\rangle- |\varphi_{-\pmb k}\rangle\right)/\delta {\pmb k}  \\
&= &-A_{-\pmb  k}-\mathscr A_{-\pmb k},
\Eeq
where $\mathscr A_{-\pmb k}$ is pure gauge term having no contribution to the Berry curvature.

\begin{figure}[htbp]
\includegraphics[width=8.5cm]{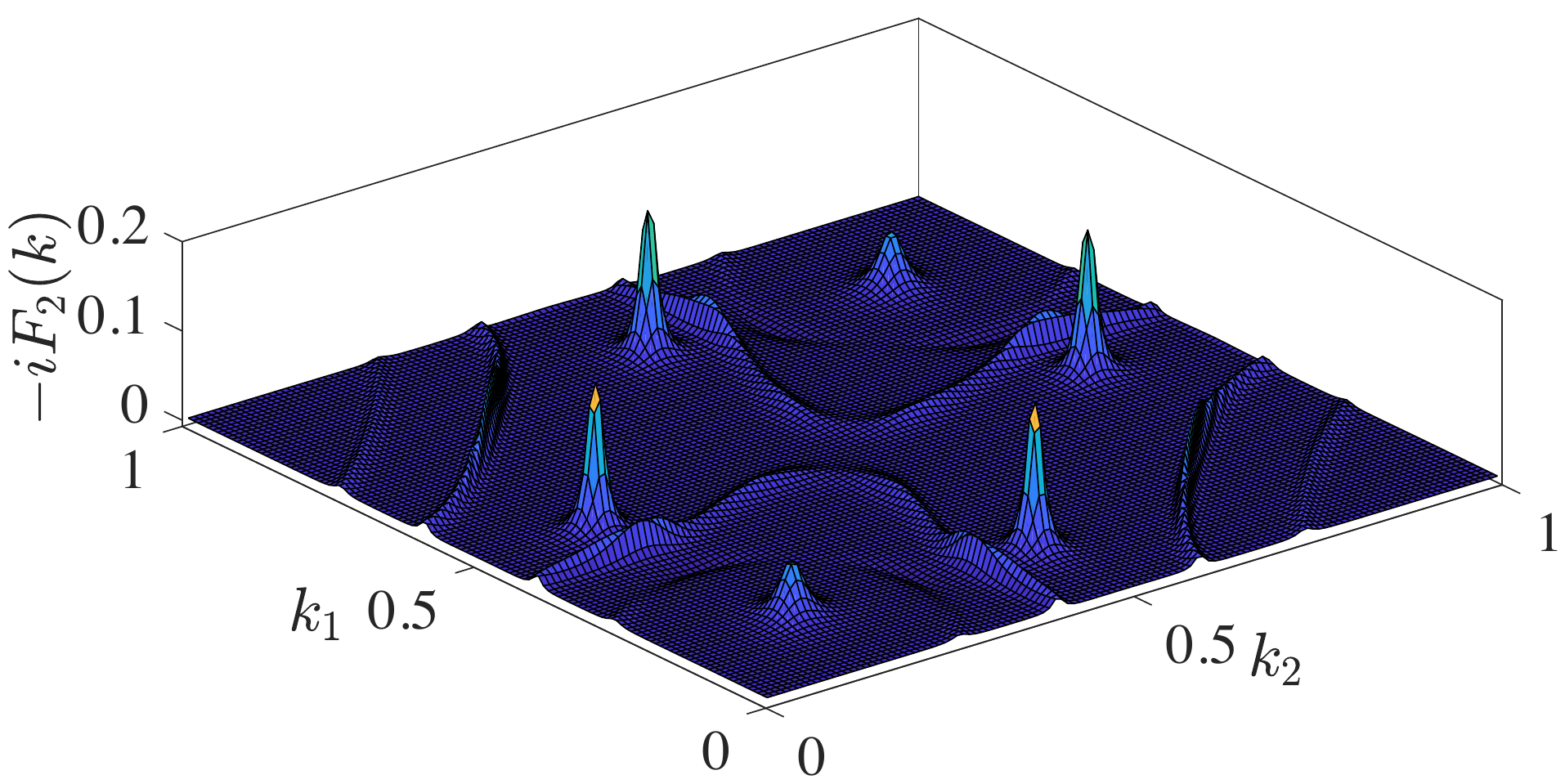} \
\caption{Berry curvature of the chiral spin liquid descending from the gSL-III in a weak field along $(111)$ direction. Only the data of the second occupied band are shown.}
\label{fig:BerryCurvature}
\end{figure}

It can be further shown that the Berry curvature $F_{xy}=\partial_{k_x} A_{k_y}-\partial_{k_y}A_{k_x}$ in the first BZ is symmetric under inversion, $F_{xy}(-\pmb k)=F_{xy}(\pmb k)$ for every band. This result is verified numerically, as shown in Fig.~\ref{fig:BerryCurvature}.
We note that the Berry curvatures of each band are also symmetric under inversion operation.
Therefore, the cones related by inversion symmetry have the same contribution to the Chern number. The proof of the other symmetry, $\sigma_mT$, is very similar and will not be repeated here.

\subsection{Chern number in a generic magnetic field}\label{SM:gM}
We claim that the spin liquid states with different number of Majorana cones belong to different phases. This conclusion is partially confirmed by the fact that in a weak magnetic field $\pmb B\parallel (111)$ the resultant Chern numbers for states with different number of cones are generally different.  However, there are exceptions.  Notice that the gSL-I, gSL-II, gSL-V and gSL-VI contain 10,6,10,2 cones, respectively, but their Chern numbers are all equal to 1.

This issue is solved by changing the field orientations. From the MMD information of the cones listed in Table.\ref{tab:Mass_II_VIII}, one can obtain the chiralities of the cones with respect to arbitrarily oriented weak magnetic fields, and thus the total Chern number can be read out. For instance, with applied field $\pmb B\parallel (1,-1,5)$,  $\pmb B\parallel (1, -1,-5)$  and $\pmb B\parallel (10,1,1)$,  the resultant Chern numbers are shown in Table.~\ref{tab:generalfield}. It can be seen that the states containing different numbers of cones indeed behave differently.

Especially, the gSL-VIII phase may obtain a nontrivial Chern number in a generic magnetic field although it has a trivial Chern number in the case $\pmb B\parallel (111)$. For example, for the weak field $\pmb B\parallel (1,1,4)$ one obtains Chern number $\nu=-2$ and for the weak field $\pmb B\parallel (1,2,4)$ one obtains Chern number $\nu=2$.

\begin{table}[htpb]
\centering
\begin{tabular}{c|ccccc}
\hline
\hline
field orientation & gSL-I &  gSL-II  &  gSL-V  &  gSL-VI  \\
\hline
$\pmb B\parallel (1,-1,5)$  & $\nu=1$ & $\nu=-1$ &$\nu=1$ & $\nu=1$  \\
$\pmb B\parallel (1,-1,-5)$ & $\nu=-1$ & $\nu=1$ &$\nu=-1$ & $\nu=-1$  \\
$\pmb B\parallel (10,1,1)$  & $\nu=-3$ & $\nu=-1$ &$\nu=-1$ & $\nu=1$  \\
\hline
\hline
\end{tabular}
\caption{Chern numbers in differently oriented (weak) magnetic fields.}\label{tab:generalfield}
\end{table}

\begin{table}[t]
\centering
\begin{tabular}{c|c||c|c|c|c|c}
\hline
\hline
Parent state & $\, \nu \,$ & $\rho_1$ & $\rho_2$ & $\rho_3$ & ($\rho_4$) & GSD \\
\hline
$Z_2$ QSL     & $ 0 $  &  $\,$1.4$\times$10$^{-5}$ &  0.4557 & 1.1081   & 2.4362 & 4 \\
\hline
KSL          & $1$    & 0.9544   & 0.9866  & 1.0590   &                         & 3  \\
\hline
PKSL         & $5$    & 0.6773   & 1.0163  & 1.3064   &                         & 3  \\
\hline
gSL-I        & $1$    & 0.4273   & 1.0005  & 1.5722   &                         & 3  \\
\hline
gSL-II       & $1$    & 0.3085   & 1.0004  & 1.6912   &                         & 3 \\
\hline
gSL-III      &$-1$    & 0.0377   &  1.0012 & 1.9611   &                         & 3 \\
\hline
gSL-IV       & $ 3$   &  0.3004  &  0.9782 & 1.7215   &                         & 3 \\
\hline
gSL-V        & $ 1$   &  0.2120  &  0.9814 & 1.8066   &                         & 3 \\
\hline
gSL-VI       & $ 1$   &  0.0316  &  0.8871 & 2.0812   &                         & 3 \\
\hline
gSL-VII      & $ -2$  &  0.0013  &  0.7287 & 1.2701   & 1.9999                  & 4 \\
\hline
gSL-VIII     & $ 0$   &  $\,$1.3$\times$10$^{-7}$     &  0.5228 & 1.0080   & 2.4692  & 4 \\
\hline
Dimer        & $ 0$   &  $\,$7.3$\times$10$^{-7}$     &  $\,$2.8$\times$10$^{-6}$ & $\,$8.8$\times$10$^{-6}$   & 4.0000  & 1 \\
\hline
\hline
\end{tabular}
\caption{Eigenvalues of the overlap matrices of the ground states of gapped states on a torus. $\nu$ is mean-field Chern number of the gapless states which are gapped by a small magnetic field $\pmb B \parallel {1\over\sqrt3}(\pmb x+\pmb y+\pmb z)$. The system size we adopt is 8$\times$8$\times$2. The data for the "$Z_2$ QSL" and the "dimer" states are calculated without applying magnetic fields.
}\label{tab:GSD}
\end{table}

\subsection{Ground state degeneracy} \label{app: GSD}
To see if the Gutzwiller projected wave functions are indeed nontrivial, namely, to check if the $Z_2$ gauge fields are deconfined after projection, we calculate the ground state degeneracy (GSD) on a torus. To this end, we change the boundary conditions and calculate the overlap matrix between the Gutzwiller-projected states, namely,
$\rho_{\alpha\beta} = \langle P_G \psi_\alpha | P_G \psi_\beta \rangle = \rho_{\beta\alpha}^*$, where $\alpha,\beta \in \{++,+-,-+,--\}$ are the boundary conditions ($+$ stands for periodic boundary condition and $-$ stands for anti-periodic boundary condition) along $x,y$-direction, respectively. If the Chern number of the mean-field ground state is odd, then the state $|\psi_{++}\rangle$ vanishes after Gutzwiller projection\cite{PKSL}, therefore there are at most 3-fold degenerate ground states on a torus and the $\rho$ matrix is 3 by 3. Otherwise, if the Chern number is even, then there are at most 4-fold degenerate ground states and the $\rho$ matrix is 4 $\times$ 4.

If the matrix $\rho$ has only one significant eigenvalue and all the others are vanishingly small, then the GSD is 1 which means that the Z$_2$ gauge field is confined. On the other hand, if $\rho_{\alpha\beta}$ has more than one (nearly degenerate) nonzero eigenvalues, then the GSD is nontrivial and hence the Z$_2$ gauge fluctuations are deconfined. The data for the eigenvalues of $\rho$ for projected mean-field states carrying different Chern numbers are summarized in Table.~\ref{tab:GSD}.

The first eigenvalue of gSL-VIII $1.3\times 10^{-7}$ is very small, which seems to indicate that the GSD is not 4. However, the situation is quite similar in the $Z_2$ spin liquid state in the pure Kitaev model, where the smallest eigenvalue is $1.4\times 10^{-5}$ and is much smaller than other eigenvalues. Recalling that the Kitaev model is exactly solvable, where the gapped states belongs to the toric code phase and are $Z_2$ deconfined\cite{Kitaev}. Therefore, we also believe that the gSL-VIII is also $Z_2$ deconfined and nontrivial.

Particularly, we calculate the GSD of the $\pi$-flux state discussed in Appendix \ref{gutzVMC}. For a torus with $12\times12$ unit cells, the eigenvalues of the overlap matrix $\rho$ are given by $2.5\times 10^{-5}, 2.9\times 10^{-5}, 6.4\times 10^{-5}, 3.99988$. The last eigenvalue is by far larger than the remaining three ones, therefore the GSD is 1 and therefore the projected $\pi$-flux state is trivial. Actually, we have performed a finite-size scaling calculation (not shown), which indeed indicates that the GSD of this state is 1 in the large-size limit.\\

\section{$K$-$\Gamma$ chains}\label{sm:chain}
In the $\delta_{zz}=1$ limit, the system becomes decoupled spin chains. A single chain is shown to exhibit a hidden $O_h$ symmetry and supports a gapless phase  described by an emergent $SU(2)_1$ Wess-Zumino-Witten model\cite{Affleck}. Under a six-sublattice rotation\cite{HYKee}, the spin chain can be mapped into a model with three sublattices,
\begin{gather}\label{3sub}
H^{1D} = \sum_{\langle i,j \rangle \in(\alpha\beta)\gamma} -K S_i^\gamma S_j^\gamma - \Gamma (S_i^\alpha S_j^\alpha + S_i^\beta S_j^\beta),
\end{gather}
where $\gamma=x,y,z$. Furthermore, it was shown that a three-sublattice rotation can map $(K,\Gamma)$ to $(K,-\Gamma)$ \cite{Affleck}.

\begin{figure}[t]
\includegraphics[width=8.5cm]{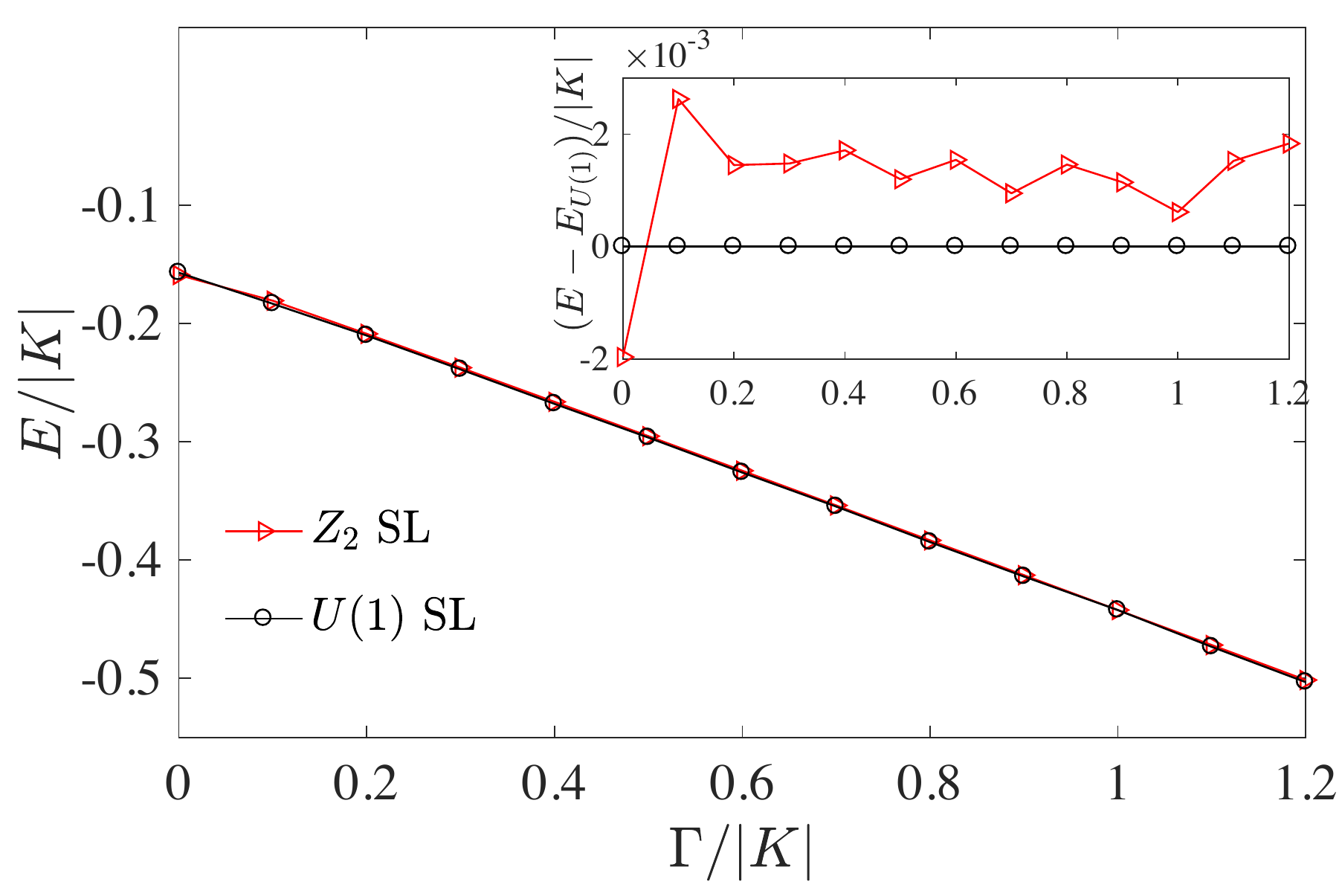} \
\caption{Energy per site for the $U(1)$ and $Z_2$ ansatz's for the $K$-$\Gamma$ chain (\ref{3sub}). The calculation is performed for a system with 120 sites under periodic boundary condition. The level crossing at small $\Gamma$ indicates a first-order phase transition between the $Z_2$ QSL and the $U(1)$ SL. The insert shows the energy difference between the two states.}
\label{fig:Chain}
\end{figure}

For the parameters $K<0$ and $\Gamma>0$ that we are studying, the transformed model (\ref{3sub}) can be further transformed into a fully anti-ferromagnetic one by a three-sublattice transformation:
\beq\label{3subAFM}
H^{1D} = \sum_{\langle i,j \rangle \in(\alpha\beta)\gamma} |K| S_i^\gamma S_j^\gamma + |\Gamma| (S_i^\alpha S_j^\alpha + S_i^\beta S_j^\beta).
\eeq

We adopt two different types of ansatz's. In the first one, we treat the original $K$-$\Gamma$ interactions directly. The mean-field Hamiltonian of the $K$-$\Gamma$ chain is descending from the two-dimensional ansatz (\ref{Hmf}). Since the mean-field decoupling contains spinon-pairing terms, the projected state is a $Z_2$ spin liquid state. In the second one, we start from the transformed three-sublattice model (\ref{3subAFM}) since it is equivalent to the $K$-$\Gamma$ chain up to unitary transformations. Noticing that $|K| S_i^\gamma S_j^\gamma + |\Gamma| (S_i^\alpha S_j^\alpha + S_i^\beta S_j^\beta)=|\Gamma| \pmb S_i\cdot\pmb S_j + (|K|-|\Gamma|)S_i^\gamma S_j^\gamma $, we can introduce the following $U(1)$ ansatz,
\Beq
H_{U(1)} = \sum_{\langle i,j \rangle \in \gamma}  \chi (c_{i,\uparrow}^\dagger c_{j,\uparrow} + c_{i,\downarrow}^\dagger c_{j,\downarrow})  +  \chi^\gamma C_i^\dagger \sigma^\gamma C_j  + H.c.
\Eeq
where $C_i^\dagger=(c_{i,\uparrow}^\dagger,c_{i,\downarrow}^\dagger)$ and 4 variational parameters, namely $\chi, \chi^x, \chi^y, \chi^z$ appear in above equation. Finally, a $U(1)$ spin liquid state for the original $K$-$\Gamma$ model can be obtained from above $U(1)$ ansatz by performing the inverse of the three-sublattice and six-sublattice transformations.

We have also considered magnetically ordered states but find that they are not energetically favored. Our VMC calculation indicates that there exists a finite $Z_2$ phase at small $|\Gamma|$ which is separated from the $U(1)$ state by a first-order phase transition at $|\Gamma|/|K| \simeq 0.05$, see Fig.~\ref{fig:Chain}.

Our results are qualitatively consistent with previous results\cite{Affleck}, except that our $Z_2$ phase shrinks to a single first-order phase transition point at $\Gamma=0$ in Ref.\onlinecite{Affleck}.\\

%%%%%%%%%%%%%%%%%%%%%%%%%%%%%%%%%%%%%%%%%%%%%%%%%%%%%%%%%%%%%%%%

\ \\


\begin{thebibliography}{99}
\bibitem{Balents}
L. Balents,
%Spin liquids in frustrated magnets,
Nature (London) {\bf 464}, 199 (2010).

\bibitem{zhouyi}
Y. Zhou, K. Kanoda, and T.-K. Ng,
Rev. Mod. Phys. {\bf 89}, 025003 (2017).

\bibitem{XGWen}
M. Levin and X.-G. Wen,
%String-net condensation: a physical mechanism for topological phases
Phys. Rev. B {\bf 71}, 045110 (2005).


\bibitem{Kitaev}
A. Kitaev, Ann. Phys. {\bf 321}, 2 (2006).

\bibitem{Anderson}
I. Affleck, Z. Zou, T. Hsu, and P. W. Anderson,
% SU(2) gauge symmetry of the large-U limit of the Hubbard model
Phys. Rev. B {\bf 38}, 745 (1988).


\bibitem{rjk}
G. Jackeli and G. Khaliullin,
%Mott Insulators in the Strong Spin-Orbit Coupling Limit: From Heisenberg to
%a Quantum Compass and Kitaev Models,
Phys. Rev. Lett. {\bf 102}, 017205 (2009).

\bibitem{rcjk}
J. Chaloupka, G. Jackeli, and G. Khaliullin,
%Kitaev-Heisenberg Model on a Honeycomb Lattice: Possible Exotic Phases in
%Iridium Oxides ${A}_{2}{\mathrm{IrO}}_{3}$.
Phys. Rev. Lett. {\bf 105}, 027204 (2010).

\bibitem{rfgft}
J. M. Fletcher, W. E. Gardner, A. C. Fox, and G. Topping,
%X-Ray, infrared, and magnetic studies of \ensuremath{\alpha}- and
%\ensuremath{\beta}-ruthenium trichloride.
J. Chem. Soc. A 1038 (1967).

\bibitem{YJKim}
K. W. Plumb, J. P. Clancy, L. J. Sandilands, V. V. Shankar, Y. F. Hu, K. S. Burch, H.-Y. Kee, and Y.-J. Kim,
Phys. Rev. B {\bf 90}, 041112(R) (2014).

\bibitem{rsea}
J. A. Sears, M. Songvilay, K. W. Plumb, J. P. Clancy, Y. Qiu, Y. Zhao,
D. Parshall, and Y.-J. Kim,
%Magnetic order in $\ensuremath{\alpha}-{\text{RuCl}}_{3}$:
%A honeycomb-lattice quantum magnet with strong spin-orbit coupling.
Phys. Rev. B {\bf 91}, 144420 (2015).

\bibitem{rjea}
R. D. Johnson, S. C. Williams, A. A. Haghighirad, J. Singleton, V. Zapf, P.
Manuel, I. I. Mazin, Y. Li, H. O. Jeschke, R. Valent\'{\i}, and R. Coldea,
%Monoclinic crystal structure of $\ensuremath{\alpha}-{\mathrm{RuCl}}_{3}$
%and the zigzag antiferromagnetic ground state.
Phys. Rev. B {\bf 92}, 235119 (2015).

\bibitem{rcaoetal}
H.-B. Cao, A. Banerjee, J.-Q. Yan, C. A. Bridges, M. D. Lumsden, D. G.
Mandrus, D. A. Tennant, B. C. Chakoumakos, and S. E. Nagler,
%Low-temperature crystal and magnetic structure of
%$\ensuremath{\alpha}-{\mathrm{RuCl}}_{3}$.
Phys. Rev. B {\bf 93}, 134423 (2016).

\bibitem{incomm}
S. C. Williams, R. D. Johnson, F. Freund, S. Choi, A. Jesche, I. Kimchi, S. Manni, A. Bombardi, P. Manuel, P. Gegenwart, and R. Coldea,
%Incommensurate counterrotating magnetic order stabilized by Kitaev interactions in the layered honeycomb α-Li2IrO3
Phys. Rev. B {\bf 93}, 195158 (2016)

\bibitem{singh}
Y. Singh and P. Gegenwart,
Phys. Rev. B {\bf 82}, 064412 (2010).

\bibitem{XLiu}
X. Liu, T. Berlijn, W.-G. Yin, W. Ku, A. Tsvelik, Y.-J. Kim, H. Gretarsson, Y. Singh, P. Gegenwart, and J. P. Hill,
Phys. Rev. B {\bf 83}, 220403(R) (2011).

\bibitem{ryea}
F. Ye, S.-X. Chi, H.-B. Cao, B. C. Chakoumakos, J. A. Fernandez-Baca,
R. Custelcean, T.-F. Qi, O. B. Korneta, and G. Cao,
%Direct evidence of a zigzag spin-chain structure in the honeycomb lattice:
%A neutron and x-ray diffraction investigation of single-crystal
%Na${}_{2}$IrO${}_{3}$.
Phys. Rev. B {\bf 85}, 180403(R) (2012).

\bibitem{rchoietal}
S. K. Choi, R. Coldea, A. N. Kolmogorov, T. Lancaster, I. I. Mazin, S. J.
Blundell, P. G. Radaelli, Y. Singh, P. Gegenwart, K. R. Choi, S.-W. Cheong,
P. J. Baker, C. Stock, and J. Taylor,
%Spin Waves and Revised Crystal Structure of Honeycomb Iridate
%${\mathrm{Na}}_{2}{\mathrm{IrO}}_{3}$.
Phys. Rev. Lett. {\bf 108}, 127204 (2012).

\bibitem{abra}
M. Abramchuk, C. Ozsoy-Keskinbora, J. W. Krizan, K. R. Metz, D. C. Bell, and F. Tafti,
%Cu2IrO3 : A New Magnetically Frustrated Honeycomb Iridate,
J. Am. Chem. Soc. {\bf 139}, 15371 (2017).


\bibitem{rhliiro}
K. Kitagawa, T. Takayama, Y. Matsumoto, A. Kato, R. Takano, Y. Kishimoto,
S. Bette, R. Dinnebier, G. Jackeli, and H. Takagi,
%A spin–orbital-entangled quantum liquid on a honeycomb lattice,
Nature (London) {\bf 554}, 341 (2018).


% \bibitem{trebst}
% S. Trebst, arXiv:1701.07056.

\bibitem{lefran}
E. Lefrancois, M. Songvilay, J. Robert, G. Nataf, E. Jordan, L. Chaix, C. V. Colin, P. Lejay, A. Hadj-Azzem, R. Ballou, and V. Simonet,
Phys. Rev. B {\bf 94}, 214416 (2016).

\bibitem{bera}
A. K. Bera, S. M. Yusuf, A. Kumar, and C. Ritter, Phys.
Rev. B {\bf 95}, 094424 (2017).

\bibitem{LiYuan}
W. Yao and Y. Li,
%Ferrimagnetism and anisotropic phase tunability by magnetic fields in Na2Co2TeO6
arXiv:1908.09427.


\bibitem{HSKim}
H.-S. Kim and H.-Y. Kee,
%Crystal structure and magnetism in $\alpha$-RuCl$_3$: An ab initio study
Phys. Rev. B {\bf 93}, 155143 (2016).

\bibitem{Li_theo}
W. Wang, Z.-Y. Dong, S.-L. Yu, and J.-X. Li,
%Theoretical investigation of magnetic dynamics in $\alpha$-RuCl$_3$,
Phys. Rev. B {\bf 96}, 115103 (2017).

\bibitem{Jianxin_nuetron}
K. Ran, J. Wang, W. Wang, Z.-Y. Dong, X. Ren, S. Bao, S. Li, Z. Ma, Y. Gan, Y. Zhang, J. T. Park, G. Deng, S. Danilkin, S.-L. Yu, J.-X. Li, and J. Wen,
%Spin-Wave Excitations Evidencing the Kitaev Interaction in Single Crystalline $\alpha$-RuCl$_3$,
Phys. Rev. Lett. {\bf 118}, 107203 (2017).



\bibitem{lukas}
L. Janssen, E. C. Andrade, and M. Vojta,
%Magnetization processes of zigzag states on the honeycomb lattice: Identifying spin models for α-RuCl3 and Na2IrO3
Phys. Rev. B {\bf 96}, 064430 (2017).


\bibitem{Ji_neutron}
S.-H. Do, S.-Y. Park, J. Yoshitake, J. Nasu, Y. Motome, Y. S. Kwon, D. T. Adroja, D. J. Voneshen, K. Kim, T.-H. Jang, J.-H. Park, K.-Y. Choi and S. Ji,
Nat Phys {\bf 13}, 1079 (2017).


\bibitem{rins_field}
A. Banerjee, P. Lampen-Kelley, J. Knolle, C. Balz, A. A. Aczel, B. Winn,
Y. Liu, D. Pajerowski, J.-Q. Yan, C. A. Bridges, A. T. Savici, B. C.
Chakoumakos, M. D. Lumsden, D. A. Tennant, R. Moessner, D. G. Mandrus,
and S. E. Nagler,
%Excitations in the field-induced quantum spin liquid state of
%$\alpha$-RuCl$_3$,
npj Quantum Materials {\bf 3}, 8 (2018).


\bibitem{Bastien_raman_exp}
L. J. Sandilands, Y. Tian, K. W. Plumb, Y.-J. Kim, and K. S. Burch,
% Scattering Continuum and Possible Fractionalized Excitations in α-RuCl3.
Phys. Rev. Lett. {\bf 114}, 147201(2015).

\bibitem{Nasu_raman_theo}
J. Nasu, J. Knolle, D. L. Kovrizhin, Y. Motome, and R. Moessner,
%Fermionic response from fractionalization in an insulating two-dimensional magnet.
Nat Phys {\bf 12}, 912 (2016).

\bibitem{Hirobe_thermal}
D. Hirobe, M. Sato, Y. Shiomi, H. Tanaka, and E. Saitoh,
%Magnetic thermal conductivity far above the Néel temperature in the Kitaev-magnet candidate α-RuCl3
Phys. Rev. B {\bf 95}, 241112(R) (2017).

\bibitem{rzea}
J. Zheng, K. Ran, T. Li, J. Wang, P.-S. Wang, B. Liu, Z.-X. Liu, B. Normand, J. Wen, and W. Yu,
%Gapless Spin Excitations in the Field-Induced Quantum Spin Liquid Phase
%of $\alpha$-RuCl$_3$,
Phys. Rev. Lett. {\bf 119}, 227208 (2017).

\bibitem{thermal}
I. A. Leahy, C. A. Pocs, P. E. Siegfried, D. Graf, S.-H. Do, K.-Y. Choi, B. Normand, and M. Lee,
%Anomalous Thermal Conductivity and Magnetic Torque Response in the Honeycomb Magnet $\alpha$-RuCl$_3$,
Phys. Rev. Lett. {\bf 118}, 187203 (2017).

\bibitem{Masuda}
Y. Kasahara, T. Ohnishi, Y. Mizukami, O. Tanaka, Sixiao Ma, K. Sugii, N. Kurita, H. Tanaka, J. Nasu, Y. Motome, T. Shibauchi, and Y. Matsuda,
%Majorana quantization and half-integer thermal quantum Hall effect in a Kitaev spin liquid
Nature (London) {\bf 559}, 227 (2018).

\bibitem{Baek}
S.-H. Baek, S.-H. Do, K.-Y. Choi, Y. S. Kwon, A. U. B. Wolter, S. Nishimoto, J. van den Brink, and B. B\"{u}chner,
%Evidence for a Field-Induced Quantum Spin Liquid in $\alpha$-RuCl$_3$,
Phys. Rev. Lett. {\bf119}, 037201 (2017).

\bibitem{Jansa}
N. Jan\v{s}a, A. Zorko, M. Gomil\v{s}ek, M. Pregelj, K. W. Kr\"{a}mer, D. Biner, A. Biffin, C. R\"{u}egg, and M. Klanj\v{s}ek,
%Observation of two types of fractional excitation in the Kitaev honeycomb magnet,
Nat Phys {\bf14}, 786 (2018).


\bibitem{Liu_KG}
%Dirac and Chiral Quantum Spin Liquids on the Honeycomb Lattice in a Magnetic Field
Z.-X. Liu and B. Normand,
Phys. Rev. Lett. {\bf 120}, 187201 (2018).



\bibitem{pressure-tuned}
R. Yadav, S. Rachel, L. Hozoi, J. van den Brink, and G. Jackeli,
%Strain- and pressure-tuned magnetic interactions in honeycomb Kitaev materials
Phys. Rev. B {\bf 98}, 121107(R) (2018).


\bibitem{sun_pressure}
Z. Wang, J. Guo, F. F. Tafti, A. Hegg, S. Sen, V. A. Sidorov, L. Wang, S. Cai, W. Yi, Y. Zhou, H. Wang, S. Zhang, K. Yang, A. Li, X. Li, Y. Li, J. Liu, Y. Shi, W. Ku, Q. Wu, R. J. Cava, and L. Sun,
Phys. Rev. B {\bf 97}, 245149 (2018).


\bibitem{yu_pressure}
Y. Cui, J. Zheng, K. Ran, J. Wen, Z.-X. Liu, B. Liu, W. Guo, and W. Yu,
Phys. Rev. B {\bf 96}, 205147 (2017).



\bibitem{Bastien_pressure}
G. Bastien, G. Garbarino, R. Yadav, F. J. Martinez-Casado, R. B. Rodr\'{\i}guez, Q. Stahl, M. Kusch,
S. P. Limandri, R. Ray, P. Lampen-Kelley, D. G. Mandrus, S. E. Nagler, M. Roslova, A. Isaeva, T. Doert, L. Hozoi, A. U. B. Wolter, B. B\"{u}chner, J. Geck, and J. van den Brink,
Phys. Rev. B {\bf 97}, 241108(R) (2018).

\bibitem{Mei_pressure}
G. Li, X. Chen, Y. Gan, F. Li, M. Yan, S. Pei, Y. Zhang, L. Wang, H. Su, J. Dai, Y. Chen, Y. Shi, X. Wang, L. Zhang, S. Wang, D. Yu, F. Ye, J.-W. Mei, and M. Huang,
% Raman spectroscopy evidence for dimerization and Mott collapse in α-RuCl3 under pressures,
Phys. Rev. Materials {\bf3}, 023601 (2019).

\bibitem{HermannLiIrO}
V. Hermann, J. Ebad-Allah, F. Freund, A. Jesche, A. A. Tsirlin, P. Gegenwart, and C. A. Kuntscher,
%Optical signature of the pressure-induced dimerization in the honeycomb iridate $\alpha$-Li$_2$IrO$_3$
Phys. Rev. B {\bf 99}, 235116 (2019).


\bibitem{zigchain}
V. Hermann, S. Biswas, J. Ebad-Allah, F. Freund, A. Jesche, A. A. Tsirlin, M. Hanfland, D. Khomskii,
P. Gegenwart, R. Valent\'{\i}, and C. A. Kuntscher,
%Pressure-induced formation of rhodium zigzag chains in the honeycomb rhodate Li2RhO3
Phys. Rev. B {\bf 100}, 064105 (2019).


\bibitem{singleQ}
J. G. Rau, E. K.-H. Lee, and H.-Y. Kee,
Phys. Rev. Lett. {\bf 112}, 077204 (2014).

\bibitem{Pollmann}
M. Gohlke, G. Wachtel, Y. Yamaji, F. Pollmann, and Y. B. Kim,
%Quantum spin liquid signatures in Kitaev-like frustrated magnets
Phys. Rev. B {\bf 97}, 075126 (2018).

\bibitem{Kee}
A. Catuneanu, Y. Yamaji, G. Wachtel, Y. B. Kim, and H.-Y. Kee,
%Path to stable quantum spin liquids in spin-orbit coupled correlated materials
npj Quantum Materials {\bf 3}, 23 (2018).

\bibitem{PKSL}
J. Wang, B. Normand, and Z.-X. Liu,
Phys. Rev. Lett. {\bf 123}, 197201 (2019).


\bibitem{igg}
X.-G. Wen, Phys. Rev. B {\bf 65}, 165113 (2002).

\bibitem{You_PSG}
Y.-Z. You, I. Kimchi, and A. Vishwanath,
Phys. Rev. B {\bf 86}, 085145 (2012).




\bibitem{Motome}
Y. Motome and J. Nasu,
%Hunting Majorana Fermions in Kitaev Magnets
arXiv:1909.02234.




\bibitem{Affleck}
W. Yang, A. Nocera, T. Tummuru, H.-Y. Kee, and I. Affleck,
%Phase diagram of the spin-1/2 Kitaev-Gamma chain and emergent SU(2) symmetry
arXiv:1910.14304.


\bibitem{Wen_class}
X.-G. Wen,
Phys. Rev. B {\bf 85}, 085103 (2012).

\bibitem{Kitaev_class}
%A. Kitaev, AIP Conf.Proc. {\bf 1134}, 22 (2009).
A. Kitaev,
in Advances in Theoretical Physics: Landau Memo- rial Conference, edited by V. Lebedev and M. Feigel’man, AIP Conf. Proc. No. 1134 (AIP, New York, 2009), p. 22.

\bibitem{ZhaoYuXin}
Y.-X. Zhao, A. P. Schnyder, and Z.-D. Wang,
%Unified theory of PT and CP invariant topological metals and nodal superconductors
Phys. Rev. Lett. {\bf 116}, 156402 (2016).


\bibitem{YouXu_PRX18} Y.-Z. You, Y.-C. He, C. Xu, and A. Vishwanath,
Phys. Rev. X {\bf 8}, 011026 (2018).

\bibitem{SymmetricMass}
In principle,  interactions between the cones may also gap out the cones without breaking any symmetry. However, this only occurs at strong interactions and even so the gap opening process is still a phase transition \cite{YouXu_PRX18}.

\bibitem{fluxcrystal}
S.-S. Zhang, C. D. Batista, and G. B. Hal\'{a}sz,
%Toward Kitaev's sixteenfold way in a honeycomb lattice model
arXiv:1910.00601.


\bibitem{Song}
X.-Y. Song, Y.-Z. You, and L. Balents,
Phys. Rev. Lett. {\bf 117}, 037209 (2016).

\bibitem{note}
As shown in Ref.\onlinecite{PKSL}, the Chern number may change when increasing the intensity of the magnetic field.


\bibitem{Qianghua}
S. Liang, M.-H. Jiang, W. Chen, J.-X. Li, and Q.-H. Wang,
%Intermediate gapless phase and topological phase transition of the Kitaev model in a uniform magnetic field
Phys. Rev. B {\bf 98}, 054433 (2018).

\bibitem{Nasu}
J. Nasu, Y. Kato, Y. Kamiya, and Y. Motome,
%Successive Majorana topological transitions driven by a magnetic field in the Kitaev model
Phys. Rev. B {\bf 98}, 060416(R) (2018).

\bibitem{Hemele}
A. M. Essin and M. Hermele,
%Classifying fractionalization: symmetry classification of gapped Z2 spin liquids in two dimensions
Phys. Rev. B {\bf 87}, 104406 (2013).
%M. Hemele, 2010,


\bibitem{RanYing}
A. Mesaros and Y. Ran,
%Classification of symmetry enriched topological phases with exactly solvable models
Phys. Rev. B {\bf 87}, 155115 (2013).


\bibitem{ChengMengQiYang}
Y. Qi, M. Cheng, and C. Fang,
%Symmetry fractionalization of visons in Z2 spin liquids
arXiv:1509.02927.


\bibitem{ChengWang}
M. Barkeshli, P. Bonderson, M. Cheng, and Z. Wang,
%Symmetry Fractionalization, Defects, and Gauging of Topological Phases
Phys. Rev. B {\bf100}, 115147 (2019).
%M. Bakashili, Meng Cheng, Z, Wang, PRB, 2019


\bibitem{ChenXie}
X. Chen, X. Chen, F. J. Burnell, A. Vishwanath, and L. Fidkowski,
%Anomalous symmetry fractionalization and surface topological order
Phys. Rev. X {\bf 5}, 041013 (2015).


\bibitem{NiNi}
J. Xing, H. Cao, E. Emmanouilidou, C. Hu, J. Liu, D. Graf, Arthur P. Ramirez, G. Chen, and N. Ni,
arXiv:1903.03615.

\bibitem{Sala}
G. Sala, M. B. Stone, B. K. Rai, A. F. May, D. S. Parker, G. B. Hal\'{a}sz, Y.-Q. Cheng, G. Ehlers, V. O. Garlea, Q. Zhang, M. D. Lumsden, and A. D. Christianson,
arXiv:1907.10627.


\bibitem{JQYan}
J.-Q. Yan, S. Okamoto, Y. Wu, Q. Zheng, H.-D. Zhou, H.-B. Cao, and M. A. McGuire,
Phys. Rev. Materials {\bf 3}, 074405 (2019).


\bibitem{SSZhang}
S.-S. Zhang, Z. Wang, G. B. Hal\'{a}sz, and C. D. Batista,
% Vison Crystals in an Extended Kitaev Model on the Honeycomb Lattice
Phys. Rev. Lett. {\bf 123}, 057201 (2019).


\bibitem{HYKee}
P. P. Stavropoulos, A. Catuneanu, and H.-Y. Kee,
Phys. Rev. B {\bf 98}, 104401 (2018).



\end{thebibliography}
\end{document}